%% file: Hamdan_Fazi_Focusing_Phenomena_in_Linear_Inverse_Problems_in_Acoustics_v2.tex
\pdfoutput=1
\documentclass[3p, twoside, onecolumn, 11pt, sort&compress, times, fleqn]{elsarticle}
\makeatletter
\def\ps@pprintTitle{%
	\let\@oddhead\@empty
	\let\@evenhead\@empty
	\def\@oddfoot       {\hbox to \textwidth%
		{\ifnopreprintline\relax\else
			\@myfooterfont%
			\ifx\@elsarticlemyfooteralign\@elsarticlemyfooteraligncenter%
			\hfil\@elsarticlemyfooter\hfil%
			\else%
			\ifx\@elsarticlemyfooteralign\@elsarticlemyfooteralignleft%
			\@elsarticlemyfooter\hfill{}%
			\else%
			\ifx\@elsarticlemyfooteralign\@elsarticlemyfooteralignright%
			{}\hfill\@elsarticlemyfooter%
			\else%
			Preprint submitted to \ifx\@journal\@empty%
			Elsevier%
			\else\@journal\fi\hfill21st October 2020\fi%
			\fi%
			\fi%
			\fi%
		}
	}%
	\let\@evenfoot\@oddfoot}
\makeatother
\usepackage{lineno}
\usepackage{hyperref}
\hypersetup{
	colorlinks = true,
	linkcolor = blue,
	anchorcolor = blue,
	citecolor = blue,
	filecolor = blue,
	urlcolor = blue
}
\usepackage{amsfonts}
\usepackage{amsmath}
\usepackage{upgreek}
\usepackage[british]{babel}
\usepackage{csquotes}
\usepackage{mathtools}
\usepackage{caption}
\usepackage{subcaption}
\usepackage{graphicx}
\usepackage{bm}
\usepackage{color} % in Tex live as 'graphics'
\usepackage{soul}
\usepackage[capitalise]{cleveref}

\crefname{appendix}{}{}

\input{MathCommands}

\journal{Journal of Sound and Vibration}

\begin{document}
\begin{frontmatter}

\title{Focusing Phenomena in Linear Discrete Inverse Problems \\in Acoustics}

\author{Eric C. Hamdan\corref{cor1}}
\cortext[cor1]{Corresponding author.}
\ead{e.hamdan@soton.ac.uk}
\author{Filippo Maria Fazi}
\ead{filippo.fazi@soton.ac.uk}

\address{Institute of Sound and Vibration Research, University of Southampton, University Road, Southampton, SO17 0LG, United Kingdom}

\begin{abstract}
The focusing operation inherent to the linear discrete inverse problem is formalised. The development is given in the context of sound-field reproduction where the source strengths are the inverse solution needed to recreate a prescribed pressure field at discrete locations. The behaviour of the system is fundamentally tied to the amount of acoustic crosstalk at each control point as a result of the focusing operation inherent to the pseudoinverse. The maximisation of the crosstalk at just one point leads to linear dependence in the system. On the other hand, its minimisation leads to the ideal focusing state wherein the sources can selectively focus at each point, while a null is created at all other points. Two theoretical case studies are presented that demonstrate ideal and super ideal focusing, wherein the latter the condition number is unitary. First, the application of binaural audio reproduction using an array of loudspeakers is examined and several cases of ideal focusing are presented. In the process, the Optimal Source Distribution is re-derived and shown to be a case of super ideal focusing. Secondly, the application of recreating multiple sound zones is examined using a uniform linear array. The conditions are derived to achieve ideal focusing at control points positioned arbitrarily in the far-field. In all cases, the ability to maintain ideal focusing as a function of frequency requires proportional changes in the source or control point geometry.
\end{abstract}


\end{frontmatter}

%% main 
\section{Introduction}
Linear discrete inverse problems (LDIPs) have a wide range of applicability in acoustics, for example, they have been studied and exploited in the applications of acoustic holography, e.g., \cite{Veronesi1989, Williams2001, Chardon2012}, source strength identification, e.g. \cite{Nelson2000, Kim2004, Holland2012, Holland2013}, sound-field reproduction, e.g., \cite{Bauck1992,Kirkeby1993, Nelson1995, Fazi2010,Olivieri2016, Galvez2019, Hoffmann2019}, sound-field acquisition, e.g., \cite{Hoffmann2015}, active noise control, e.g., \cite{Nelson1992a, Nelson1996multichannel, House2020}, aeroacoustics, e.g., \cite{Pignier2017, Leclere2017}, and ultrasound, e.g., \cite{Tanter2000, Tanter2001}. Summaries of inverse problems in acoustics given by Nelson and Wu in \cite{Wu2008} and \cite{Nelson2001a}, respectively, still hold relevance today. The commonality between these applications is the underlying mathematics of the problem that has been studied in great detail through various lenses, e.g., \cite{Hansen1987, Meyer2001, AdiBen-Israel2003}. On the other hand, the differences in these applications lie in the intrinsic physical properties of each problem which then gives rise to unique physical interpretations to the same mathematics. The concern here lies with a physical interpretation of the inverse solution in the context of sound-field reproduction at discrete points in space. Namely, it is assumed that the analysis takes place in the frequency domain, i.e., $\mathbb{C}$, and that the coefficient matrix of the system, referred to as the \textit{plant} matrix, is composed of electro-acoustic transfer functions between a finite set of $L$ radiating and $M$ receiving acoustic transducers, e.g., the transfer functions relating a set of loudspeakers and microphones. A physical analysis framework is developed that relates the frequency-dependent stability of the plant matrix pseudoinverse, its condition number, and the necessary requirements for optimal conditioning to the focusing phenomena in the inverse solution. However, since the transfer functions can be any complex numbers, the general results can be applied to any LDIP of arbitrary complex data.

The first part of the present work forms the foundation of the analysis framework and puts specific emphasis on rigorously establishing the focusing phenomena in the pseudoinverse, both mathematically and conceptually. Embedded in the pseudoinverse is the abstract concept of focusing filters, also recognised as conjugation or time-reversal, e.g., \cite{Fink1997}. In the beamforming literature these filters are often associated with the \textit{beamforming gain}, e.g., \cite{Theodoridis2013, Yan2019}. These filters may be \textit{temporally matched} filters, e.g., \cite{Turin1960}, and are \textit{spatially matched}, e.g., \cite{Tanter2000}, where the latter is a concept introduced in the time-reversal literature. Indeed, researchers studying time-reversal techniques have been active in formalising the links between the focusing operator used by \textit{time-reversal mirrors} and the inverse solution, e.g., Tanter et.\ al \cite{Tanter2000, Tanter2001}. In \cite{Montaldo2004} and \cite{Vignon2006} it was shown that iterative time-reversal techniques can converge to the matrix inverse, shedding further light on this relationship. In contrast, the emphasis here is on formalising how the focusing operation underpinning the time-reversal mirror determines the condition number of the plant matrix and the amplification factor of its pseudoinverse on the input signals, i.e., stability. These intrinsic quantities are of paramount importance in practical applications and have largely been studied and interpreted in the past through other means such as the singular value decomposition (SVD) e.g., \cite{Nelson2000, Fazi2010, Hamdan2021}. A novel, yet intrinsically related, physical interpretation of these quantities is developed here. 

The pseudoinverse is conceptually decomposed into two stages: a focusing and inversion stage. This concept has been previously observed in the literature, e.g., by Gauthier et.\ al in \cite{Gauthier2011}, exploited by G\'alvez et.\ al in \cite{Galvez2019} from a structural stand point to increase computational efficiency in real-time sound-field control, and physically analysed by the authors in \cite{Hamdan2021} in the context of multichannel CTC systems. Here the physical consequences of this structure are explored for the general $M\times L$ linear system. In the first stage, matched focusing filters are applied to maximise the pressure at each control point. The forward focusing operation, or operator, is represented by the Gram matrix associated with the plant matrix, e.g., \cite{Fink1993}. However, due to the physical limitations imposed by spatial sampling, e.g., \cite{Fazi2010}, the focusing operation generally results in unwanted leakage at other points while focusing at a given control point, referred to as \textit{focusing crosstalk}. The focusing crosstalk elements are specifically the off-diagonals of the Gram matrix. The crosstalk can also originate from any reflections or unwanted noise at the time of measurement as the acoustic environment is typically not anechoic (this is why focusing crosstalk is not referred to generally as side lobes or grating lobes). In the second stage the exact inverse of the focusing operation eliminates the crosstalk and normalises the target point pressures to one. This stage is the inverse of the Gram matrix. It is shown that the system becomes linearly dependent when the focusing crosstalk is maximised at any one control point. On the other hand, the consequence of natural minimisation of the crosstalk at all points is shown to be a necessary yet insufficient requirement for a system to achieve a unitary condition number, i.e., become optimally conditioned.

The minimisation of the focusing crosstalk has been studied previously by Prada in \cite{Prada1994, Prada1996} in the context of iterative time-reversal mirrors assuming ideally (or well) resolved sources (also referred to by Prada as the ability to perform \textit{selective focusing}). A set of ideally resolved sources was theoretically defined to mean, at a given frequency, the scatterers can be focused on by an array of transducers through means of the iterative time reversal mirror without imparting any leakage to other scatterers in the same space. In the mathematical sense a set of ideally resolved sources mean that the columns and rows of the Gram matrix are pairwise orthogonal, i.e, the Gram matrix is diagonal, and thus the focusing crosstalk is naturally zero. Prada recognised that when multiple sources are ideally resolved the eigenvectors of the Gram matrix provide the exact inputs to the iterative time-reversal mirror that allow for selective focusing at multiple points \cite{Prada1996}. Here the definition of selective focusing is sharpened to mean a pressure null is created exactly at the other control points while focusing on one (for all points considered), but does not imply any specific spatial extent of the null or attempt to describe the field at other points in space. Furthermore, the nomenclature introduced by Prada to denote ideally resolved sources is married into the single term \textit{ideal focusing}.

Not long after Prada had explored ideal focusing in time-reversal applications, Nelson and Kim, in \cite{Nelson1999, Nelson2001b, Kim2003}, had determined and studied two analytical cases when the $M\times L$ plant matrix achieves a unitary condition number, one for a model uniform linear array (ULA) and the other for the uniform rectangular array (URA). In each case, the sources were assumed to be simple sources in either 2D or 3D space and the control points were assumed to placed in the far-field at positions that aligned exactly with the regularly spaced source array geometry. The results of their analysis established a inter-element spacing for which the plant matrix becomes a scalar multiple of a block Fourier matrix, or discrete Fourier transform (DFT) matrix, whose columns and rows are pairwise orthogonal and condition number is unity, i.e., optimally conditioned \cite{Kim2003}. A more recent case of when the plant can be described by a scalar multiple of a block Fourier matrix was established in the aeroacoustics context by Hirono in \cite{Hirono2018}. Thus, Nelson and Kim gave what seems to be the first analytical example of what is referred to herein as \textit{super ideal focusing}, the state wherein ideal focusing is achieved with a unitary condition number. Furthermore, they gave the geometry restrictions in two theoretical cases that allow for super ideal focusing, which is a topic of central importance in this work. However, Nelson and Kim did not interpret the results in terms of focusing phenomena nor was the explicit connection made to Prada's previous analysis on ideally resolved sources. 

Around the same time Nelson and Kim were exploring analytical cases of optimally conditioned systems, Takeuchi and Nelson developed the theory of Optimal Source Distribution (OSD), e.g., \cite{Takeuchi2002, Takeuchi2007, Takeuchi2008}. Using a monopole in free-field and shadowless head model, those researchers analysed the two-channel crosstalk cancellation (CTC) system, i.e., the $2\times 2$ system, for the case of a symmetric geometry. In the same way that Nelson and Kim established a geometry requirement for ULAs and URAs to maintain optimal conditioning, Takeuchi and Nelson established that the angular span of the symmetric source pair, under the stated acoustic model, must vary with frequency in order to maintain the unitary condition number \cite{Takeuchi2002}. Their analysis relied directly on the SVD of the plant and inspection of the singular values to determine how the loudspeaker geometry must change in order to maintain optimal conditioning \cite{Takeuchi2002}. Thus, Takeuchi and Nelson seem to have provided the first example of a physical analysis that led to an explicit frequency-dependent loudspeaker geometry method for maintaining optimal conditioning in sound-field control based on a modal analysis. Yet, in light of the current understanding, both groups of researchers were describing the exact same phenomena in different instances. It was Prada that identified the core and common element of ideal focusing, yet, these findings remained disconnected to the general behaviour of the LDIP. Furthermore, Prada's findings remained conceptually disparate from the ideal focusing cases indirectly established by Nelson and Kim, and Takuechi and Nelson, who interpreted the results through the ability to achieve optimal conditioning. A specific contribution of this work is to establish the rigorous conditions under which ideal focusing leads to the optimally conditioned system, i.e., super ideal focusing, with worked examples, thus tying these two branches of work together in a common analysis.

The minimisation of the focusing crosstalk, albeit indirect, can also be found in the work of researchers exploring optimisation algorithms for determining the optimal placement of loudspeaker and control points in sound-field reproduction. For example, in \cite{Asano1999} Asano et.\ al explored an algorithm based on the Gram-Schmidt orthogonalisation process which sought to maximise the linear independence of the plant based from a candidate set of known transfer functions (this approach was revisited recently by Koyama et.\ al in \cite{Koyama2020} wherein an extensive summary on geometry optimisation techniques for sound-field control are given). However, the underlying physical behaviour was left unexplored in these studies and they were not related to the previously mentioned findings on loudspeaker and control points geometries that lead to optimal conditioning nor to the concept of selective focusing, where these facets of the problem are all linked by the state of Gram matrix. 

Thus, it appears that there exists a need to unify these direct and indirect observations of the focusing phenomena within the LDIP into a single development. The goal of this work is to address this task, interpret the general results, and to provide analytical case studies that communicate the general findings in an intuitive fashion. The novel contributions of this work are summarised as follows:
\begin{enumerate}
	\item A novel analysis of the LDIP for underdetermined systems is given that centres around the phenomena associated with the focusing operation within the pseuodinverse. The concept of focusing crosstalk is formalised and it is shown how the system's linear independence is tied directly to the focusing behaviour. The conditions for instability and maximum stability are formalised in terms of the focusing operation. This, in turn, establishes ideal focusing.
	\item A novel modal analysis of the ideal focusing state leads to the derivation of the associated singular system. Knowledge of the singular system establishes the explicit connection between the ideal focusing state, the system condition number, and pseudoinverse amplification factor. This analysis culminates in the general and exact requirements for super ideal focusing, i.e., an optimally conditioned system. Furthermore, a physical interpretation of the optimally conditioned system is given in light of the focusing phenomena.
	\item 
	A case study is presented that examines the focusing operation in the application of CTC. The explicit physical requirements are derived for ideal focusing in the two-channel CTC problem, both for arbitrary transfer functions and for a monopole in free-field and shadowless head acoustic model (that allows for asymmetric cases). In the process, OSD is re-derived through the present framework and it is shown to be special case of super ideal focusing when the system is symmetric and a case of when compensated delay-and-sum beamforming solves the inverse sound-field reproduction problem. The results generalise the findings of Takeuchi and Nelson established in \cite{Takeuchi2002}. Finally, conditions for super ideal focusing with a far-field symmetric multichannel CTC geometry are established. The conditions are shown to rely on the number of loudspeakers.
	\item
	A second case study is presented that derives the conditions for ideal focusing considering a ULA formed of simple sources and control points in the far-field, where the control point directions can be in non-regular layouts. This extends the previous work of Nelson and Kim in \cite{Nelson2001b, Kim2003} and establishes more general conditions to achieve optimal conditioning in the case of the ULA, specifically in regards to the number of possible control point directions in which super ideal focusing can occur. It is shown that this number is directly dependent on the dimensions of the ULA.
\end{enumerate}

The remainder of this work is outlined as follows: In \cref{sec:J3-Discrete Acoustic Inverse Problem}, the LDIP is introduced in the context of the sound-field reproduction problem. In \cref{sec:J3-Focusing Phenomenon in the Inverse Solution}, the focusing operation in the pseudoinverse is established for underdetermined systems. In \cref{sec:J3-Ideal Focusing}, the ideal focusing state is formalised followed by its modal decomposition in \cref{sec:J3-Modal Analysis of the Ideal Focusing State}. The modal analysis culminates in \cref{sec:J3-Super Ideal Gram} where the super ideal focusing state is formalised. In \cref{sec:J3-Case-Studies}, theoretical case studies of ideal focusing in sound-field reproduction are presented. In \cref{sec:J3-CTC}, the focusing behaviour of the CTC system ($2\times L$, where $L\geq2$) is analysed and several examples of super ideal focusing are established. In \cref{sec:J3-Sound-Zones}, a $M\times L$ case is examined. A ULA composed of plane wave sources and with control points in arbitrary positions in the far-field is considered and the conditions for super ideal focusing are established for non-regular control point directions. Finally, concluding remarks are given in \cref{sec:J3-Conclusion}.
\section{The Linear Discrete Inverse Problem in Sound-field Reproduction}
\label{sec:J3-Discrete Acoustic Inverse Problem}
\begin{figure}
	\centering
	\includegraphics[width=0.55\linewidth]{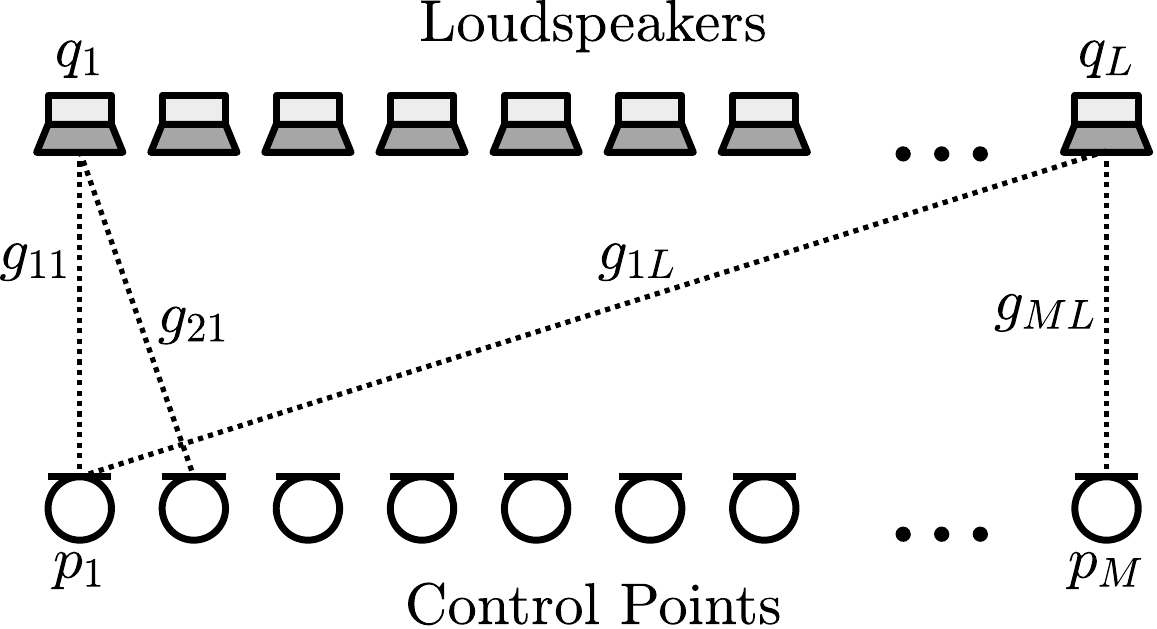}
	\caption{$M\leq L$ sound-field control geometry.}
	\label{fig:mimo-system-geometry}
\end{figure}

As depicted in \cref{fig:mimo-system-geometry}, the following considers the use of $L$ loudspeakers at positions $\sL{l}\inRv{3}$, where $l=1, \dots, L$, to control the acoustic pressure at $M$ field positions $\xM{m}\inRv{3}$, where $m=1, \dots, M$. The task at hand is to determine the inputs to the loudspeakers that are necessary to reproduce a prescribed set of target pressures at the control points. The general problem can be stated in the frequency domain as the linear system
\begin{equation}
\label{eq:discrete-inverse-problem}
\plant(\omega)\solution(\omega)\overset{!}{=}\desired(\omega) ,
\end{equation}
where $\ieq$ denotes \enquote{\dots ideally equal to \dots},
$
\desired(\omega)= \begin{bmatrix}
d_1(\omega) & \dots & d_M(\omega)
\end{bmatrix}\trans\inC{M}
$ is the vector of target pressures intended to be reproduced at the control points, 
\begin{equation}
\plant(\omega) = \begin{bmatrix}
\gM{1}\trans(\omega)\\
\vdots\\
\gM{M}\trans(\omega)
\end{bmatrix}\inCm{M}{L},
\end{equation}
is the plant matrix where
$
\bv{g}_m(\omega)= \begin{bmatrix}
g_{m1}(\omega) & \dots & g_{mL}(\omega)
\end{bmatrix}\trans\inC{L}
$ 
is the $m$th vector of acoustic transfer functions between the $L$ loudspeakers sources and the $m$th control point, and $
\solution(\omega) = \begin{bmatrix}
q_1(\omega) & \dots & q_L(\omega)
\end{bmatrix}\trans\inC{L}
$
is the vector of unknown source strengths intended as input to the loudspeakers. The source strengths ideally result in reproduction of the target pressures at the control points given $\plant(\omega)$. Application of a chosen vector of source strengths in the forward problem leads to
\begin{equation}
\label{eq:J3-forward-problem}
\plant(\omega)\solution(\omega) = \pressures(\omega),
\end{equation}
where 
$
\pressures(\omega) = \begin{bmatrix}
p_1(\omega) & \dots & p_M(\omega)
\end{bmatrix}\trans\inC{M}
$
is the vector of reproduced pressures at the control points. The difference between the target pressures and the reproduced pressures is the residual vector
\begin{align}
\bv{r}(\omega) = \desired(\omega) -\pressures(\omega).
\end{align}
Here the efficacy of a chosen vector of source strengths is measured by the Euclidean distance between the target and reproduced pressures, i.e., $\norm{\bv{r}(\omega)}$, where $\norm{\cdot}$ denotes the $\ell_2$ norm (although other norms could be used).

A \textit{solution} is defined as a vector of source strengths for which the residual vector
equals the zero vector, i.e., $\norm{\bv{r}(\omega)} = 0$. The existence of a solution or an infinite number of solutions to \cref{eq:discrete-inverse-problem} depends on the rank of the plant matrix, i.e., $\rank{\plant(\omega)}$, and the rank of its augmented matrix, i.e., $\rank{\plant(\omega) | \desired(\omega)}$, according to the Rouche-Capelli theorem \cite{Shafarevich2012}. There is at least one solution if and only if the ranks of these two matrices are equal, and that is to say $\desired(\omega)\in\range{\plant(\omega)}$, where $\range{\cdot}$ denotes the range of a matrix \cite{AdiBen-Israel2003}. When $\desired(\omega)\in\range{\plant(\omega)}$ the system is said to be \textit{consistent} \cite{AdiBen-Israel2003}. If the target vector does not lie in the range of the plant matrix the system is said to be \textit{inconsistent} and only an approximate solution can be obtained (this corresponds to when the ranks of the plant matrix and its augmented matrix are not equal) \cite{AdiBen-Israel2003}. 

The following considers solutions, whether exact or approximate, that minimise the $\ell_2$ norm of the residual vector, i.e., a solution is sought such that
\begin{equation}
\label{eq:J3-least-squares-problem}
\solution_0(\omega) \defined \mathop{\mathrm{argmin}}\limits_{\solution(\omega)} \norm{\bv{r}(\omega)}.
\end{equation}
This class of solutions are referred to as \textit{least-squares} solutions, e.g., \cite{AdiBen-Israel2003}. \cref{eq:J3-least-squares-problem} is trivially minimised when the solution is exact. When the solution is approximate, it is a \textit{best-fit} solution in that it minimises the Euclidean distance between the two vectors. 

This analysis further restricts the solution space by assuming $M\leq L$, i.e., the number of controls points is equal to or less than the number of loudspeakers. When $M<L$ there are more unknowns than equations and the system is said to be underdetermined \cite{Golub2013}. If the system is underdetermined \textit{and} consistent, then there are an infinitude of exact solutions \cite{Burrus2013}. In this case, the system has $L-\rank{\plant(\omega)}$ extra degrees of freedom. Of these infinite exact solutions the one of minimum $\ell_2$ norm is given by
\begin{equation}
\label{eq:J3-pseudo-inverse-solution}
\solution_0(\omega) = \filters(\omega)\desired(\omega),
\end{equation}
where $\filters(\omega) \defined \plant^+(\omega)$ is defined as the Moore-Penrose pseudoinverse, e.g., \cite{AdiBen-Israel2003}. \cref{eq:J3-pseudo-inverse-solution} is referred to as the \textit{minimum norm least-squares} solution as it simultaneously satisfies \cref{eq:J3-least-squares-problem} and minimises the $\ell_2$ norm of the solution, i.e., $\norm{\bv{q}_0(\omega)}\leq\norm{\bv{q}(\omega)}$ for all possible solutions \cite{AdiBen-Israel2003}. Note also that 
$
\norm{\solution_0(\omega)}=\norm{\filters(\omega)\desired(\omega)}\leq\fronorm{\filters(\omega)}\norm{\desired(\omega)}
$
which implies, of all matrices in place of $\filters(\omega)$ that could equally satisfy \cref{eq:J3-identity-eq}, the pseudoinverse is of minimum Frobenius norm (denoted by $\fronorm{\cdot}$). The solution given by \cref{eq:J3-pseudo-inverse-solution} will be referred to throughout as the \textit{inverse solution}.  Furthermore, when the plant matrix is square and non-singular only one exact solution to \cref{eq:discrete-inverse-problem} exists given by the standard matrix inverse, i.e.,
$
M=L\rightarrow\solution_0(\omega)=\plant(\omega)^{-1}\desired(\omega)
$,
where $(\cdot)^{-1}$ denotes the matrix inverse.

Note that in practice a modelling delay $\delay$, where $\Delta t$ is a delay in seconds, must be included to ensure causality, e.g., \cite{Nelson1992}. Inclusion of the modelling delay results in
\begin{equation}
\label{eq:J3-identity-eq}
\plant(\omega)\filters(\omega)\delay = \bv{I}\delay,
\end{equation}
where $\bv{I}\inRm{M}{M}$ is an identity matrix. In this context, each column of the identity matrix represents the reproduction of a delayed impulse at one control point while a null is reproduced at all other points. Substitution of the inverse solution \cref{eq:J3-pseudo-inverse-solution} into \cref{eq:J3-forward-problem}, including the modelling delay, results in reproduced pressures
\begin{equation}
\pressures(\omega) = \desired(\omega)\delay,
\end{equation}
i.e., the pressures measured at the control points will be a delayed copy of the target signals. Furthermore $\norm{\bv{r}(\omega)} = 0$ (when neglecting the delay), i.e, the solution is exact. 

Note that the following will omit the delay $\delay$ and dependence on $\omega$ for the sake of brevity.
\section{Focusing Phenomena in the Inverse Solution}
\label{sec:J3-Focusing Phenomenon in the Inverse Solution}
The following examines the pseudoinverse for underdetermined systems in greater detail in the context of the discrete loudspeaker reproduction problem described in \cref{sec:J3-Discrete Acoustic Inverse Problem}. The underlying focusing phenomena are formalised and this subsequently leads to the concept of the ideal focusing state introduced in \cref{sec:J3-Ideal Focusing}.

Tantamount to the analysis is understanding the two primary stages of the pseudoinverse. These stages can be seen more clearly in an expanded form. When $M\leq L$ the pseudoinverse is given by
\begin{equation}
\label{eq:pinv-underdetermined-adj-determ}
\pinv = \underbrace{\plant\herm}_{\text{focusing filters}}\underbrace{\inv{\ggh}}_{\text{inverse Gram}} = \plant\herm\frac{\adj{\ggh}}{\determ{\ggh}},
\end{equation}
where $\determ{\cdot}$ denotes the determinant and $\adj{\cdot}$ denotes the adjugate of a matrix. 

\cref{eq:pinv-underdetermined-adj-determ} states that the pseudoinverse can be treated as two distinct operations: the focusing stage and the inversion stage. The focusing stage refers the application of the focusing filters $\plant\herm$ while the inversion stage is given by the inverse of the Gram matrix, where the Gram matrix is $\ggh$, e.g., \cite{Rozanski2017}. Note that \enquote{the Gram} will be used throughout to refer to the Gram matrix.
\subsection{Focusing Stage}
\label{sec:J3-Focusing Stage}
The focusing stage serves to maximise pressure locally at the control points, i.e., to create the delayed impulses along the diagonal of the identity matrix in \cref{eq:J3-identity-eq}. The focusing is accomplished through the time-reversed filters $\plant\herm=\begin{bmatrix}
\gM{1}^* & \dots & \gM{M}^*
\end{bmatrix}$, e.g., \cite{Fink1993}, referred to herein as \textit{focusing filters}. A key point is that the focusing stage is concerned only with said operation and does not attempt to create simultaneous nulls at the other control points, i.e., the off-diagonal zeros in the identity matrix in \cref{eq:J3-identity-eq}.

The behaviour of the focusing filters is understood directly by the $M\times M$ Gram:
\begin{equation}
\Gram \defined \ggh = \begin{bmatrix}
\gram{1}{1} & \gram{1}{2} & \dots & \gram{1}{M} \\[10pt]	
\gram{2}{1} & \gram{2}{2} & \dots & \gram{2}{M}\\[10pt]
\vdots & \vdots & \ddots & \vdots\\[10pt]
\gram{M}{1} & \gram{M}{2} & \dots & \gram{M}{M}
\end{bmatrix},
\end{equation}
where $\gram{i}{j}\defined \gM{j}\herm\gM{i}, ~i,j=1,\dots,M$. Physically speaking, the Gram can be thought of as the forward problem having applied only the focusing filters $\plant\herm$. 
The $m$th column of the Gram represents the reproduced pressures at the control points as a result of focusing at the $m$th control point. The entries along the diagonal $\gram{i}{i}=\norm{\gM{i}}^2$ are the focus point pressures (these are normalised to one in the inversion step) while the off-diagonal entries $\gram{i}{j}$ for $i\neq j$ are the so-called \textit{focusing crosstalk} terms. All of these terms are the complex pressure that is measured at each control point as a result of focusing at each point individually. The magnitude of each focusing term can be written as
\begin{equation}
\label{eq:focusing-xt}
\abs{\gram{i}{j}} = \norm{\gM{i}}\norm{\gM{j}}\cos\Theta_{ij},
\end{equation}
where $0\leq \Theta_{ij}\leq \pi/2$ is the so-called \textit{Hermitian angle}, e.g., \cite{Scharnhorst2001}. \cref{eq:focusing-xt} has a simple geometric interpretation as the dot product between two vectors on the real plane whose angular separation does not extend beyond ninety degrees. The magnitude of the focusing crosstalk is thus one measure of collinearity, or orthogonality, between these two complex vectors. Note that when $\Theta_{ij}=0$, for example, the respective vectors are parallel and the focusing term is maximised. On the other hand, when $\Theta_{ij}=\pi/2$, the respective rows are orthogonal and the corresponding focusing term goes to zero. 

The concept of the focusing term can be extended to the magnitude response of the focused sound-field at all points in space. While this quantity has been given several names in the literature, it will be referred to here as the \textit{beamforming gain}, e.g., \cite{Theodoridis2013, Hamdan2021}, and is defined here in normalised form as
\begin{equation}
\arraygain{}{0}\defined\frac{\abs{\gM{}(\bv{x}_0)\herm\gM{}(\bv{x})}}{\norm{\gM{}(\bv{x})}\norm{\gM{}(\bv{x}_0)}},
\end{equation}
where $\bv{x}_0\inRv{3}$ is the target point at which focusing is desired, $\bv{x}\inRv{3}$ is an arbitrary measurement point in space and $\gM{}(\bv{x}_0),\gM{}(\bv{x})\inCv{L}$ are the transfer functions between the $L$ sources to the target and measurement points, respectively. That is to say $\abs{\gM{}(\bv{x}_j)\herm\gM{}(\bv{x}_i)}=\abs{\gram{i}{j}}$. Note from \cref{eq:focusing-xt}, the beamforming gain is related directly to the Hermitian angle, i.e., $\abs{\gM{}(\bv{x}_0)\herm\gM{}(\bv{x})}=\norm{\gM{}(\bv{x})}\norm{\gM{}(\bv{x}_0)}\cosTheta{}(\bv{x},\bv{x}_0)$, where $\Theta(\bv{x}, \bv{x}_0)$ is the Hermitian angle between $\gM{}(\bv{x})$ and $\gM{}(\bv{x}_0)$. It follows that $\arraygain{}{0} = \cosTheta{}(\bv{x}, \bv{x}_0)$.

\begin{figure}
	\centering
	\begin{subfigure}[t]{0.49\linewidth}
		\centering
		\includegraphics[trim={11mm 61mm 11mm 45mm},clip,width=1\linewidth]{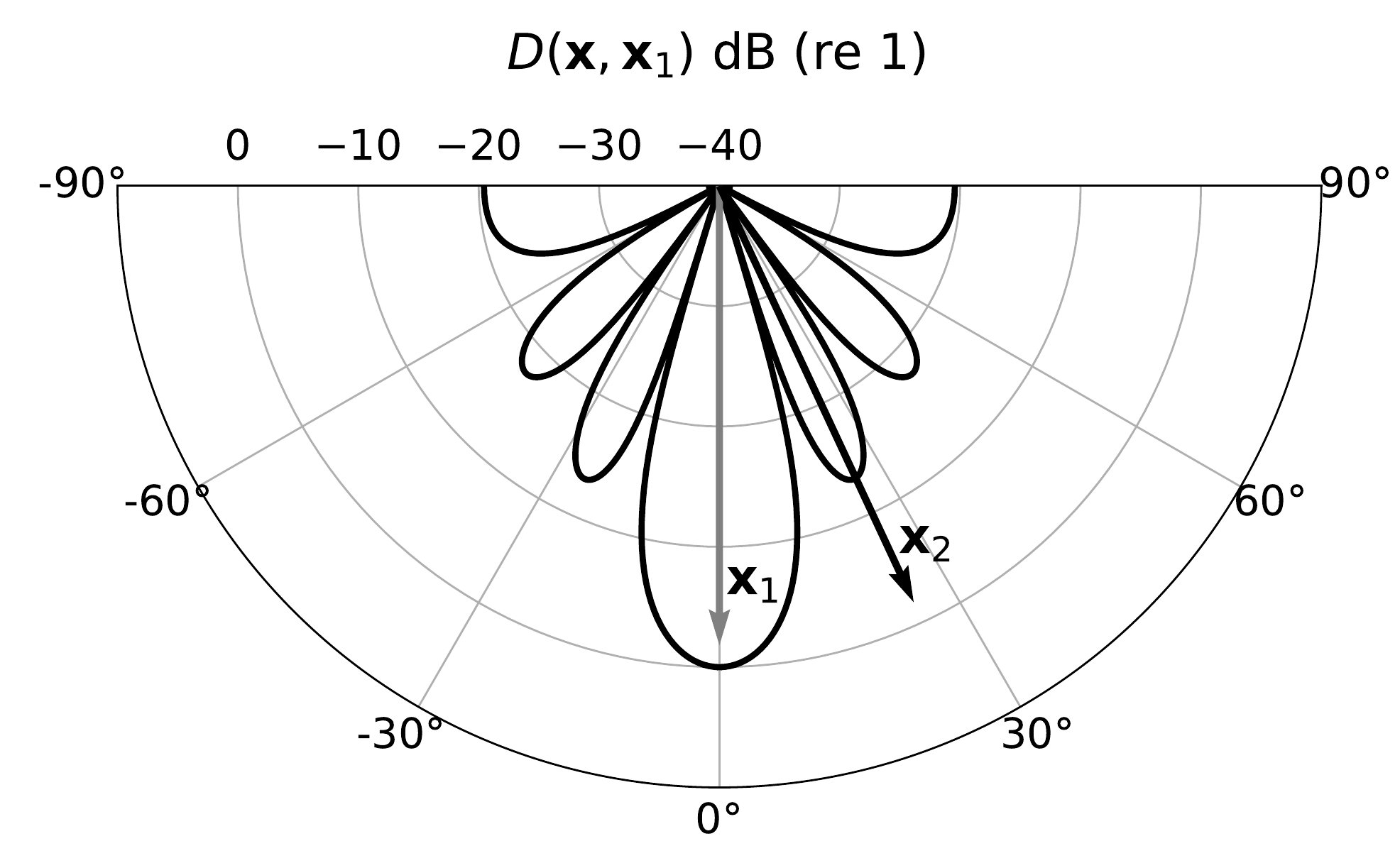}
		\caption{}
		\label{fig:J3-Single_Focusing_Filter_High_XT}
%		\vspace{30mm}
	\end{subfigure}
	\begin{subfigure}[t]{0.49\linewidth}
		\centering
		\includegraphics[trim={11mm 61mm 11mm 45mm},clip,width=1\linewidth]{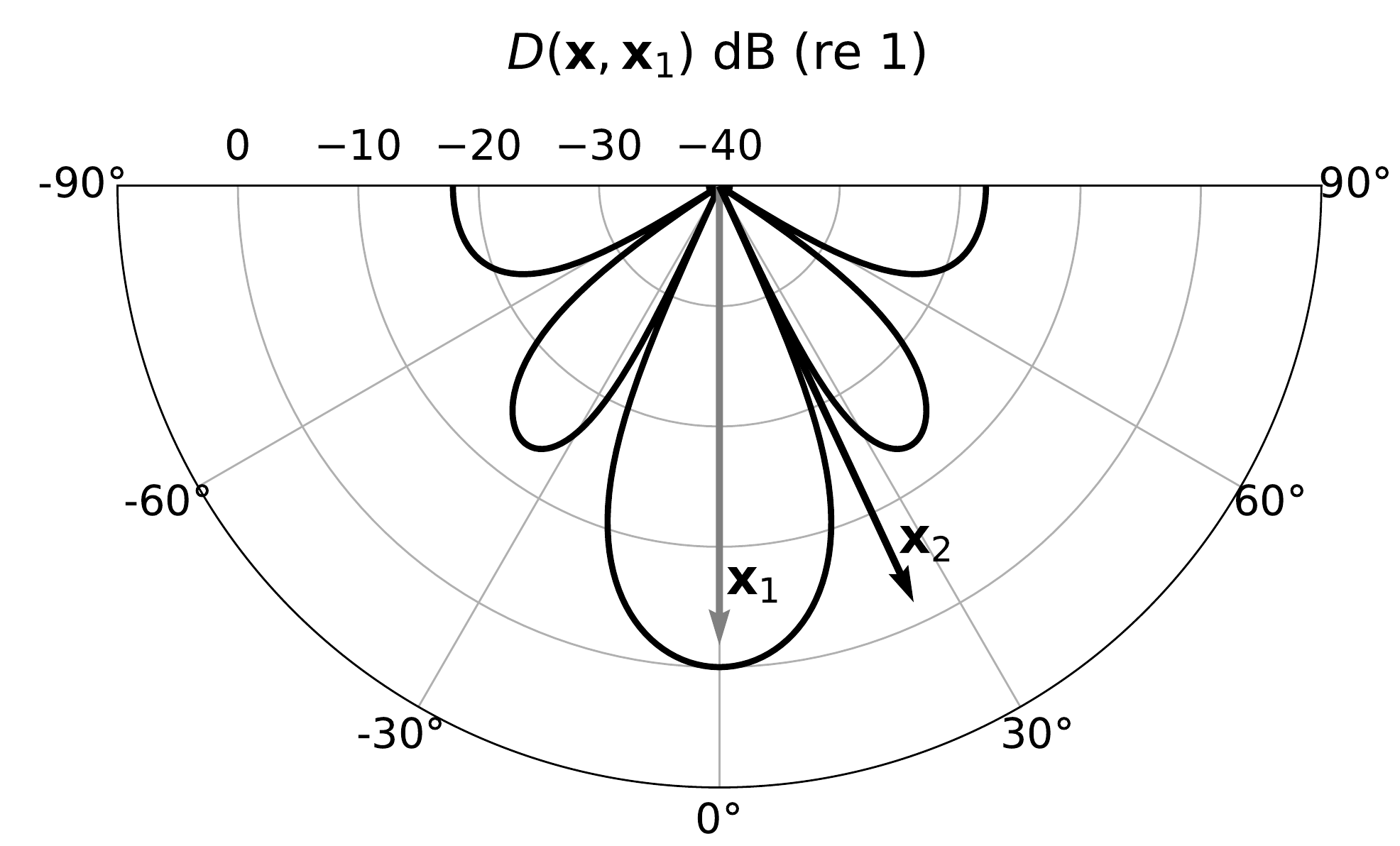}
		\caption{}
		\label{fig:J3-Single_Focusing_Filter_Low_XT}
	\end{subfigure}
	\caption{Normalised beamforming gain $\arraygain{}{1}$ measured on an arc in the far-field due to a twenty-channel ULA composed of monopole sources in free-field focused at the point $\xM{1}$ (which is positioned on the arc). a) $\xM{2}$ lies in the direction of a relatively large side lobe; b) $\xM{2}$ lies in the direction of a null.}
	\label{fig:J3-Single_Focusing_Filter}
\end{figure}

In general, the focusing crosstalk can be due to grating lobes or side lobes as a result of focusing, or due to any other disturbance that occurs in the field at the time of measurement. Thus, design considerations such as loudspeaker and control point geometry, in addition to the influence of scattering bodies in the field, all determine the resulting focusing crosstalk. For example, if the loudspeakers are assumed to be monopole sources in free-field, then the focusing operation can be considered a weighted delay-and-sum operation and the focusing crosstalk would be the grating or side lobe pressure that is radiated to the control points as a result of the $m$th focusing operation. An example is given in \cref{fig:J3-Single_Focusing_Filter} which shows $\arraygain{}{1}$ for a compact twenty-channel uniform linear array (ULA) composed of monopole sources in free-field with an inter-element spacing of 1.2 cm. The beamforming gain was calculated for a target point $\bv{x}_1$ along an arc in the far-field relative to the centre of the array for two different frequencies in Hz. The null is desired at the point $\bv{x}_2$ which was also positioned along the arc. \cref{fig:J3-Single_Focusing_Filter_High_XT} shows that at approximately 4899 Hz the focusing crosstalk is relatively high at $\bv{x}_2$ due to a prominent side lobe. On the other hand, \cref{fig:J3-Single_Focusing_Filter_High_XT} shows that at approximately 3411 Hz the focusing crosstalk is naturally zero at $\bv{x}_2$. The former case presents a greater difficulty than the latter in the inverse problem, for in the latter there exists an unwanted pressure that must be cancelled in the inversion stage.  

It has been shown that the forward focusing operation is the first step in the application of the inverse solution under the stated assumptions. In this process the pressures are maximised locally at the focus points. However, the $m$th focusing operation targeted at the $m$th point results in focusing crosstalk at all other control points. This focusing crosstalk must be eliminated before correct reproduction of the target pressures can occur. The following section details the inversion stage designed to eliminate the focusing crosstalk. 
\subsection{Inversion Stage and Stability}
The primary purpose of the inversion stage, given by $\Gram^{-1}$, is to cancel out the focusing crosstalk at the null points and to simultaneously equalise the pressure at the focus points. In this stage the concern lies with the determinant of the Gram, i.e.,
\begin{equation}
\gramian\defined\determ{\Gram},
\end{equation}
referred to herein as the \textit{gramian}, e.g., \cite{Rozanski2017}. The stability of the system at each frequency ultimately depends on the value of the gramian, as can be seen from \cref{eq:pinv-underdetermined-adj-determ}. For example, if the Gram is singular the gramian will be zero and as a result the inverse filters will explode, i.e., a singularity is encountered. On the other hand, if the gramian is maximised, it has the least amplifying effect on the inverse filters. 

If fact, if just one of the rows of the plant matrix are collinear, i.e., the rows of the plant form a linearly dependent set, the gramian will be zero. This means
\begin{equation}
\gramian = 0 \Leftrightarrow \Theta_{ij} = 0, ~ i\neq j.
\end{equation}
The key result is that \textit{if the focusing crosstalk is maximised at any one of the control points the gramian is zero} and the inverse solution explodes. This is another way of stating that the Gram becomes singular and therefore its matrix inverse does not exist. This fact makes it clear that focusing crosstalk is undesired in general.

In summary, the focusing crosstalk has been defined and it has been shown that it can lead to system instability. The following section addresses the phenomenon of minimising the focusing crosstalk to zero at all points and the resulting behaviour of the system. 

\section{Ideal Focusing}
\label{sec:J3-Ideal Focusing}
The question now is under what conditions is the gramian maximised and what is the nature of the focusing crosstalk in this state? The answers to both of these questions lies in the work of Hadamard. Hadamard had previously investigated the gramian (and in general determinants of arbitrary complex matrices) and produced what is called Hadamard's inequality, e.g., \cite{Rozanski2017}, which is
\begin{equation}
\label{eq:hadamard-inequality}
\gramian \leq \prod_{m=1}^{M}\gram{m}{m}.
\end{equation}
Here Hadamard's inequality states that the magnitude of the gramian is less than or equal to the product of the $\ell_2$ norms of the rows of the plant matrix. Most importantly, \cref{eq:hadamard-inequality} becomes an equality \textit{if and only if the Gram is a diagonal matrix}, e.g., \cite{Futura1971}. That is to say \textit{the gramian is maximised when the rows of the plant are pairwise orthogonal}, i.e,
\begin{equation}
\gramian= \prod_{m=1}^{M}\gram{m}{m} \Leftrightarrow \Theta_{ij}=\frac{\pi}{2}~\forall i\neq j.
\end{equation} The diagonal form of the Gram, which will be referred to as the \textit{ideal Gram}, is defined as
\begin{equation}
\label{eq:diagonal-gram}
\dggh_{\mathrm{I}} \defined \diag{\gram{1}{1}, \dots, \gram{M}{M}},
\end{equation}
where the subscript $(\cdot)_{\mathrm{I}}$ denotes the ideal focusing state.
When the Gram is ideal it means that all of the focusing crosstalk terms have gone to zero, i.e., $\abs{\gram{i}{j}} = 0, \forall i\neq j$. In other words, focusing at each control point is accomplished without leakage to the other control points. This special phenomenon will be referred to as \textit{ideal focusing}. 

It is paramount to the study at hand to understand the behaviour of the inverse solution when the Gram is ideal. Firstly, the inverse of the ideal Gram follows from the properties of diagonal matrices and is
\begin{equation}
\label{eq:J3-inv-ideal-gram}
\dggh_{\mathrm{I}}^{-1} = \diag{\gram{1}{1}^{-1},\dots, \gram{M}{M}^{-1}}.
\end{equation}
\cref{eq:J3-inv-ideal-gram} indicates that in the ideal focusing state the inverse stage amounts to a normalisation at the focus points, i.e., the inverse solution's only task is to normalise the diagonal entries of the Gram to one. There is no need to cancel the focusing crosstalk in the ideal state as there is none to begin with. Thus an intuitive link has been made between maximisation of the Gram and the effort needed to accomplish the inversion (in this case effort refers to the amplification factor due to the gramian). Given \cref{eq:J3-inv-ideal-gram} it follows that the pseudoinverse ultimately simplifies to
\begin{equation}
\label{eq:normalised-focusing-filters}
\filters_{\mathrm{I}} \defined \plant\herm\dggh^{-1} = \begin{bmatrix}
\gM{1}^*/\gram{1}{1} & \dots & \gM{M}^*/\gram{M}{M}
\end{bmatrix}.
\end{equation}
The result given by \cref{eq:normalised-focusing-filters} states that the normalised focusing filters $\bv{g}_m^*/\gram{m}{m}$ satisfy the discrete inverse problem given by \cref{eq:discrete-inverse-problem} when the Gram is ideal. As such, they are referred to as \textit{ideal focusing filters}. Furthermore, the \textit{ideal inverse solution} is
\begin{equation}
\bv{q}_{\mathrm{I}} \defined \filters_{\mathrm{I}}\desired = \begin{bmatrix}
d_1\gM{1}^*/\gram{1}{1} + \dots + d_M\gM{M}^*/\gram{M}{M}
\end{bmatrix},
\end{equation} and also the minimum norm least-squares solution by definition. 

In summary, by way of Hadamard's inequality, it has been shown that when all of the focusing crosstalk terms go to zero the gramian is maximised. When this phenomenon occurs the Gram is said to be ideal and the system is in an ideal focusing state. They key point is that the array of sources is able to focus maximal pressure at each control point without imparting any leakage to the other control points and that the amplification factor due to the gramian is minimised.

\section{Modal Analysis of the Ideal Focusing State}
\label{sec:J3-Modal Analysis of the Ideal Focusing State}
The following explores the ideal focusing solution further and derives the system eigenvalues and eigenvectors when in the ideal focusing state. In doing so, the physical concept of focusing crosstalk and its influence is ultimately related to the amplification factor of the pseudoinverse and to the system condition number. 

\subsection{Eigendecomposition and Radiation Modes}
\label{sec:Singular Value Decomposition and Radiation Modes}
The eigendecomposition of the Gram, ideal or otherwise, is fundamentally related to the general modal analysis of the entire system which is given by the singular value decomposition (SVD) of $\bv{G}$. The SVD of $\bv{G}$ is defined as \cite{Meyer2001}
\begin{equation}
\label{eq:J3-svd}
\bv{G} = \svd,
\end{equation}
where $\bv{U}\in\mathbb{C}^{M\times M}$ is a unitary matrix whose columns are basis vectors referred to here as the \textit{field pressure modes}, $\bv{V}\in\mathbb{C}^{L\times L}$ is a unitary matrix whose columns are basis vectors referred to here as the \textit{source strength modes}, and $\bv{\Sigma}\in\mathbb{R}_{\geq0}^{M\times L}$ is a rectangular diagonal matrix whose elements along the main diagonal are the singular values of $\bv{G}$, i.e., $\bv{\Sigma}=\diag{\sigma_1, \dots, \sigma_r}$, where $r=\rank{\plant}$. Note that no logical ordering of the singular values is enforced here. From \cref{eq:J3-svd}, it follows
\begin{equation}
\label{eq:ggh-diagonalisation}
\Gram = \bv{U}\bv{\Lambda}\bv{U}^{-1},
\end{equation}
where $\bv{\Lambda}=\bv{\Sigma}\bv{\Sigma}\herm=\diag{\lambda_1,\dots, \lambda_r}$ is the square diagonal matrix of eigenvalues. \cref{eq:ggh-diagonalisation} is the eigendecomposition of the Gram and states that the eigenvalues of $\Gram$ are the squared singular values of the radiation matrix $\bv{G}$, i.e., $\lambda_m = \sigma_m^2$, where $m=1, \dots, r$. \cref{eq:ggh-diagonalisation} also states that the left singular vectors of $\bv{G}$ are eigenvectors of $\Gram$. The eigenvalues of the Gram are considered the \textit{mode radiation efficiencies}. Consider that 
\begin{equation}
\label{eq:J3-mode-relationship}
\plant\bv{v}_m = \sigma_m \bv{u}_m,
\end{equation}
which states that each source strength mode is uniquely related to each field pressure mode by its corresponding radiation efficiency. The radiation efficiency determines how well a given mode radiates to the control points. When the system is underdetermined only a subset of the source strength modes are utilised, i.e., 
the source strength modes with indices $\rank{\plant}<l\leq L$ span the null space of the plant matrix. These modes are not utilised by the system and are effectively \textit{unseen} at the control points, e.g., \cite{Fazi2007}.
\subsection{Ideal Singular System}
The Gram is the matrix of its eigenvalues when ideal focusing occurs, i.e.,
\begin{equation}
\label{eq:J3-ideal-eigenvalues-matrix}
\Gram = \dggh_{\mathrm{I}},
\end{equation}
where the $m$th eigenvalue is $\lambda_m = \gram{m}{m}$. \cref{eq:J3-ideal-eigenvalues-matrix} makes the key point that when in the ideal focusing state the radiation efficiency of each mode is equivalent to the maximisation of the focusing pressure at the corresponding control point. A key conclusion is that in the ideal focusing state the radiation efficiency is not related to the focusing crosstalk and is solely to do with the magnitudes of the individual transfer functions.

It follows that the eigenvectors of the ideal Gram and the left singular values of the corresponding plant are
\begin{equation}
\label{eq:J3-ideal-left-singular-vectors}
\bv{U}= \bv{I},
\end{equation}
i.e., the eigenvectors are the standard basis vectors $\bv{e}_m\inRv{M}$. Thus, when the Gram is ideal the singular values of the plant are
\begin{equation}
\label{eq:J3-ideal-singular-values}
\sigma_m = \sqrt{\gram{m}{m}}.
\end{equation}
The corresponding source strength modes can be found by rearranging \cref{eq:J3-svd} and substituting in \cref{eq:J3-ideal-eigenvalues-matrix} (taking the square root) and \cref{eq:J3-ideal-left-singular-vectors}. This yields the first $M$ source strength modes as 
\begin{equation}
\rsvectors_{l\leq M} =\begin{bmatrix}
\gM{1}^*/\sqrt{\gram{1}{1}} & \dots & \gM{M}^*/\sqrt{\gram{M}{M}}
\end{bmatrix}.
\end{equation}
When these source strength modes are input to the system, assuming the Gram is ideal, the output at the control points is given by
\begin{equation}
\plant\frac{\gM{m}^*}{\sqrt{\gram{m}{m}}} = \sqrt{\gram{m}{m}}\bv{e}_m,
\end{equation}
meaning that each source strength mode input results in ideal focusing at the corresponding control point.
\subsubsection{Ideal Amplification Factor and Condition Number}
From the derivation of the ideal singular system it follows that the $\ell_2$ norm of the pseudoinverse is given by
\begin{equation}
\label{eq:J3-ideal-pinv-norm}
\norm{\filters_{\mathrm{I}}} = \frac{1}{\min\left(\sigma_m\right)} = \frac{1}{\mathrm{min}\left(\sqrt{\gram{m}{m}}\right)}.
\end{equation}
\cref{eq:J3-ideal-pinv-norm} indicates that the squared amplification factor of the pseudoinverse in the ideal focusing state is the reciprocal of the smallest focus point pressure. In other words, the system only requires as much amplification on the input signal as dictated by the inverse of the smallest focus point pressure, therefore, it serves to have the largest focus point pressure possible. It must also be noted that in general the determinant of a matrix is equal to the product of its eigenvalues. In light of this fact, it follows that when the Gram is ideal and the determinant is maximised, the amplification factor given by \cref{eq:J3-ideal-pinv-norm} is, by definition, minimised given the relative norms of the plant rows. This result is another way of seeing that the system requires the least effort when the focusing crosstalk is minimised.

The condition number of the ideal singular system is
\begin{equation}
\label{eq:J3-ideal-cond-number}
\kappa\left(\plant\right) = \frac{\mathrm{max}\left(\sigma_m\right)}{\mathrm{min}\left(\sigma_m\right)} = \frac{\mathrm{max}\left(\sqrt{\gram{m}{m}}\right)}{\mathrm{min}\left(\sqrt{\gram{m}{m}}\right)}.
\end{equation}
\subsubsection{Super Ideal Focusing}
\label{sec:J3-Super Ideal Gram}
The system is optimally conditioned when the focus point pressures are equal, as seen from inspection of \cref{eq:J3-ideal-cond-number}. In this case, the Gram is the scalar matrix
\begin{equation}
\label{eq:J3-unit-gram}
\bv{\Lambda}_{opt} \defined \lambda_{opt}\bv{I},
\end{equation}
where $\lambda_{opt} \defined \gram{1}{1} = \gram{2}{2} = \dots = \gram{M}{M}$, and the subscript $(\cdot)_{opt}$ denotes the super ideal focusing state. \cref{eq:J3-unit-gram} is the optimal form of the Gram because it corresponds to the minimisation of the $\ell_2$ norm of the pseudoinverse \textit{and} the system condition number, where both quantities become
\begin{equation}
\norm{\filters_{opt}} = \frac{1}{\sqrt{\lambda_{opt}}},
\end{equation}
and 
\begin{equation}
\kappa\left(\plant\right) = 1,
\end{equation}
respectively. Thus the super ideal focusing state is equivalent to an optimally-conditioned system. The key physical result is that \textit{the system is optimally-conditioned if and only if ideal focusing occurs and when the focusing is equally efficient to all control points in the field}, in which case super ideal focusing is said to occur. Stated rigorously 
\begin{equation}
\kappa(\plant)=1\Leftrightarrow\Gram = \mathbf{\Lambda}_{opt},
\end{equation}
meaning the rows of the plant are pairwise orthogonal and of equivalent $\ell_2$ norm (considering only underdetermined systems). 

When super ideal focusing occurs the eigenvectors are still given by \cref{eq:J3-ideal-left-singular-vectors} while the singular values become the scalar matrix $\bv{\Sigma}= \sqrt{\mathbf{\Lambda}_{opt}}$, the first $M$ source strength modes simplify to $\rsvectors_{l\leq M} = \lambda_{opt}^{-1/2}\plant\herm$, and the pseudoinverse simplifies to \begin{equation}
\label{eq:J3-super-ideal-focusing-pseudoinverse}
\filters_{opt}\defined  \lambda_{opt}^{-1}\plant\herm.
\end{equation}
Furthermore, the filters given by \cref{eq:J3-super-ideal-focusing-pseudoinverse} are \textit{spatially matched} at $M$ points, e.g., \cite{Tanter2000}, as shown in \cref{sec:J3-Temporarily and Spatially Matched Filters}. This means that these filters maximise the magnitude of the reproduced complex pressure at all $M$ control points simultaneously. In fact, super ideal focusing \textit{must occur} in order for the filters to be spatially matched at all points. 
\section{Case Studies}
\label{sec:J3-Case-Studies}
Two theoretical case studies of ideal focusing are given in light of the results presented in \crefrange{sec:J3-CTC}{sec:J3-Sound-Zones}. These studies consider two applications of sound-field control that use the inverse solution. In \cref{sec:J3-CTC} the application of binaural audio reproduction at two points in space using loudspeakers, e.g., $M=2$ and $L\geq2$, is examined. Special interest is put on how OSD, previously established in \cite{Takeuchi2002}, is a specific case of the more general super ideal focusing state that has been established herein. Furthermore, more general cases are considered and additional ideal focusing solutions are presented, in particular for the two-channel CTC system with arbitrary geometry (no symmetry constraint) and for a symmetric multichannel CTC case. Following, in \cref{sec:J3-Sound-Zones} the application of creating an arbitrary number of sound zones using the inverse solution is considered, e.g., \cite{Nelson1996multichannel}. It is shown under what conditions the far-field ULA beamformer is in the super ideal focusing state in the direction of $M$ control points simultaneously. Thus, a theoretical study is given of when super ideal focusing occurs in the recreation of $M$ sound zones in the far-field using a linear loudspeaker array.
\subsection{Acoustic Crosstalk Cancellation ($M=2, L\geq 2$)}
\label{sec:J3-CTC}
The following considers the discrete sound-field control problem of reproducing a stereo binaural audio signal at two control points in space, $\bv{x}_1$ and $\bv{x}_2$, using $L\geq 2$ loudspeakers. This problem is generally referred to as crosstalk cancellation (CTC). The term \enquote{CTC system} will then refer to the general $2\times L$ system throughout. In this case, the Gram is 
\begin{equation}
\label{eq:J3-two-point-gram}
\Gram_{M=2} = \begin{bmatrix}
\gram{1}{1} & \gramtwo \\[10pt]
\gramthree & \gram{2}{2}
\end{bmatrix},
\end{equation}
and the gramian is
\begin{align}
\label{eq:J3-two-point-gramian}
\gramian_{M=2}& = \gram{1}{1}\gram{2}{2}-\abs{\gramtwo}^2,\\[5pt]
\nonumber
& = \gram{1}{1}\gram{2}{2}\sin^2\Theta_{12},
\end{align}
where $\sin^2\Theta_{12} = 1-\cos^2\Theta_{12} = 1 - \abs{\gramtwo}^2(\gram{1}{1}\gram{2}{2})^{-1}$, and $\Theta_{12}$ is the Hermitian angle between the two rows of the plant matrix. In \cref{eq:J3-two-point-gramian} the effect of the focusing crosstalk can be visibly seen. The gramian takes its most simple form in that it depends on $\sin^2\Theta_{12}$. As shown in \cref{sec:J3-Focusing Stage}, this terms amounts to a measure of collinearity between the two rows measured by the Hermitian angle $\Theta_{12}$. 

In this case, the gramian has a direct geometric interpretation in that \cref{eq:J3-two-point-gramian} is the formula for the squared area of a parallelogram, e.g., \cite{Meyer2001, Gover2010}. The side lengths of the parallelogram are given by the norms of the plant rows and the angle between the sides is the Hermitian angle $\Theta_{12}$. Thus the behaviour of the gramian, and by definition the product of the eigenvalues, is described entirely by the area of a parallelogram when $M=2$.

\begin{figure}
	\centering
	\begin{subfigure}[t]{0.47\linewidth}
		\centering
		\includegraphics[width=0.9\linewidth]{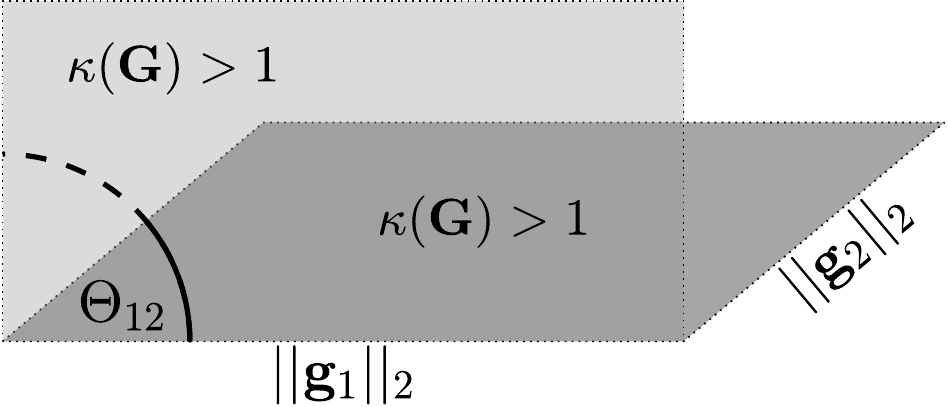}
		\caption{}
		\label{fig:Parallelograms_Rect}
	\end{subfigure}
	~~
	\begin{subfigure}[t]{0.47\linewidth}
		\centering
		\includegraphics[width=0.9\linewidth]{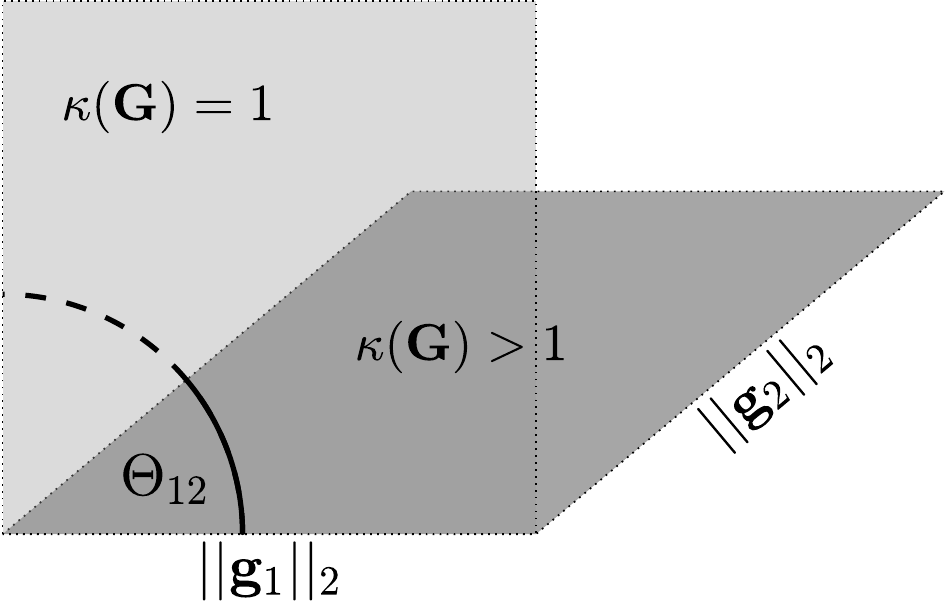}
		\caption{}
		\label{fig:Parallelograms_Rhombus}
	\end{subfigure}
	\caption{Parallelograms depicting the $2\times L$ system robustness. a) $\kappa(\plant)>1$ when the side lengths are not equal, even if $\Theta_{12}=90^\circ$; b) $\kappa(\plant)>1$ as long as $\Theta_{12}<90^\circ$, even if the side lengths are equal. $\kappa(\plant)=1$ when the side lengths are equal and $\Theta_{12}=90^\circ$.}
	\label{fig:Parallelograms}
\end{figure}

\cref{fig:Parallelograms} depicts four of the five possible states in the form of parallelograms. Not depicted is the maximally ill-conditioned state when the focusing crosstalk is maximised, i.e., $\Theta_{12} = 0$, in which case the area of the parallelogram is zero. When $0\leq\Theta_{12}\leq\pi/2$ the focusing crosstalk is non-zero, but not maximised, which corresponds to a parallelogram with opposite sides that are parallel and opposite angles that are equal, shown for non-equal and equal side lengths in dark grey in \cref{fig:Parallelograms_Rect} and \cref{fig:Parallelograms_Rhombus}, respectively. In the latter case, when the plant row norms are equal, the parallelogram is a rhombus. In both cases, the condition number is greater than one. On the other hand, when $\Theta_{12}=\pi/2$, the focusing crosstalk is zero and the parallelogram becomes a rectangle, as shown in light grey in \cref{fig:Parallelograms_Rect}. Thus, when the parallelogram is a rectangle the system has achieved the ideal focusing state. However, when the rectangle has unequal side lengths, the condition number remains greater than one because the norms of the plant rows are not equal. Finally, \textit{when the parallelogram becomes a square, the CTC system has achieved the super ideal focusing state} described in \cref{sec:J3-Super Ideal Gram}, as shown in light grey in \cref{fig:Parallelograms_Rhombus}. It is \textit{only} in this state that the CTC system is optimally-conditioned, since in all other states the eigenvalues are not equal.  
\subsubsection{Two-channel Ideal and Super Ideal Focusing}
\label{sec:General Two-channel Ideal and Super Ideal Focusing}
The following derives the ideal and super ideal focusing conditions for the general $2\times 2$ CTC system. Thus, the system is square, and it has a unique inverse solution (assuming the system is consistent). In this case, the focusing crosstalk can be written as
\begin{equation}
\label{eq:J3-gen-two-channel-focusing-xt}
\gramtwo = \Delta g_1 \euler^{\Ij\Delta\phi_1}+\Delta g_2 \euler^{\Ij\Delta\phi_2},
\end{equation} 
where $\Delta g_l \defined  \abs{g_{1l}}\abs{g_{2l}}$ and $\Delta\phi_l\defined\angle g_{1l}-\angle g_{2l}$, and $l=1,2$. \cref{eq:J3-gen-two-channel-focusing-xt} must be set to zero to achieve the ideal focusing state. 
Thus, the following is required:
\begin{align}
\label{eq:J3-two-channel-ideal-focusing-cond-1}
\Delta\phi_1-\Delta\phi_2&=(2n-1)\pi,\\[10pt]
\label{eq:J3-two-channel-ideal-focusing-cond-3}
\Delta g_1 &= \Delta g_2,~~\Delta g_l > 0,
\end{align}
where $n\in\mathbb{Z}$. When the conditions given by \cref{eq:J3-two-channel-ideal-focusing-cond-1} and \cref{eq:J3-two-channel-ideal-focusing-cond-3} are met the two-channel CTC system achieves ideal focusing in general, i.e., \textit{for arbitrary transfer functions}. Recall, the super ideal focusing state has not been achieved necessarily as $\gram{1}{1}\neq\gram{2}{2}$, in general. However, the norms \textit{are} equal in the case when the $2\times 2$ plant is
\begin{equation}
\label{eq:J3-two-by-two-plant}
\plant = \begin{bmatrix}
g_{11} & g_{21} \\[10pt]
g_{21} & g_{11}
\end{bmatrix},
\end{equation}
in which case the conditions \cref{eq:J3-two-channel-ideal-focusing-cond-3} and $\gram{1}{1}=\gram{2}{2}$ are satisfied (note as well that $\gram{1}{2}=\gram{2}{1}$ under symmetry). When the $2 \times 2$ plant is symmetric the focusing crosstalk simplifies to 
\begin{equation}
\label{eq:J3-two-channel-symmetric-focusing-xt}
\gram{2}{1} = 2\Delta g_1\cos\Delta\phi_1.
\end{equation}
Thus, the general symmetric $2\times 2$ system achieves super ideal focusing when  
\begin{equation}
\Delta \phi_1 = (2n-1)\frac{\pi}{2}.
\end{equation}
\subsubsection{Two-channel Ideal and Super Ideal Focusing: Monopoles in Free-field and Shadowless Head}
\label{sec:Two-channel Ideal and Super Ideal Focusing: Monopoles in Free-field and Shadowless Head}
\crefrange{eq:J3-two-channel-ideal-focusing-cond-1}{eq:J3-two-channel-ideal-focusing-cond-3} provide the general requirements for the two-channel CTC system to achieve ideal and super ideal focusing, where in the latter case, the system is optimally-conditioned. However, the results are rather difficult to interpret physically as no assumptions have been made on the acoustics. The following gives a physical interpretation by assuming that the sources are monopoles in free-field and that the control points are point microphones, i.e., a shadowless head model. In this case, a single element of the plant is given by 
\begin{equation}
\label{eq:J1-monopole}
g_{ml}=\frac{\mathrm{e}^{-\J kR_{ml}}}{R_{ml}},
\end{equation}
where $k=2\pi f/c$ is the wavenumber ($f$ is frequency in Hz, $c$ is the speed of sound (m$\cdot$s$^{-1}$)), $R_{ml}\defined\norm{\bv{s}_l-\bv{x}_m}$ is the distance in metres from the $l$th source to the $m$th control point and $l=1,2,\dots,L$ and $m=1,2$ (the typical scaling factor $1/4\pi$ is implied). The ears of the listener are considered to be located at the two control point positions $\bv{x}_1$ and $\bv{x}_2$.

Note that the analysis begins without assumptions on the geometry. In this case the focusing crosstalk 
is 
\begin{equation}
\label{eq:J3-CTC-focusing-xt}
\gramtwo = \frac{\euler^{\Ij k\pld{1}}}{\plp{1}}+\frac{\euler^{\Ij k\pld{2}}}{\plp{2}},
\end{equation}
where $\pld{l} \defined \Rml{2}{l} - \Rml{1}{l}$ is the $l$th path length difference and $\plp{l}\defined\Rml{1}{l}\Rml{2}{l}$ is the $l$th path length product. From \crefrange{eq:J3-two-channel-ideal-focusing-cond-1}{eq:J3-two-channel-ideal-focusing-cond-3} the following conditions must be met for ideal focusing:
\begin{align}
\label{eq:J3-two-channel-ideal-focusing-monopole-cond-1}
\pld{1} - \pld{2} &= (2n-1)\frac{\Lambda}{2}, \\[10pt]
\label{eq:J3-two-channel-ideal-focusing-monopole-cond-3}
\plp{1} & = \plp{2},~~\plp{l} > 0,
\end{align}
where $\Lambda= c/f$ is the acoustic wavelength in metres. \crefrange{eq:J3-two-channel-ideal-focusing-monopole-cond-1}{eq:J3-two-channel-ideal-focusing-monopole-cond-3} give the conditions for two-channel ideal focusing given the acoustic model at hand. \cref{eq:J3-two-channel-ideal-focusing-monopole-cond-1} states that the path length relationships need to change for each frequency in order to maintain ideal focusing. This implies that the system geometry must change with each frequency. If the listener is assumed to be fixed, i.e., the control points are fixed, then the loudspeakers must change position for each frequency. The same result can also interpreted with a fixed loudspeaker geometry and variable listener position, or fixed geometry and variable frequency. Furthermore, as the plant elements are scaled delays, ideal focusing means the inverse problem is satisfied by a two-channel compensated delay-and-sum operation, with filters of the form given by \cref{eq:normalised-focusing-filters}.
\subsubsection{Symmetric Two-channel Example - Optimal Source Distribution}
\label{sec:Symmetric Two-channel Example - Optimal Source Distribution}
\begin{figure}
	\centering
	\includegraphics[width=0.35\linewidth]{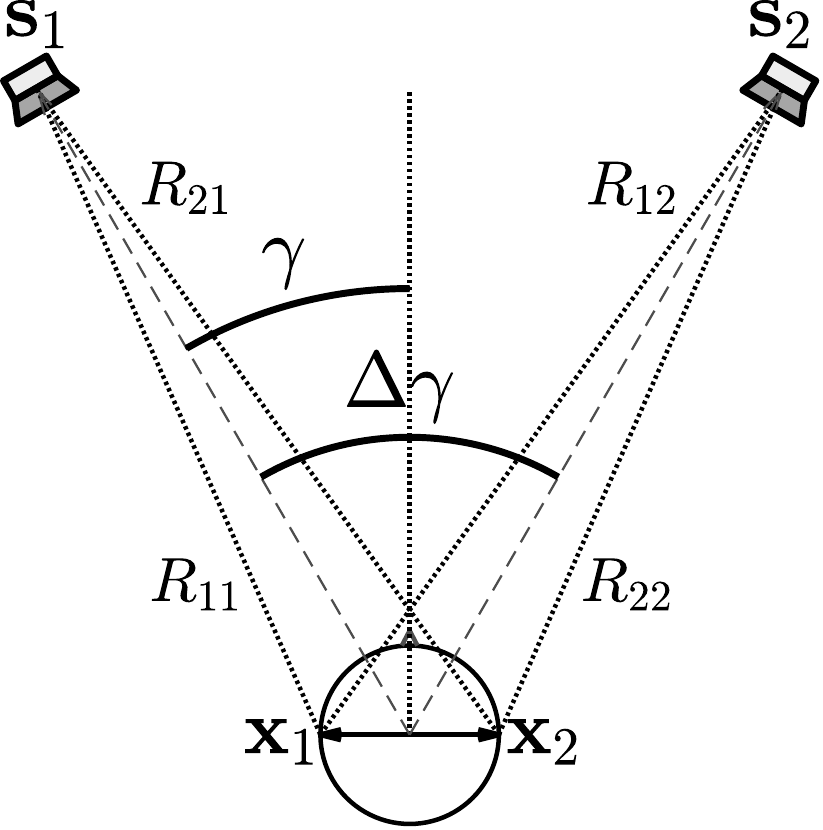}
	\caption{Symmetric two-channel geometry schematic.}
	\label{fig:two-channel-ctc-schematic}
\end{figure}

Continuing with the acoustic model established in \cref{sec:Two-channel Ideal and Super Ideal Focusing: Monopoles in Free-field and Shadowless Head}, a symmetric example can be formed by setting $R_{11} = R_{22}$ and $R_{12}=R_{21}$, as depicted in \cref{fig:two-channel-ctc-schematic}. Super ideal focusing is now possible as the norms of the plant rows are equal. Under this symmetric assumption \cref{eq:J3-two-channel-ideal-focusing-monopole-cond-1} can be used to derive the resulting super ideal focusing requirement:
\begin{equation}
\label{eq:J3-two-channel-symmetric-ideal-path-length-requirement}
\pld{opt} = (2n-1)\frac{\lambda_{opt}}{4},
\end{equation}
where $\pld{opt}\defined\pld{1}=-\pld{2} $, and $\Lambda_{opt}\defined c/f_{opt}$. The subscript $(\cdot)_{opt}$ denotes that these quantities are associated with the super ideal focusing state (in line with the notation introduced in \cref{sec:J3-Super Ideal Gram}). \cref{eq:J3-two-channel-symmetric-ideal-path-length-requirement} states that the path length difference must be an odd integer multiple of a quarter of the given wavelength in order to maintain super ideal focusing under the stated assumptions. Furthermore, \cref{eq:J3-two-channel-symmetric-ideal-path-length-requirement} states that the path length difference must change with frequency, as with the more general result given by \cref{eq:J3-two-channel-ideal-focusing-monopole-cond-1}. Most importantly, the path length difference increases and decreases monotonically with wavelength to maintain the super ideal focusing state. The result given by \cref{eq:J3-two-channel-symmetric-ideal-path-length-requirement} was previously discovered in \cite{Takeuchi2002} using the same acoustic model and symmetry assumption (with different notation). However, their original result was obtained by a direct analysis of the SVD of the plant matrix and inspection of the resulting singular values, not by analysis of the focusing crosstalk as it has been done here. It was recognised in \cite{Takeuchi2002} that \cref{eq:J3-two-channel-symmetric-ideal-path-length-requirement} must be satisfied under the stated assumptions to achieve a system condition number of one and flat pseudoinverse amplification factor (in light of the results presented here the amplification factor was only flat in their analysis because of the chosen acoustic model). Those researchers made the important link between the result given by \cref{eq:J3-two-channel-symmetric-ideal-path-length-requirement} and the requirement on the loudspeaker geometry. 

To make this explicit link to the loudspeaker geometry, a far-field assumption is applied such that
\begin{equation}
\label{eq:J3-far-field-requirement}
\pld{opt} \approx 2a\sin\gamma_{opt},
\end{equation}
where $a\defined \norm{\xM{1}}$ is the control point radius, and $\gamma_{opt}$ is the loudspeaker angle with corresponding angular span $\Delta \gamma_{opt} \defined 2\abs{\gamma_{opt}}$. Note that the control points are now assumed to be diametrically opposed such that $\bv{x}_1=-\bv{x}_2$. \cref{eq:J3-far-field-requirement} is a good approximation when $\norm{\sL{1}}\gg\norm{\xM{1}}$, i.e., the sources are a sufficient distance away to be considered in the far-field, e.g., \cite{Williams1999}. Substituting \cref{eq:J3-far-field-requirement} into \cref{eq:J3-two-channel-symmetric-ideal-path-length-requirement} yields the important result
\begin{equation}
\label{eq:J3-two-channel-opt-angle}
\gamma_{opt} = \sin^{-1}\left((2n-1)\frac{\Lambda_{opt}}{8a}\right),
\end{equation}
which requires
\begin{equation}
\label{eq:J3-two-channel-opt-lambda-restriction-symmetric}
\abs{(2n-1)\Lambda_{opt}}\leq 8a.
\end{equation}
\cref{eq:J3-two-channel-opt-angle} is the resulting requirement on the frequency-dependent loudspeaker angle which was alluded to by the more general requirement for ideal focusing given by \cref{eq:J3-two-channel-ideal-focusing-monopole-cond-1}. \cref{eq:J3-two-channel-opt-angle} states that the two-channel loudspeaker span must change \textit{continuously} with frequency in order to maintain the super ideal focusing state, i.e., pairwise orthogonality between rows. As the wavelength decreases, the loudspeaker span must decrease. On the other hand, as the wavelength increases the span must accordingly increase. Thus, a continuum of loudspeakers is required to maintain the super ideal focusing state continuously as a function of frequency, as concluded previously in \cite{Takeuchi2002}.

\begin{figure}[t]
	\centering
	\begin{subfigure}[t]{0.49\linewidth}
		\centering
		\includegraphics[width=1\linewidth]{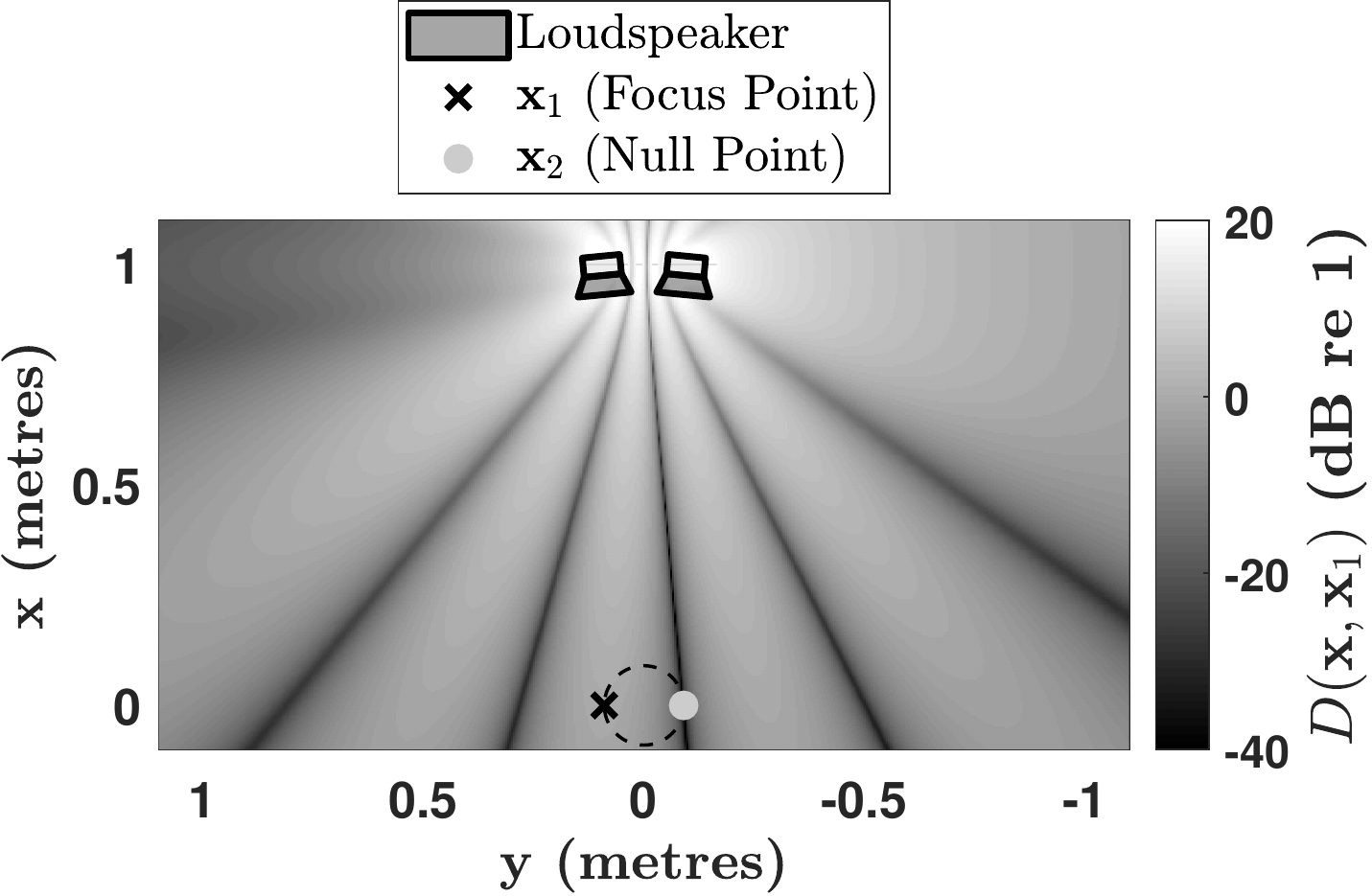}
		\caption{}
		\label{fig:Two-channel-CTC-Focus-Sound-Field-5000Hz}
	\end{subfigure}
\hfill
	\begin{subfigure}[t]{0.49\linewidth}
		\centering
		\includegraphics[trim={0mm 0mm 0mm 0mm},clip,width=1\linewidth]{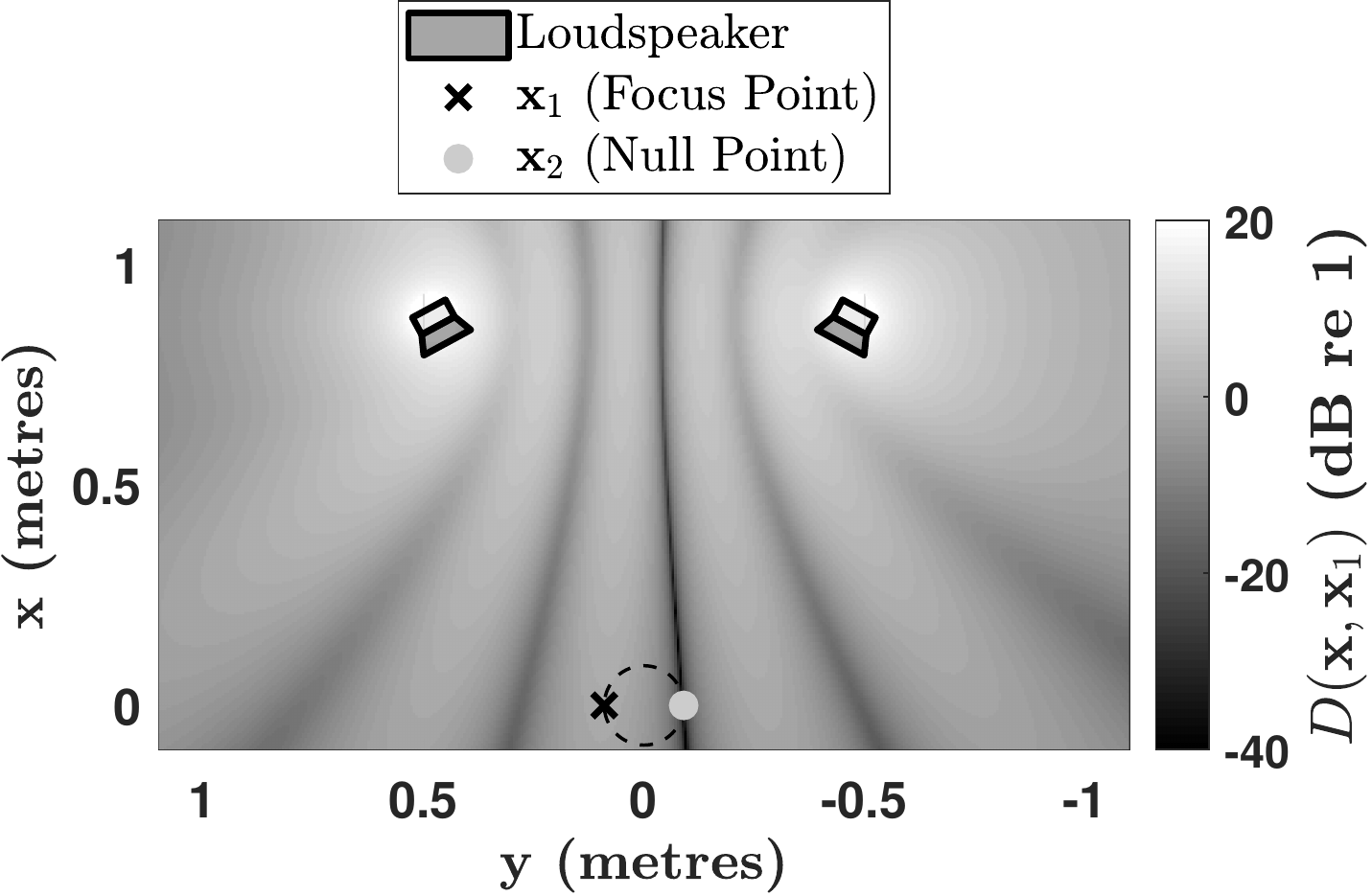}
		\caption{}
		\label{fig:Two-channel-CTC-Focus-Sound-Field-1000Hz}
	\end{subfigure}
	\begin{subfigure}[t]{0.49\linewidth}
		\centering
		\includegraphics[trim={0mm 0mm 0mm 0mm},clip,width=1\linewidth]{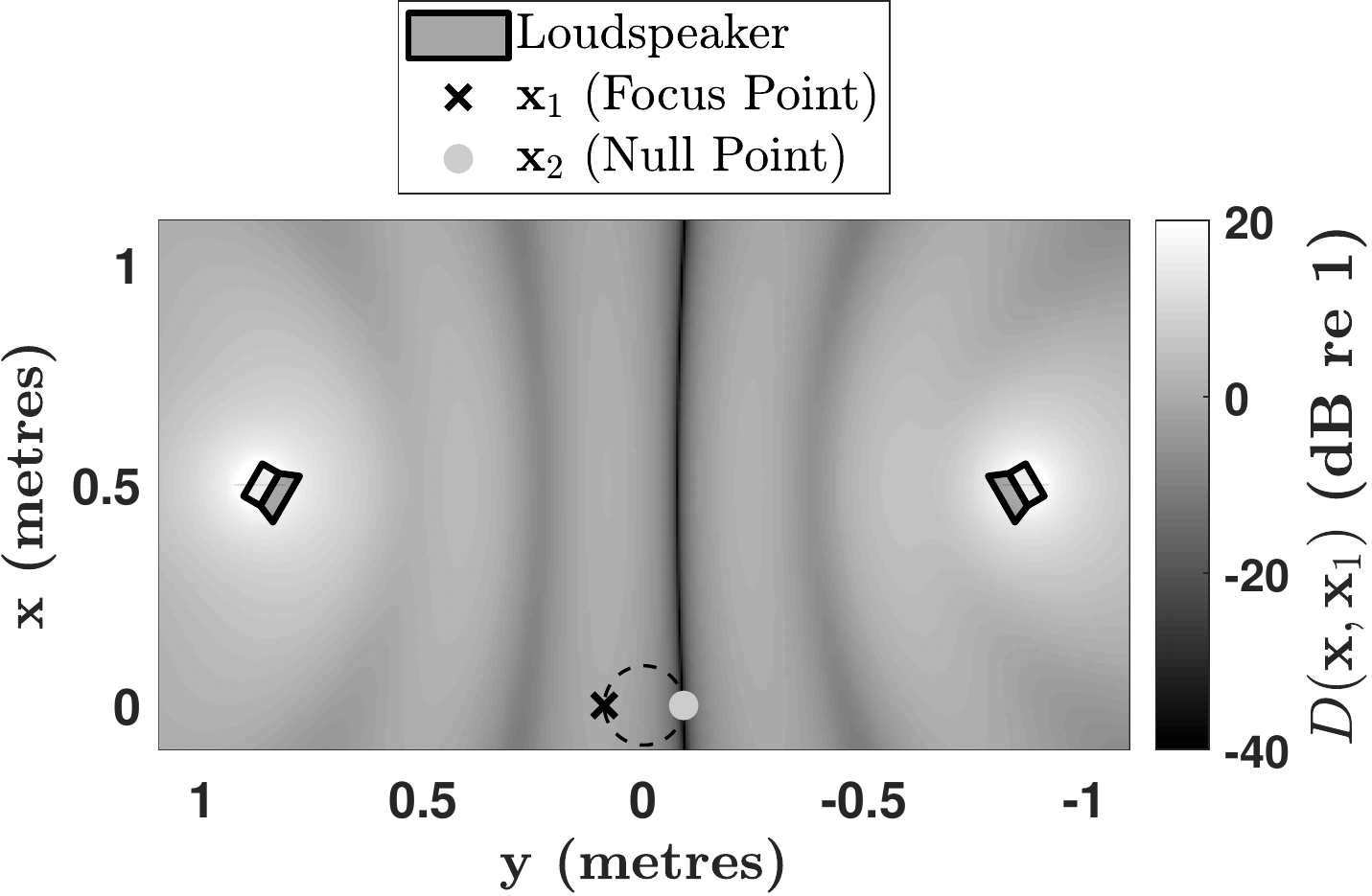}
		\caption{}
		\label{fig:Two-channel-CTC-Focus-Sound-Field-550Hz}
	\end{subfigure}
\hfill
	\begin{subfigure}[t]{0.49\linewidth}
		\centering
		\includegraphics[trim={0mm 0mm 0mm 0mm},clip,width=1\linewidth]{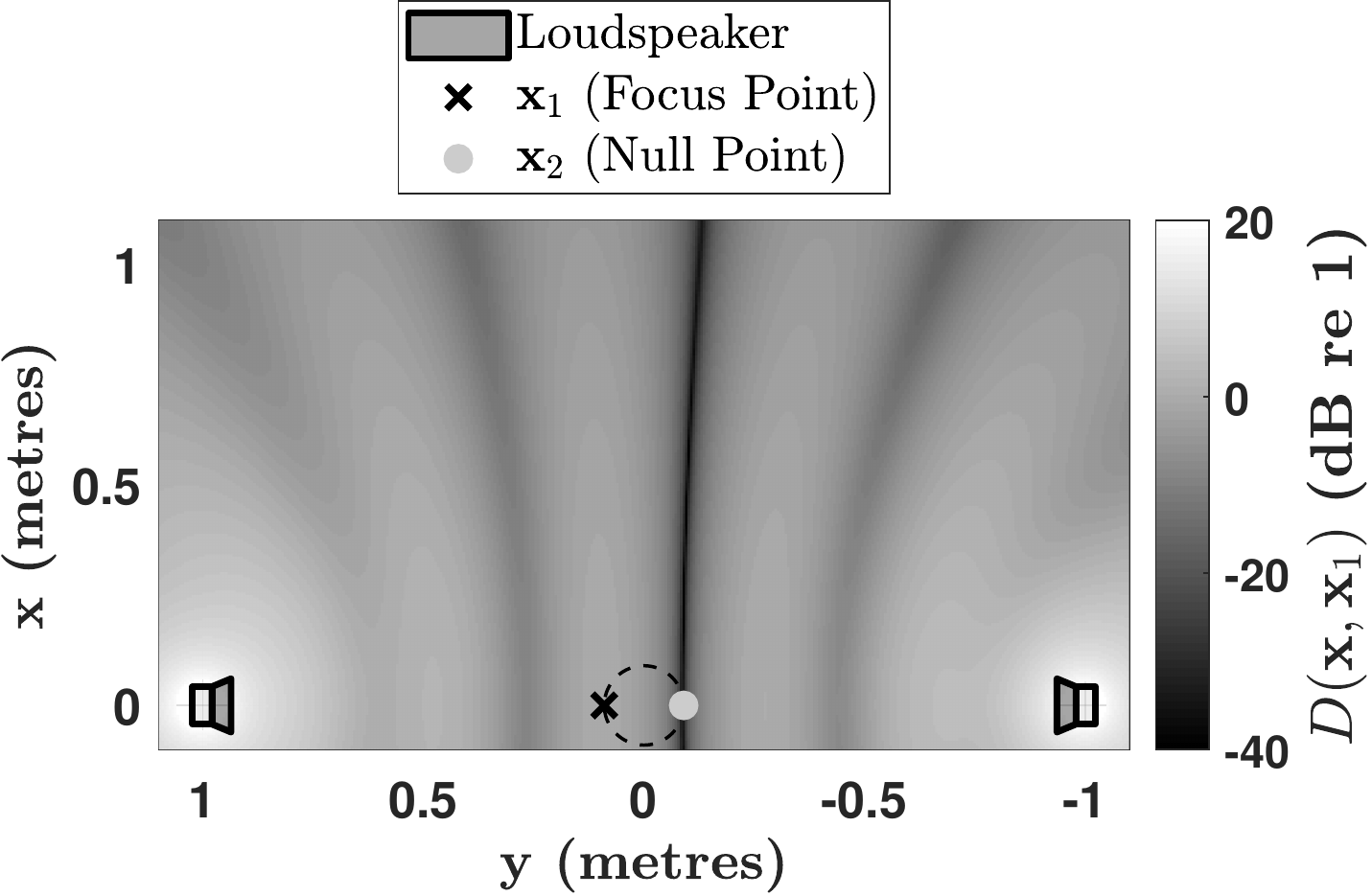}
		\caption{}
		\label{fig:Two-channel-CTC-Focus-Sound-Field-Low-Frequency}
	\end{subfigure}
	\caption{Symmetric two-channel super ideal focusing at various non-dimensional frequencies $\mu$ with changing angular span $\Delta \gamma$. a) $\mu\approx8.24$ and $\Delta\gamma =11^\circ$; b) $\mu\approx1.69$ and $\Delta\gamma =57^\circ$; c) $\mu\approx0.91$ and $\Delta\gamma =120^\circ$; d) $\mu\approx0.78$ and $\Delta\gamma =180^\circ$.}
	\label{fig:Two-channel-CTC-Focus-Sound-Field}
\end{figure}

The effect of the frequency-dependent angular span can be seen in \cref{fig:Two-channel-CTC-Focus-Sound-Field} which shows the magnitude response of the normalised beamforming gain in the horizontal plane for four non-dimensional frequencies, where non-dimensional frequency is defined as
\begin{equation}
	\mu\defined ka.
\end{equation}
The monopole sources have been placed one metre from the origin (centre of the shadowless head). Even at this close distance the conditions for ideal focusing established herein remain accurate. Note that $c=343$ m$\cdot$s$^{-1}$ and $a=0.09$ metres are used for all CTC examples throughout. 

For non-dimensional frequency $\mu\approx8.24$ the angular span is relatively narrow at $\Delta \gamma_{opt} \approx 11^\circ$, as shown in \cref{fig:Two-channel-CTC-Focus-Sound-Field-5000Hz}. As frequency decreases to $\mu\approx 1.69$ and further down to $\mu\approx 0.91$, the angular span increases accordingly to $\Delta\gamma_{opt} \approx 57^\circ$ and $\Delta\gamma_{opt} \approx 120^\circ$, respectively, as shown in \crefrange{fig:Two-channel-CTC-Focus-Sound-Field-1000Hz}{fig:Two-channel-CTC-Focus-Sound-Field-550Hz}. Finally, as shown in \cref{fig:Two-channel-CTC-Focus-Sound-Field-Low-Frequency}, when the frequency reaches the lowest possible at which ideal focusing can be maintained, here approximately $\mu\approx0.78$, the span is maximised to $\Delta \gamma_{opt} = 180^\circ$.   

\begin{figure}
	\centering
	\includegraphics[width=0.55\linewidth]{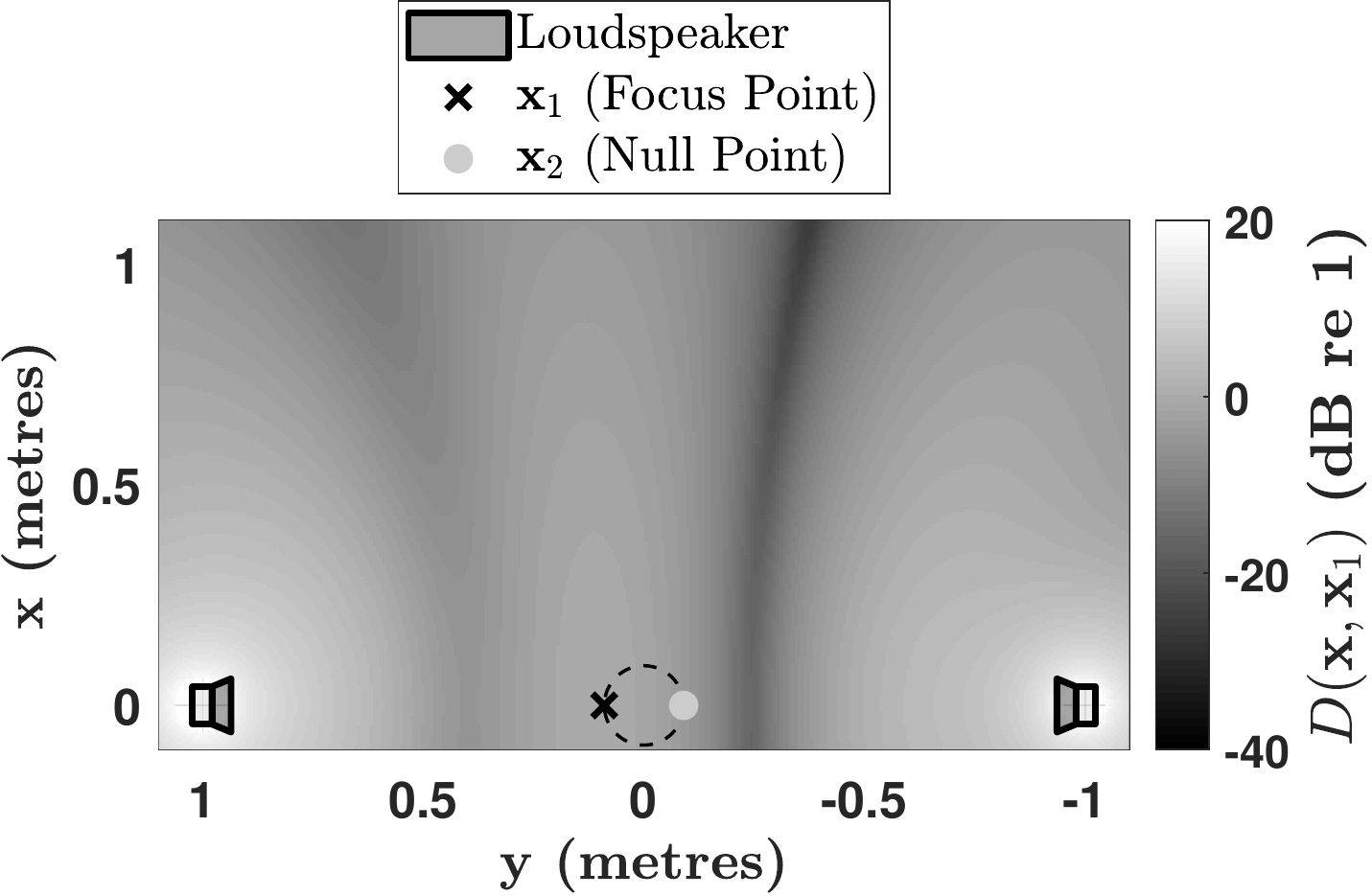}
	\caption{Example of the symmetric two-channel CTC system's inability to achieve ideal focusing at $\mu\approx 0.41$.}
	\label{fig:Two-channel-CTC-Focus-Sound-Field-250Hz}
\end{figure}

However, there is a frequency below which super ideal focusing cannot occur as shown by \cref{eq:J3-two-channel-opt-lambda-restriction-symmetric}. The maximum wavelength at which super ideal focusing can occur is
\begin{equation}
\label{eq:J3-two-channel-OSD-low-frequency}
\Lambda_{low} = 8a,
\end{equation}
i.e., when the wavelength is four times the control point separation. This corresponds to the maximum loudspeaker span $\Delta \gamma_{opt} = 180^\circ$. \cref{fig:Two-channel-CTC-Focus-Sound-Field-250Hz} shows, using the same geometry in \cref{fig:Two-channel-CTC-Focus-Sound-Field-Low-Frequency}, when the system is no longer able to achieve ideal focusing at low frequencies, in this example at $\mu\approx0.41$.

The result given by \cref{eq:J3-two-channel-opt-angle} was originally derived in \cite{Takeuchi2002}, albeit with different notation, which led to the concept and practical realisation of OSD. The same theoretical conclusions have been re-derived independently here and it has been shown that OSD is a special case of ideal focusing, specifically the more strict super ideal focusing state. With the geometric interpretation of the gramian given in \cref{sec:General Two-channel Ideal and Super Ideal Focusing} in mind, in OSD the parallelogram is a square and has a frequency-independent area for wavelengths less than $\Lambda_{low}$. 

Furthermore, these results provide an example of when compensated delay-and-sum beamforming, e.g., \cite{Olivieri2016}, satisfies the CTC inverse problem exactly and shows that OSD is a corresponding case (in its original theoretical formulation). This fact is due to the chosen acoustic model and that the pseudoinverse is given by the first two columns of \cref{eq:normalised-focusing-filters}, which are scaled delays under these acoustic assumptions. Thus, as it was derived, the theoretical OSD system is a compensated delay-and-sum beamformer that satisfies the inverse problem at two control points using a frequency-dependent two-channel loudspeaker span. Details on the practical discretisation and realisation of the OSD system can be found in \cite{Takeuchi2001, Takeuchi2001a, Takeuchi2002, Takeuchi2007, Takeuchi2008}.

The following section presents two further cases of super ideal focusing in CTC.
\subsubsection{Asymmetric Two-channel Example} 
\label{sec:Asymmetric Two-channel Example}
The following continues with the acoustic model used in \crefrange{sec:General Two-channel Ideal and Super Ideal Focusing}{sec:Symmetric Two-channel Example - Optimal Source Distribution} and presents the conditions for super ideal focusing when the geometry is asymmetric. In contrast to the symmetric two-channel case, this solution allows the loudspeakers to be placed in any direction in three-dimensional space and the two control points may rotate together in any orientation, so long as they remain diametrically opposed and the loudspeakers are equidistant from the origin. A far-field distance approximation is introduced, e.g., \cite{Hamdan2021}, to ensure that condition \cref{eq:J3-two-channel-ideal-focusing-cond-3} is satisfied. When $\norm{\sL{l}}\gg\norm{\xM{}}$ the sources are considered to be in the far-field and the plant matrix elements are approximately
\begin{align}
g_{1l} &\approx \delta \euler^{\Ij\bv{k}_l\trans\bv{x}_1},\\[5pt]
g_{2l} &\approx \delta \euler^{-\Ij\bv{k}_l\trans\bv{x}_1},
\end{align}
where $\delta \defined \euler^{-\Ij k R}/R$ is the common far-field propagation delay and attenuation term and $\bv{k}_l \defined k\hat{\bv{n}}_l$ is the $l$th wavevector, where $\hat{\bv{n}}_l\defined \sL{l}/\norm{\sL{l}}$ is the unit vector in the direction of the $l$th source.

With the stated assumptions, and according to \cref{eq:J3-two-channel-ideal-focusing-cond-1}, the following condition must be satisfied in this instance to maintain ideal focusing:
\begin{equation}
\label{eq:J3-two-channel-asymmetric-ideal-focusing-requirement}
\Delta\hat{\bv{n}}\trans\xM{1} = (2n-1)\frac{\Lambda_{opt}}{4},
\end{equation}
where $\Delta\hat{\bv{n}} \defined \hat{\bv{n}}_1-\hat{\bv{n}}_2$. \cref{eq:J3-two-channel-asymmetric-ideal-focusing-requirement} provides the condition for ideal focusing given the source direction and control point vectors. Note that the dot product can be expressed in terms of the angle between the two vectors, i.e., $\Delta\hat{\bv{n}}\trans\xM{1} = \norm{\Delta\hat{\bv{n}}}\norm{\xM{1}}\cos\theta$, where $\theta$ is the angle between $\Delta\hat{\bv{n}}$ and $\x{1}$. The relationship between wavelength and this angle is clearer in the form
\begin{equation}
\label{eq:J3-two-channel-asymmetric-lambda}
\Lambda_{opt} = \abs{\frac{4\norm{\Delta\hat{\bv{n}}}\norm{\xM{1}}\cos\theta}{(2n-1)}},
\end{equation}
where the absolute value is necessary as only positive wavelengths are permissible. Furthermore, the angle between the vectors $\Delta\hat{\bv{n}}$ and $\xM{1}$ is then
\begin{equation}
\label{eq:J3-two-channel-asymmetric-angle}
\theta = \cos^{-1}\left((2n-1)\frac{\Lambda_{opt}}{4\norm{\Delta\hat{\bv{n}}}\norm{\xM{1}}}\right),
\end{equation}  
with the restriction
\begin{equation}
\label{eq:J3-two-channel-asymmetric-lambda-restriction}
\abs{(2n-1)\Lambda_{opt}}\leq 4\norm{\Delta\hat{\bv{n}}}\norm{\xM{1}}.
\end{equation}

There is a direct relationship between $\theta$ and the wavelength at which ideal focusing can occur. Consider the case when $n=1$ and assume that the loudspeaker positions are fixed but rotation of the control points is allowed. As $\theta\rightarrow 0$ the wavelength increases until reaching its maximum of
\begin{equation}
\Lambda_{low} = 4\norm{\Delta\hat{\bv{n}}}\norm{\xM{1}}.
\end{equation}
On the other hand, as $\theta\rightarrow \pi/2$ the wavelength tends to zero. When $\theta = \pi/2$ the system becomes unstable and corresponds to an infinite frequency. 

\begin{figure}
	\centering
	\begin{subfigure}[t]{1\linewidth}
		\centering
		\includegraphics[width=0.65\linewidth]{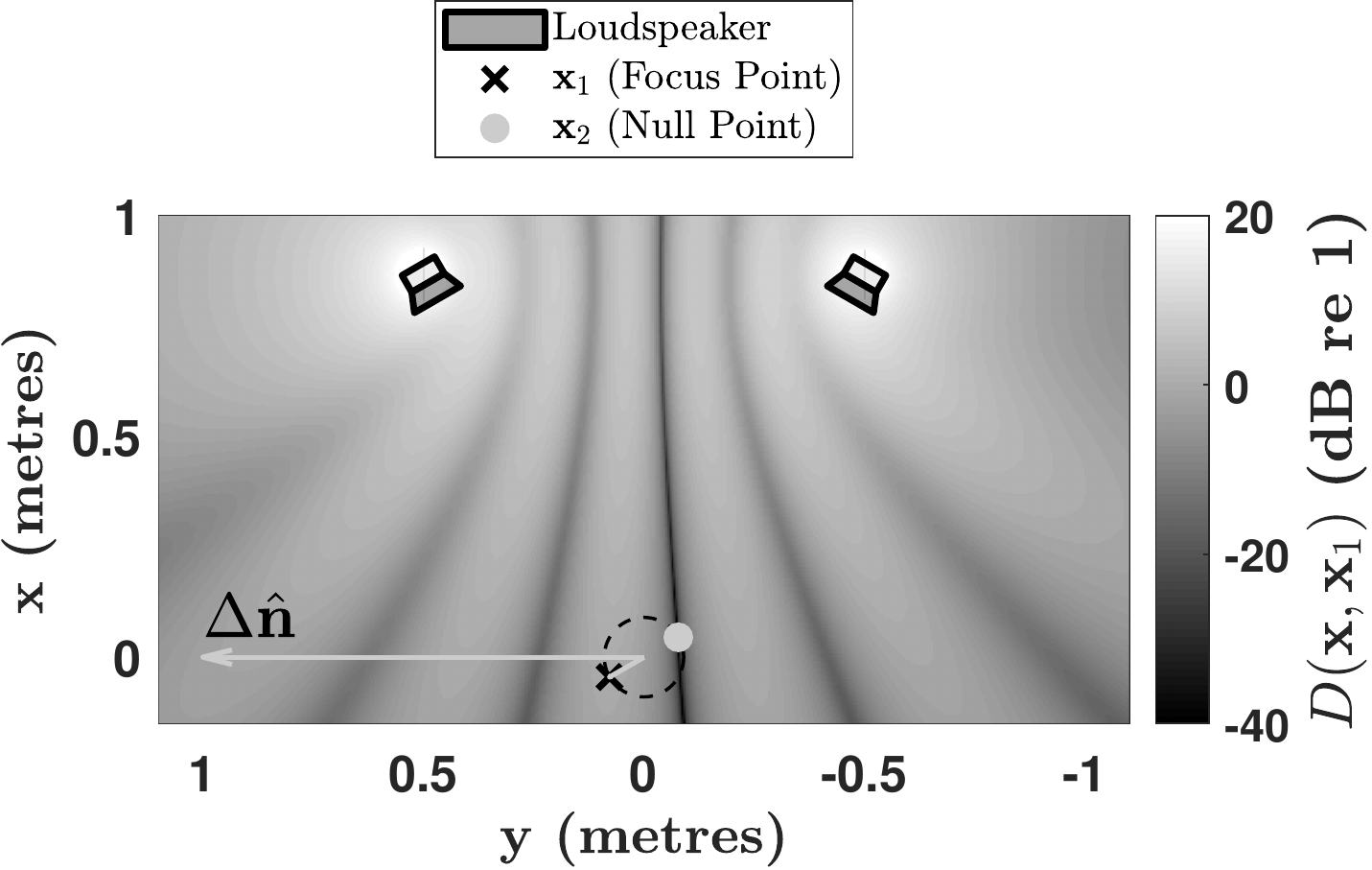}
		\caption{}
		\label{fig:Two-channel-CTC-Focus-Sound-Field-30deg-Head-Rot}
	\end{subfigure}
	\begin{subfigure}[t]{1\linewidth}
		\centering
		\includegraphics[trim={0mm 0mm 0mm 20.75mm},clip,width=0.65\linewidth]{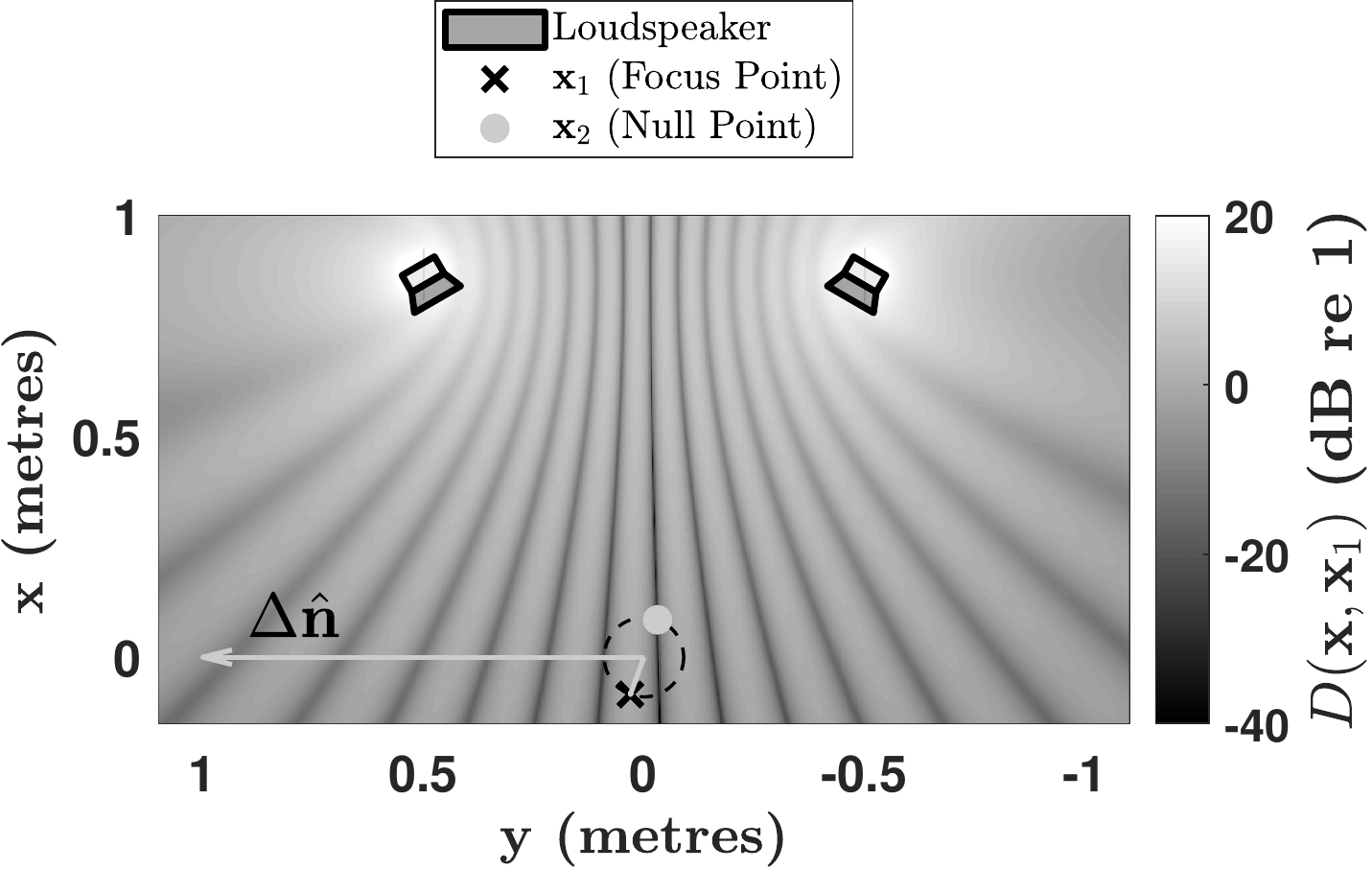}
		\caption{}
		\label{fig:Two-channel-CTC-Focus-Sound-Field-70deg-Head-Rot}
	\end{subfigure}
	\begin{subfigure}[t]{1\linewidth}
		\centering
		\includegraphics[trim={0mm 0mm 0mm 20.75mm},clip,width=0.65\linewidth]{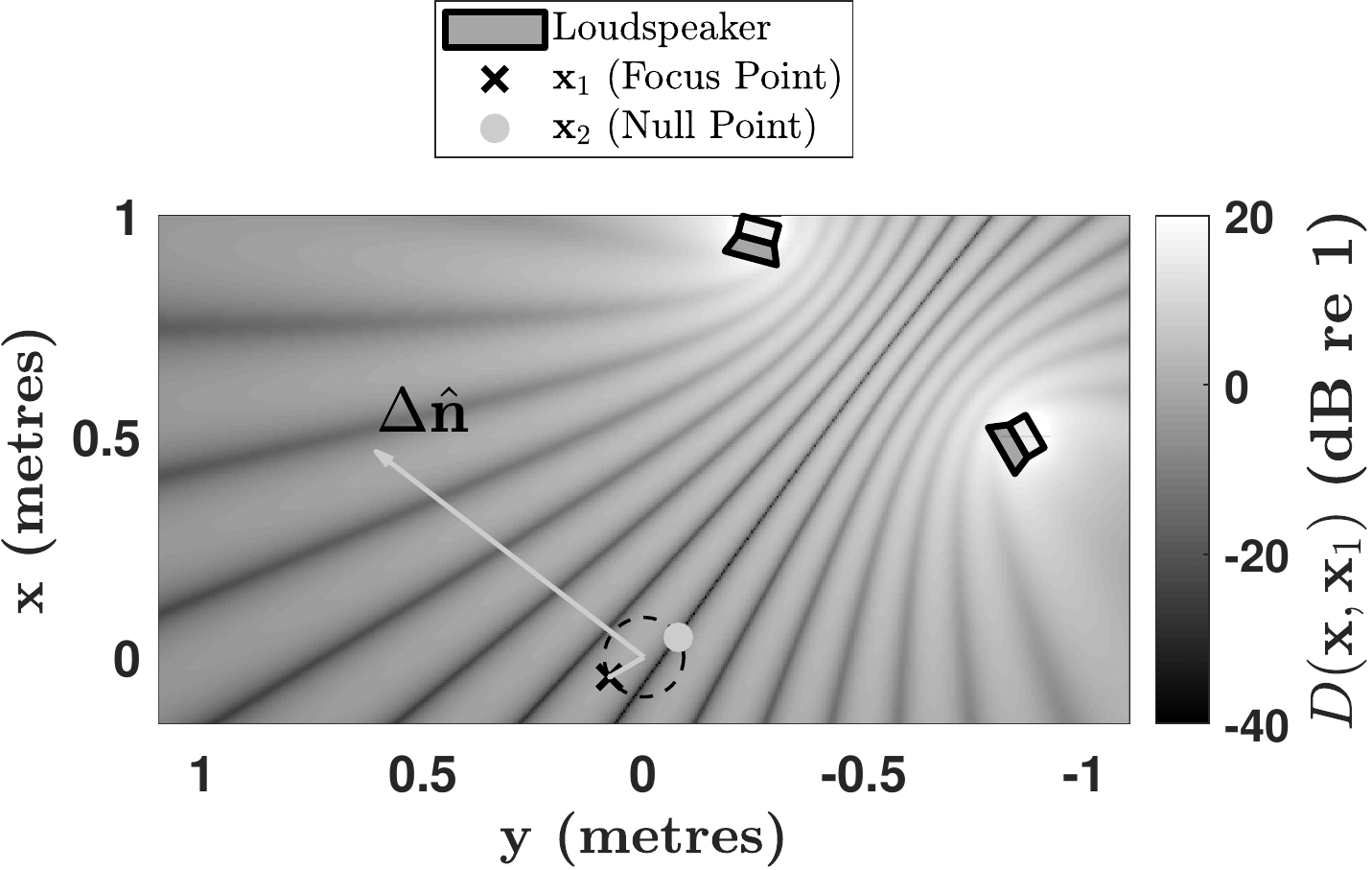}
		\caption{}
		\label{fig:Two-channel-CTC-Focus-Sound-Field-Asymmetric}
	\end{subfigure}
	\caption{Two-channel asymmetric super ideal focusing for: a) head rotation $30^\circ$ and $\mu\approx1.81$; b) head rotation $70^\circ$ and $\mu\approx4.59$; c) completely asymmetric, $\mu\approx 5.36$.}
	\label{fig:Two-channel-CTC-Focus-Sound-Field-Asymmetric-Examples}
\end{figure}

The dependence of ideal focusing on control point rotation can be understood beginning with the symmetric example. \cref{fig:Two-channel-CTC-Focus-Sound-Field-Asymmetric-Examples} shows the magnitude response of the normalised beamforming gain in the horizontal plane for three exemplary cases. \cref{fig:Two-channel-CTC-Focus-Sound-Field-30deg-Head-Rot} and \cref{fig:Two-channel-CTC-Focus-Sound-Field-70deg-Head-Rot} show a two-channel system with sources symmetric about the $x$ axis with angular span $\Delta \gamma = 60^\circ$ while the control points are rotated $30^\circ$ and $70^\circ$, respectively. \cref{eq:J3-two-channel-asymmetric-lambda} was used to calculate the frequencies at which the given geometries achieve super ideal focusing, which can be observed in both cases. In these two cases the vector $\Delta\hat{\bv{n}}$ lies along the $y$ axis, i.e., the interaural axis when in the centred forward-facing position. Thus in both \crefrange{fig:Two-channel-CTC-Focus-Sound-Field-30deg-Head-Rot}{fig:Two-channel-CTC-Focus-Sound-Field-70deg-Head-Rot} the angle $\theta$ is exactly the head rotation angle. As predicted by \cref{eq:J3-two-channel-asymmetric-lambda}, the frequency is lower in \cref{fig:Two-channel-CTC-Focus-Sound-Field-30deg-Head-Rot} than it is in \cref{fig:Two-channel-CTC-Focus-Sound-Field-70deg-Head-Rot} as a result of the smaller head rotation value, where in the former the frequency is approximately $\mu\approx 1.81$ and in the latter approximately $\mu\approx 4.59$. However, the sources need not lie about any particular axis. \cref{fig:Two-channel-CTC-Focus-Sound-Field-Asymmetric} provides an example of both head rotation and an off-centre loudspeaker placement where the frequency is approximately $\mu\approx5.36$. Regardless of the exact placement of the sources and head rotation the same principles apply, as in the symmetric examples, in that the angle $\theta$ is key to determining what frequencies at which the system can achieve ideal focusing. 

\cref{eq:J3-two-channel-asymmetric-ideal-focusing-requirement} can also yield the angles between the loudspeaker unit vectors and the first control point vector by rearranging to
\begin{equation}
\label{eq:J3-two-channel-asymmetric-angles}
\cos\theta_{11} = \cos{\theta_{12}} + (2n-1)\frac{\Lambda_{opt}}{4a},
\end{equation}
where $\theta_{11}$ and $\theta_{12}$ are the angles between the two loudspeaker direction vectors, $\hat{\bv{n}}_1$ and $\hat{\bv{n}}_2$, and the control point $\bv{x}_1$, respectively. \cref{eq:J3-two-channel-asymmetric-angles} requires that
\begin{equation}
\label{eq:J3-two-channel-asymmetric-angles-constraint}
\abs{\cos{\theta_{12}} + (2n-1)\frac{\Lambda_{opt}}{4a}}\leq 1.
\end{equation}
Using \cref{eq:J3-two-channel-asymmetric-angles} one angle can be specified given the constraint given by \cref{eq:J3-two-channel-asymmetric-angles-constraint}, which then determines the other angle. 
\subsubsection{$L\geq 2$ Example - Symmetric Uniform Path Length Difference Array}
\label{sec:Symmetric Uniform Path Length Difference Array}
\begin{figure}
	\centering
	\includegraphics[trim=125 300 125 290,clip,width=0.5\linewidth]{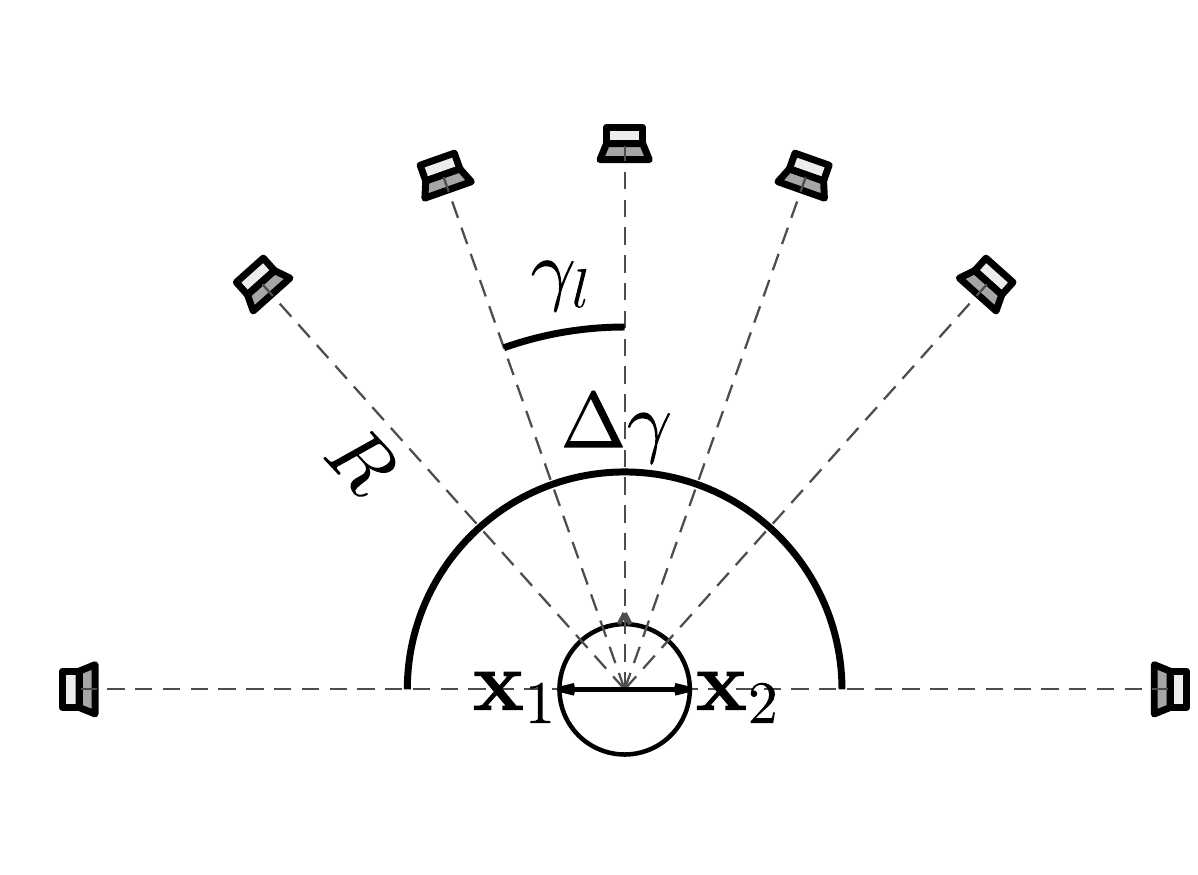}
	\caption{Top-down view of a symmetric \uplda geometry with $L=7$.}
	\label{fig:J2-upda-schematic}
\end{figure}

Consider the symmetric multichannel array geometry where the path length differences are uniformly distributed, i.e., $\pld{i+1}=\pld{i}+\Delta\pld{}$, where 
\begin{equation}
	\label{eq:uniform-path-length-difference}
	\Delta\eta \defined \frac{2\pld{max}}{L-1} \approx \frac{4a\sin\gamma_{max}}{L-1}.
\end{equation} 
Note that $\pld{max}\defined \mathrm{max}(\abs{\pld{i}})$ is the maximum absolute path length difference (corresponding to the widest symmetric two-channel pair) and $\gamma_{max}\defined\Delta\gamma/2$ is half the angular span of the array. This geometry is called the uniform path length difference array (UPDA). An example schematic of a UPDA with angular span $\Delta\gamma = 180^\circ$ and $L=7$ is shown from the top-down perspective in \cref{fig:J2-upda-schematic}. Note that the control points are assumed to be diametrically opposed. The focusing crosstalk was previously derived under the same stated symmetric far-field acoustic model (without assuming the UPDA geometry specifically) by the authors in \cite{Hamdan2021}. For the exact details of this derivation see Appendix A.2 in \cite{Hamdan2021}. Using this previously derived focusing crosstalk expression and considering the UPDA geometry requirement given by \cref{eq:uniform-path-length-difference}, the symmetric UPDA focusing crosstalk can be written as
\begin{align}
	\label{eq:UPDA-focusing-xt-sum}
	\gram{1}{2}=\gram{2}{1}&\approx\frac{1}{R^2}\sum_{l=-\frac{L-1}{2}}^{\frac{L-1}{2}}\cos\left(kl\Delta\pld{}\right), \\[10pt]
	\label{eq:UPDA-focusing-xt-simplified}
	&= \frac{\sin\left(\frac{Lk\pld{max}}{L-1}\right)}{R^2\sin\left(\frac{k\pld{max}}{L-1}\right)}.
\end{align}
The derivation of \cref{eq:UPDA-focusing-xt-simplified} is given in \cref{sec:Derivation of UPDA Focusing Crosstalk}.
Inspection shows that the zeros of \cref{eq:UPDA-focusing-xt-simplified} occur when
\begin{equation}
\label{eq:J3-UPLDA-ideal-focusing-lambda}
\Lambda_{opt} = \abs{\frac{4La\sin\gamma_{max}}{n(L-1)}},
\end{equation}
where $n\in\mathbb{Z}, n\neq zL, z\in\mathbb{Z}$. 

The low frequency limit on the symmetric UPDA ideal focusing can be found by considering \cref{eq:J3-UPLDA-ideal-focusing-lambda} when $n=1$. For a fixed head radius and number of loudspeakers \cref{eq:J3-UPLDA-ideal-focusing-lambda} is maximised when the angular span is maximised to $180^\circ$, yielding
\begin{equation}
\label{eq:J3-UPLDA-ideal-focusing-lambda-low}
\Lambda_{low} = \frac{4La}{L-1}.
\end{equation} 
When $L$ is allowed to vary \cref{eq:J3-UPLDA-ideal-focusing-lambda-low} states that the largest wavelength at which ideal focusing occurs is when $L=2$, i.e., $\Lambda_{low}=8a$. On the other hand, in the limit of infinite source density within a fixed angular span the largest possible wavelength becomes
\begin{equation}
\label{eq:UPDA-cont-low-limit}
\lim\limits_{L\rightarrow\infty}\Lambda_{low} = 4a,
\end{equation}
which is half the wavelength of the largest possible wavelength of the symmetric two-channel CTC system under the stated acoustic assumptions. \cref{eq:J3-UPLDA-ideal-focusing-lambda-low} states that for a fixed head radius the low frequency limit on ideal focusing monotonically increases in frequency as the source density increases (within the assumed $180^\circ$ angular span) with a limit given by \cref{eq:UPDA-cont-low-limit}.

\begin{figure}[t]
	\centering
	\begin{subfigure}[t]{0.49\linewidth}
		\centering
		\includegraphics[width=1\linewidth]{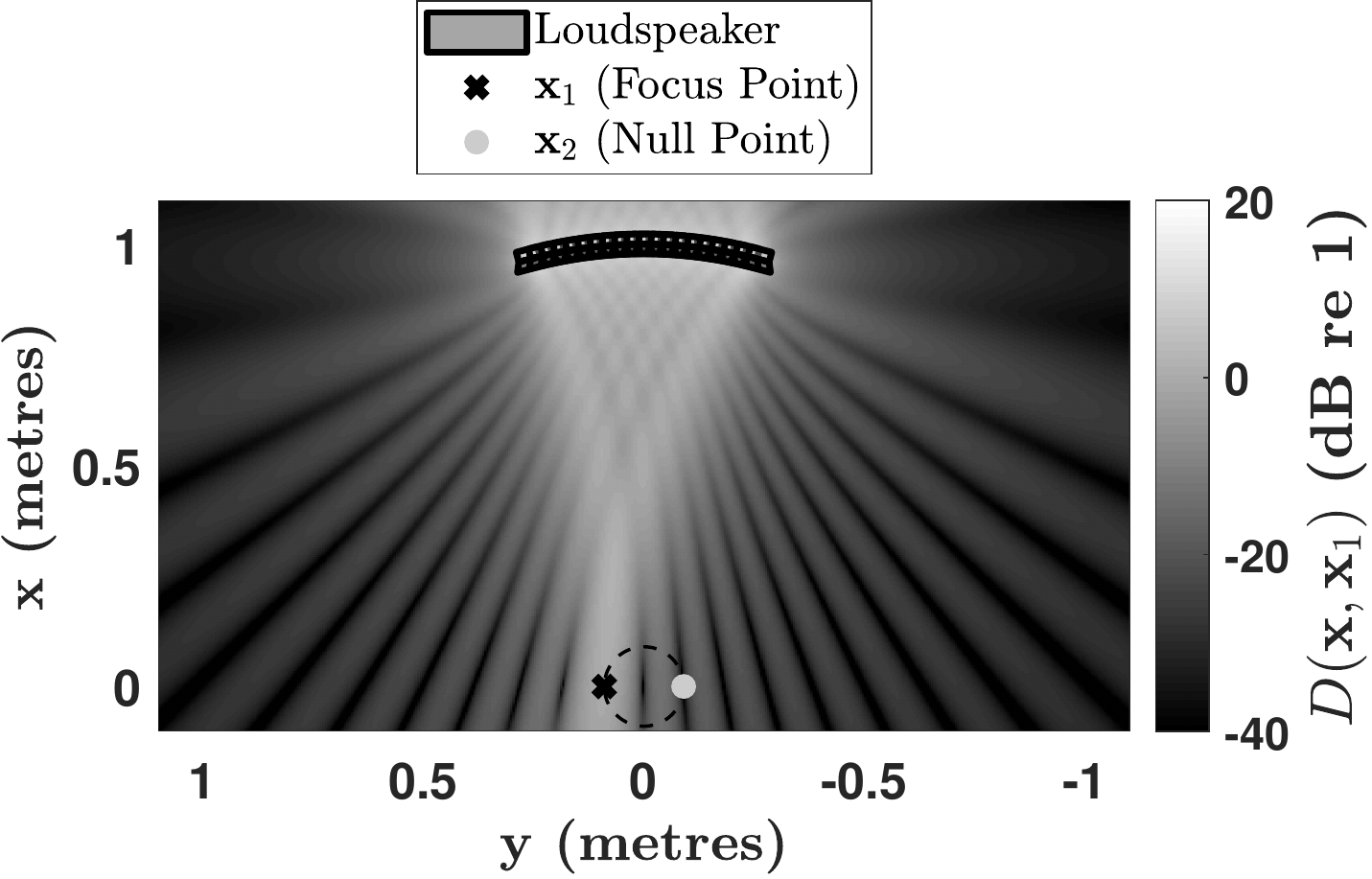}
		\caption{}
		\label{fig:UPLDA-Twenty-Channel-CTC-Focus-Sound-Field}
	\end{subfigure}
\hfill
	\begin{subfigure}[t]{0.49\linewidth}
		\centering
		\includegraphics[trim={0mm 0mm 0mm 0mm},clip,width=1\linewidth]{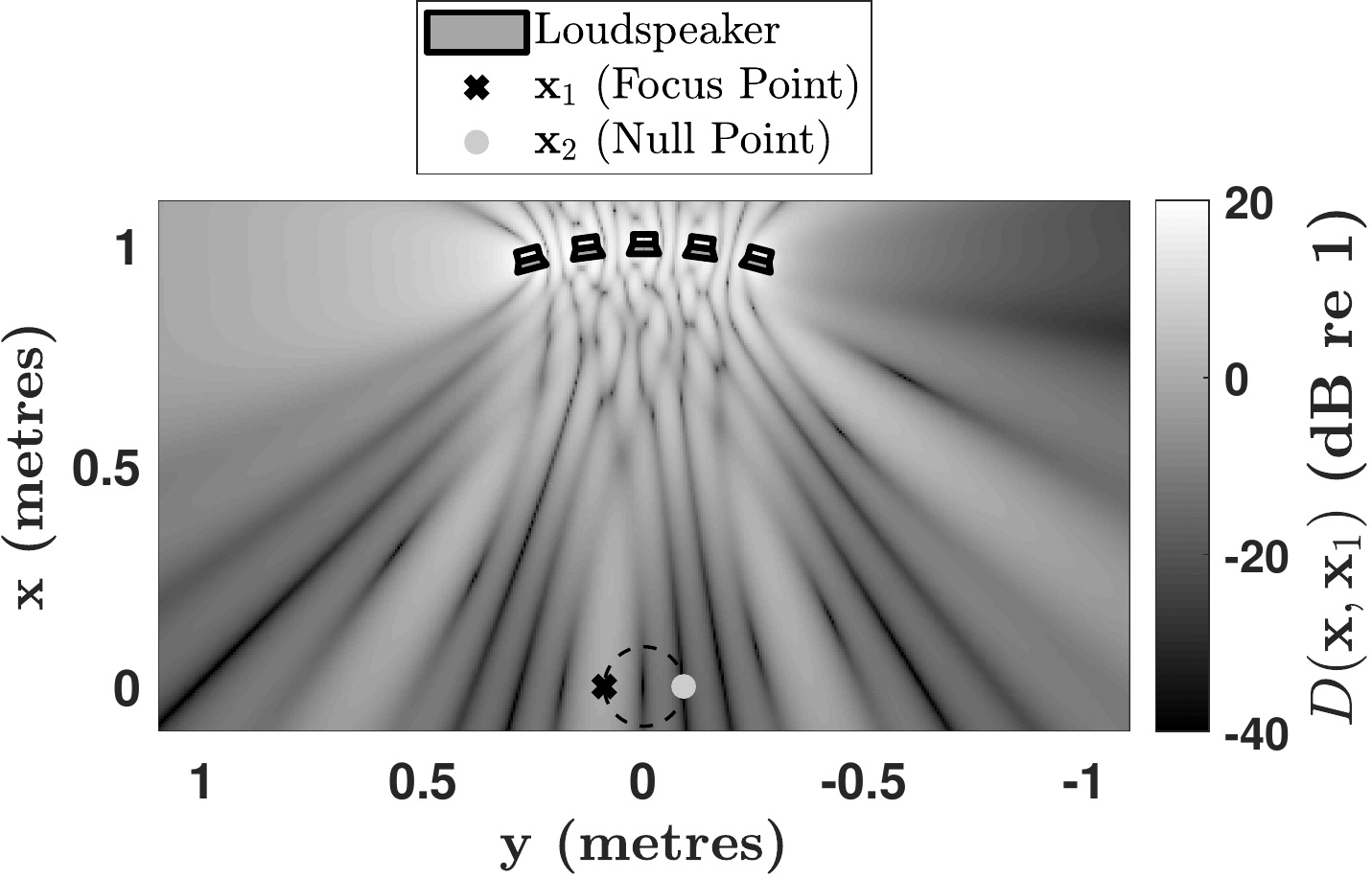}
		\caption{}
		\label{fig:UPLDA-Five-Channel-CTC-Focus-Sound-Field}
	\end{subfigure}
	\caption{Symmetric UPDA super ideal focusing for: a) $\mu\approx11.53$ and $L=20$; b) $\mu\approx9.71$ and $L=5$.}
	\label{fig:UPLDA-CTC-Focus-Sound-Field}
\end{figure}

While the conditions for ideal focusing for the UPDA have been given, the benefits over the two-channel system are not immediately clear from this perspective. However, as shown by the authors in \cite{Hamdan2021}, broadband multichannel CTC systems with a high density of sources can improve focusing quality and thereby reduce the focusing crosstalk naturally at certain frequencies. This increased focusing quality, in addition to the ability to achieve ideal focusing, is exemplified in \cref{fig:UPLDA-Twenty-Channel-CTC-Focus-Sound-Field} for an example twenty-channel UPDA with angular span $\Delta \gamma=60^\circ$. In this example the UPDA achieves ideal focusing at approximately $\mu\approx 11.53$. Furthermore, it can be seen that the focusing quality has improved over the two-channel system (compare to \crefrange{fig:Two-channel-CTC-Focus-Sound-Field-5000Hz}{fig:Two-channel-CTC-Focus-Sound-Field-Low-Frequency}). A clear beam has been formed and the radiation to other points in the space has been greatly reduced in magnitude. Thus, the improved beamforming capability has an important practical benefit in this example, in that, the area surrounding the null point is also relatively low in focusing crosstalk magnitude, where this is generally not the case with the two-channel system. However, if the UPDA source density is not sufficiently large, grating lobes will occur (as they do periodically with the two-channel system). This can be seen in \cref{fig:UPLDA-Five-Channel-CTC-Focus-Sound-Field} with a five-channel UPDA (with the same angular span as in \cref{fig:UPLDA-Twenty-Channel-CTC-Focus-Sound-Field}) where ideal focusing occurs at approximately $\mu\approx 9.71$. It can be seen that at other points in the space grating lobes occur, which can lead to unwanted focusing crosstalk at the null point in a practical scenario, i.e., within a volume with reflections.
\subsection{Sound Zones ($M\times L$)}
\label{sec:J3-Sound-Zones}
The application of creating $M$ arbitrary sound zones in the far-field using a uniform linear array is examined in light of the theoretical development given in the previous sections. The intent is to establish when ideal focusing is achieved at $M$ prescribed points. 
\subsubsection{$M$-point Super Ideal Focusing: Far-field ULA}
\begin{figure}
	\centering
	\includegraphics[width=0.4\linewidth]{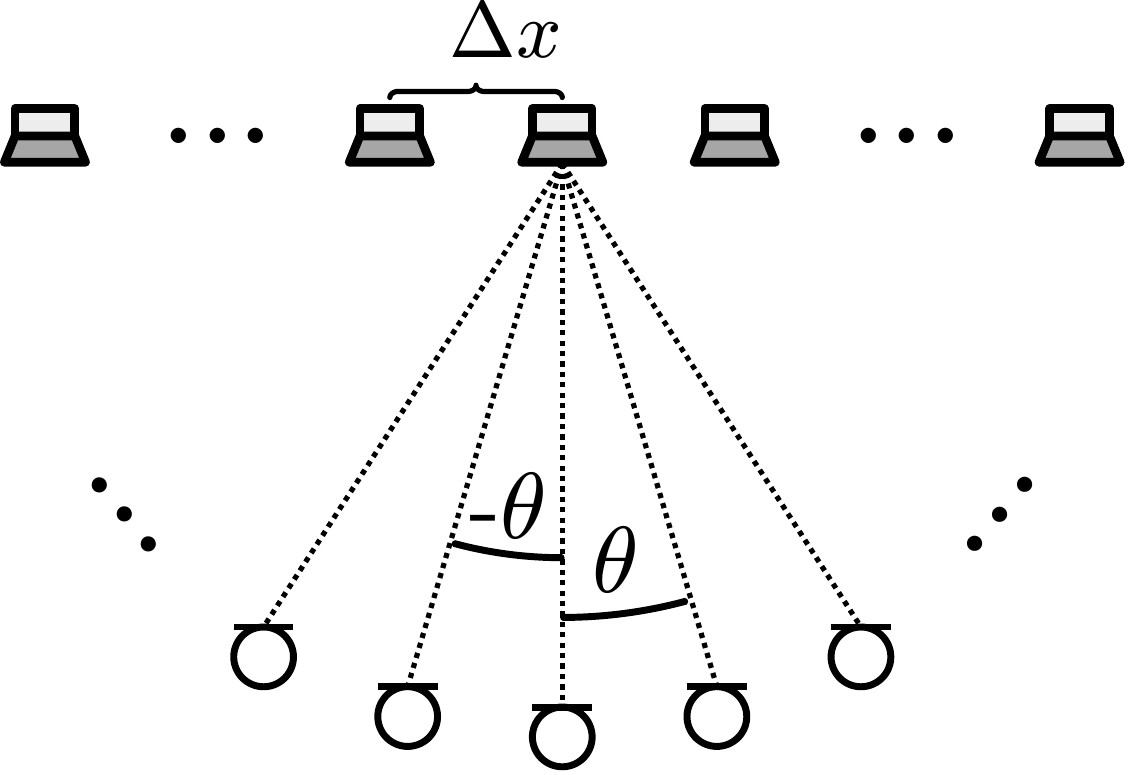}
	\caption{ULA geometry with inter-element spacing $\Delta x$. $M$ control points are positioned in the far-field at various angles.}
	\label{fig:ula-symmetric-m-point-schematic}
\end{figure}

The geometry under consideration is shown in \cref{fig:ula-symmetric-m-point-schematic} (although the points need not be symmetric about zero degrees as depicted in the figure). The ULA is comprised of $L$ monopole sources with inter-element spacing $\Delta x$ (in metres). The $M$ control points are assumed to lie in the far-field. Furthermore, with this far-field assumption the distance attenuation and propagation is neglected, thus the sources are treated as unit plane waves. For simplicity, the normalised beamforming gain is used to represent the magnitude of the focused sound-field. Under the stated assumptions, the normalised beamforming gain is the far-field ULA directivity pattern, e.g., \cite{Theodoridis2013}, written here as
\begin{align}
D(\theta, \theta_0) &= \frac{\abs{\gM{}(\theta_0)\herm\gM{}(\theta)}}{\norm{\gM{}(\theta_0)}^2}, \\[10pt]\nonumber
&= \abs{\frac{\sin\left(\frac{\pi}{\alpha}\left(\sin\theta-\sin\theta_0\right)\right)}{L\sin\left(\frac{\pi}{L\alpha}\left(\sin\theta-\sin\theta_0\right)\right)}},
\end{align}
where 
\begin{equation}
\alpha \defined \Lambda/L\Delta x,
\end{equation}
and $\theta$ and $\theta_0$ are the measurement and steering angles, respectively. Also note that $\norm{\gM{}(\theta)}^2=\norm{\gM{}(\theta_0)}^2=L$, thus, the norm requirement for super ideal focusing has been met. Now, consider $M$ discrete control point angles as directions into the far-field where ideal focusing is sought. The directivity pattern for focusing at one of these points while measuring at any other is then $\uladirect$, where $\theta = \theta_i$ and $\theta_0 = \theta_j$, and $i,j=1, \dots, M$.

\begin{figure}
	\centering
	\includegraphics[trim={11mm 60mm 11mm 45mm},clip,width=0.55\linewidth]{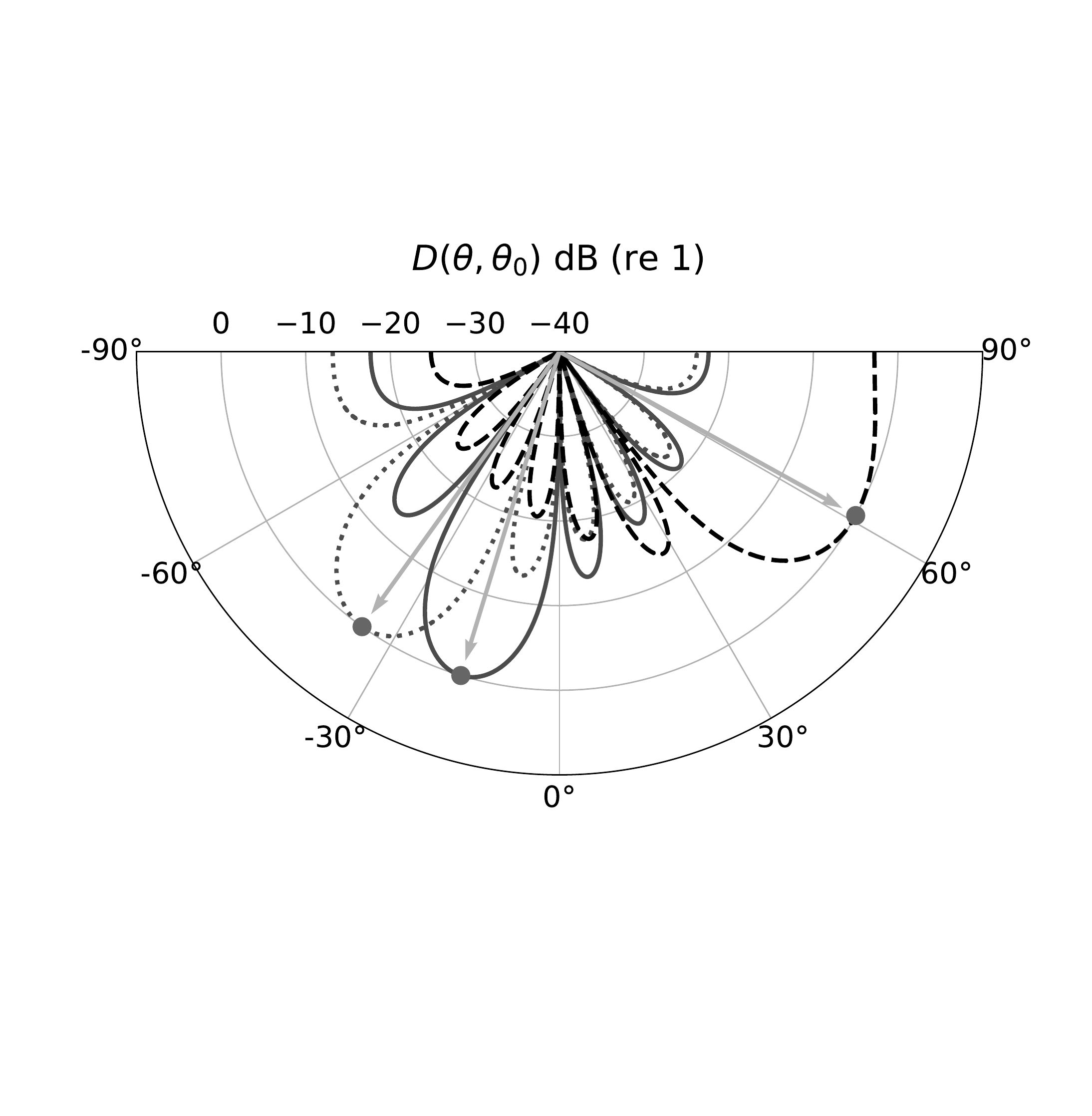}
	\caption{Far-field ULA super ideal focusing when the control points (grey circles) are asymmetric. $M=3$ and $f \approx 4899$ Hz.}
	\label{fig:ula-3-point-ideal-focusing-asymmetric2}
\end{figure}

It is important to note that $D(\theta_i, \theta_j)$ represents the normalised magnitude of the $ij$th entry of the Gram in this scenario. Thus the conditions on the control point angles that result in $\uladirect=0, i\neq j$ are sought and are found by considering when the zeros of $\uladirect$ occur. This investigation leads to the condition
\begin{align}
\label{eq:J3-ULA-super-ideal-condition}
\sin\theta_i-\sin\theta_j 
&= n_{ij}\alpha,
\end{align}
where $n_{ij}\in\mathbb{Z}$, $n_{ij}\neq zL, z\in\mathbb{Z}$, and $i\neq j$. When $n_{ij}=zL$ grating lobes occur, i.e., $\uladirect= 1$. Note that the sine terms can be written as
\begin{align}
\label{eq:J3-ula-sinei}
\sin\theta_i &= \mu_i \alpha, \\[5pt]
\label{eq:J3-ula-sinej}
\sin\theta_j &= \mu_j \alpha,
\end{align}
where $\mu_i,\mu_j\inRv{}$. Since $n_{ij}$ is an integer, and the arcsine's domain is limited, the following condition must be met:
\begin{align}
\label{eq:J3-ula-sine-integer-constraint}
\mu_i-\mu_j = n_{ij}, \\[5pt]
\label{eq:J3-ula-arcsine-constraint}
\abs{\mu_{i,j}\alpha} \leq 1.
\end{align}
Thus the conditions for the far-field ULA directivity to achieve super ideal focusing at $M$ points are given by \crefrange{eq:J3-ULA-super-ideal-condition}{eq:J3-ula-arcsine-constraint}. Under these conditions the system will be optimally-conditioned. An example of the resulting directivity for an example twenty-channel ULA with $\Delta x = 1.2$ cm  is shown in \cref{fig:ula-3-point-ideal-focusing-asymmetric2} for a control frequency of approximately $4899$ Hz. It can be seen that super ideal focusing occurs at all three points simultaneously. 

It is important to note that \cref{eq:J3-ULA-super-ideal-condition} can be rearranged in various ways to see how a change in one physical parameter requires a change in the others involved. To gain a clearer understanding of these physical relationships, the following section analyses the case for a symmetric set of control points centred about $\theta = 0^\circ$.
\subsubsection{Symmetric $M=3$} 
\label{sec:Symmetric 3 Points Centred About 0}
\begin{figure}
	\centering
	\includegraphics[trim={11mm 60mm 11mm 45mm},clip,width=0.55\linewidth]{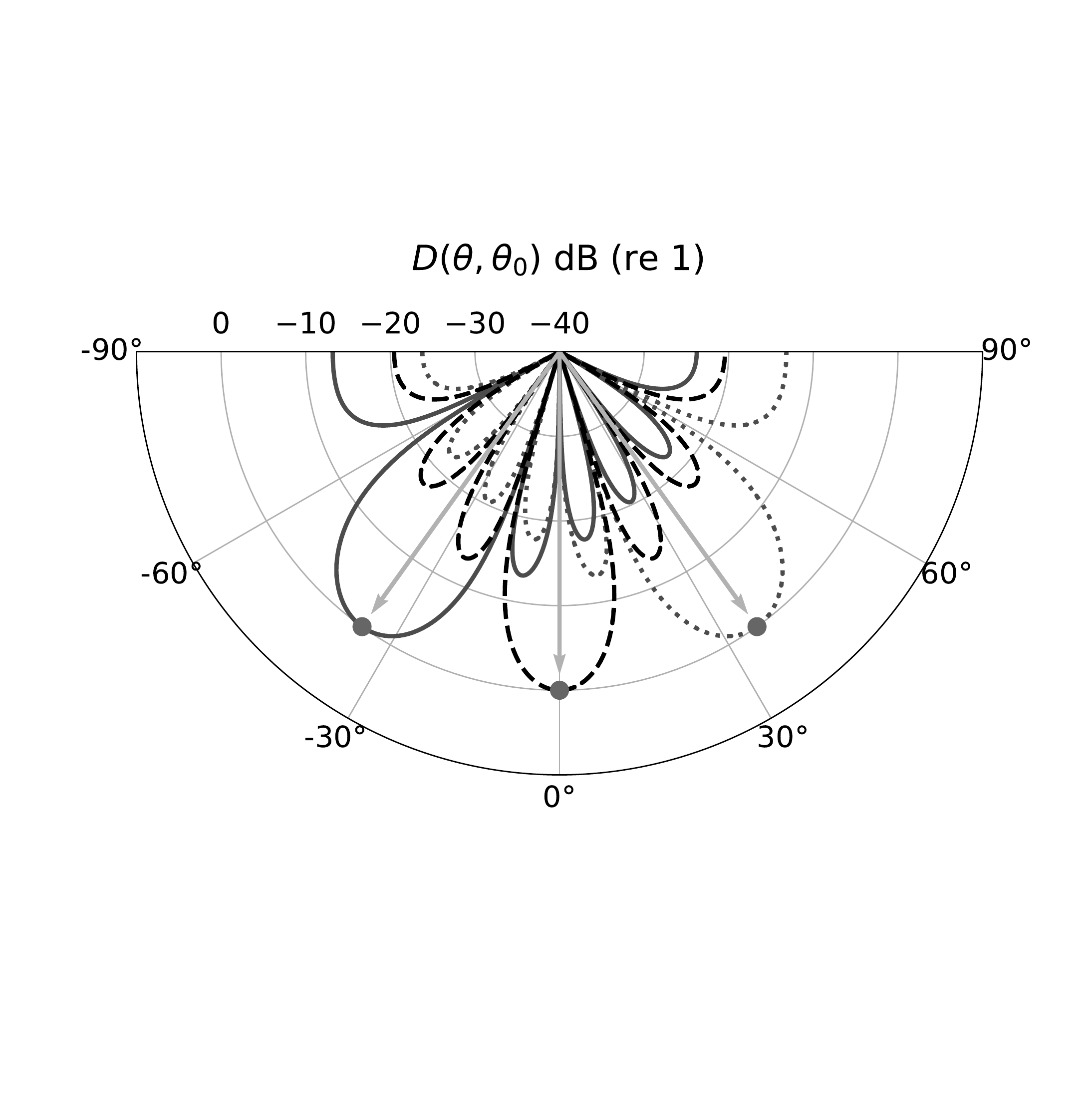}
	\caption{Far-field ULA super ideal focusing when the control points (grey circles) are symmetric about zero degrees. $M=3$ and $f \approx 4899$ Hz.}
	\label{fig:ula-3-point-ideal-focusing-symmetric}
\end{figure}

The following considers the case when the $M$ controls points are symmetric about zero degrees, as depicted in \cref{fig:ula-symmetric-m-point-schematic}. First, consider the exemplary case when $M=3$ and one of the control point angles is chosen to be $\theta_1 = 0^\circ$, i.e., $\mu_1 = 0$. Let the other two points be symmetric about zero degrees such that $\theta_2 = -\theta_3$. Note that the requirement given by \cref{eq:J3-ula-sine-integer-constraint} is strict {for any pair of angles} so it is simplest to choose $\mu_2=1$ and $\mu_3=-1$ (note that if $M$ were an even integer fractional values could be used). In this case $\theta_2 = -\theta_3= \sin^{-1}(\alpha)$. An example of the resulting directivity pattern is given in \cref{fig:ula-3-point-ideal-focusing-symmetric}, for the same twenty-channel ULA used in \cref{fig:ula-3-point-ideal-focusing-asymmetric2}, with symmetric control point angles $\theta_2 =-\theta_3\approx 35.7^\circ$. The control frequency is approximately 4899 Hz. It can be seen that the zeros in the directivity patterns for each point align exactly with the focus directions of the other points such that super ideal focusing is achieved at all three points.

From \cref{eq:J3-ula-arcsine-constraint} it can be seen that the maximum wavelength at which ideal focusing can occur (in this scenario) is limited by $L\Delta x$, i.e., the array length plus $\Delta x$. This yields the inequality
\begin{equation}
\label{eq:J3-ula-focusing-wavelength-inequality}
\Lambda\leq L\Delta x.
\end{equation}
\cref{eq:J3-ula-focusing-wavelength-inequality} also relates the number of required loudspeakers to the wavelength and inter-element spacing. When the wavelength increases the inter-element spacing must increase proportionally for a fixed number of loudspeakers or the number of loudspeakers must increase for a fixed spacing. On the other hand, when the wavelength decreases, the spacing must decrease proportionally for a fixed number of loudspeakers or the number of loudspeakers must decrease for a fixed spacing. Considering \cref{eq:J3-ula-sinei} and \cref{eq:J3-ula-sinej}, the array length can be determined by
\begin{align}
\label{eq:J3-ula-ideal-array-length}
L\Delta x &= \frac{\Lambda}{\sin\theta_{2}},
\end{align}
where only $\sin\theta_2$ need be considered (thus $0^\circ<\theta_2\leq90^\circ$).
Alternatively, one can begin with a given ULA dimension and a set of desired control frequencies and use \cref{eq:J3-ula-ideal-array-length} to calculate what angles the control points should be positioned at to achieve ideal focusing.

From \cref{eq:J3-ula-focusing-wavelength-inequality} and \cref{eq:J3-ula-ideal-array-length} it can be seen that there is a low frequency limit below which ideal focusing is physically impossible, given by 
\begin{equation}
\label{eq:ula-low-limit}
\Lambda_{low} = L\Delta x.
\end{equation}
\cref{eq:ula-low-limit} requires the symmetric control points to be positioned at the maximum angular positions of $\pm90^\circ$. Below the corresponding frequency (assuming the array length is fixed), the ULA is incapable of ideal focusing, i.e., when the wavelength is longer than the array length plus $\Delta x$ ideal focusing is physically impossible (as seen from \cref{eq:J3-ula-focusing-wavelength-inequality}). If the wavelength were allowed to be arbitrarily large, i.e., $\Lambda\rightarrow \infty$, then the number of required sources approaches infinity, i.e., $L\rightarrow \infty$.

\subsubsection{Symmetric $M$ Points} 
\begin{figure}[t]
	\centering
	\begin{subfigure}[t]{0.49\linewidth}
		\centering
		\includegraphics[trim={11mm 60mm 11mm 45mm},clip,width=1\linewidth]{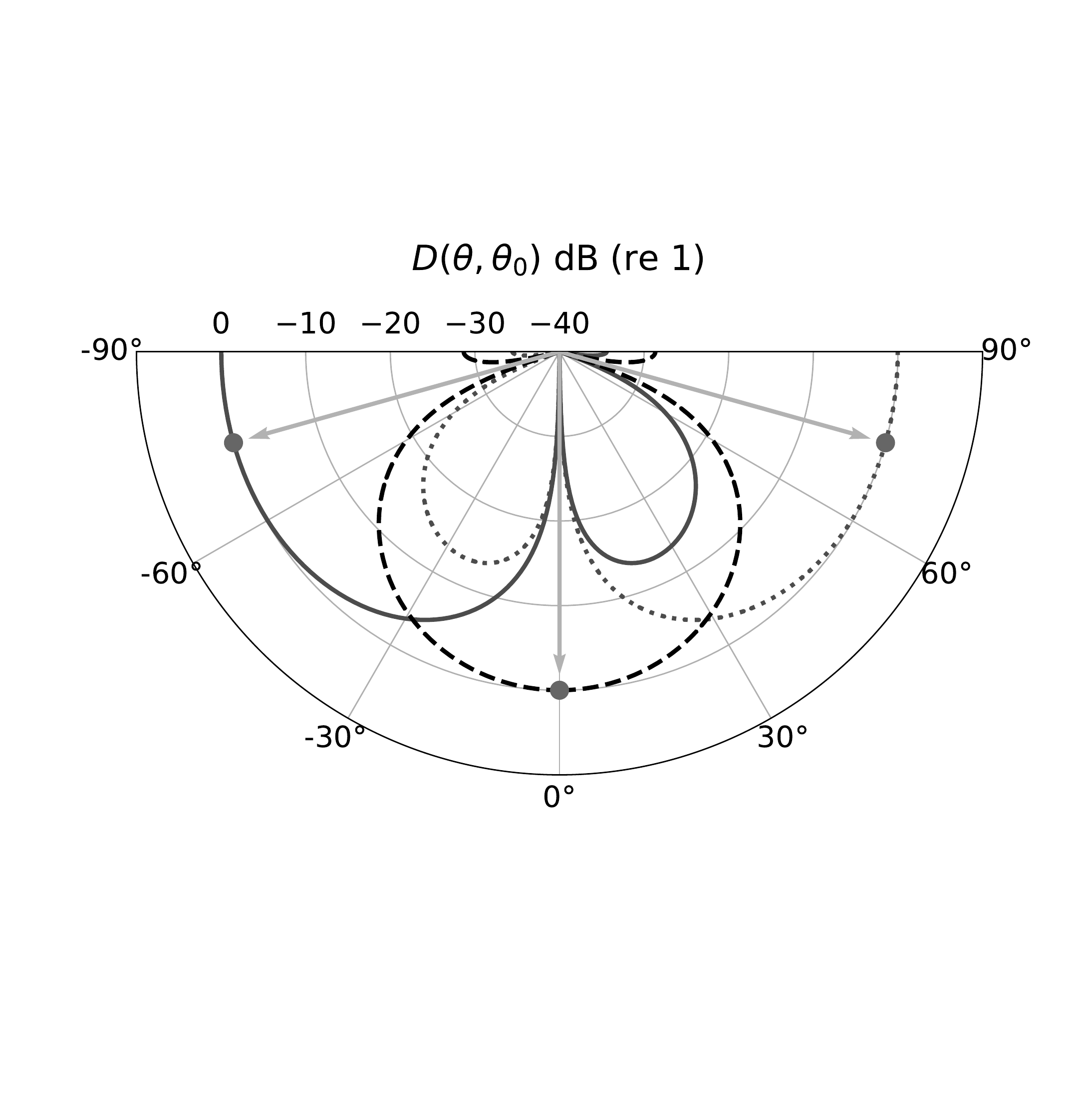}
		\caption{}
		\label{fig:ULA-M-Symmetric-1484Hz}
		\vspace{5mm}
	\end{subfigure}
	\begin{subfigure}[t]{0.49\linewidth}
		\centering
		\includegraphics[trim={11mm 60mm 11mm 45mm},clip,width=1\linewidth]{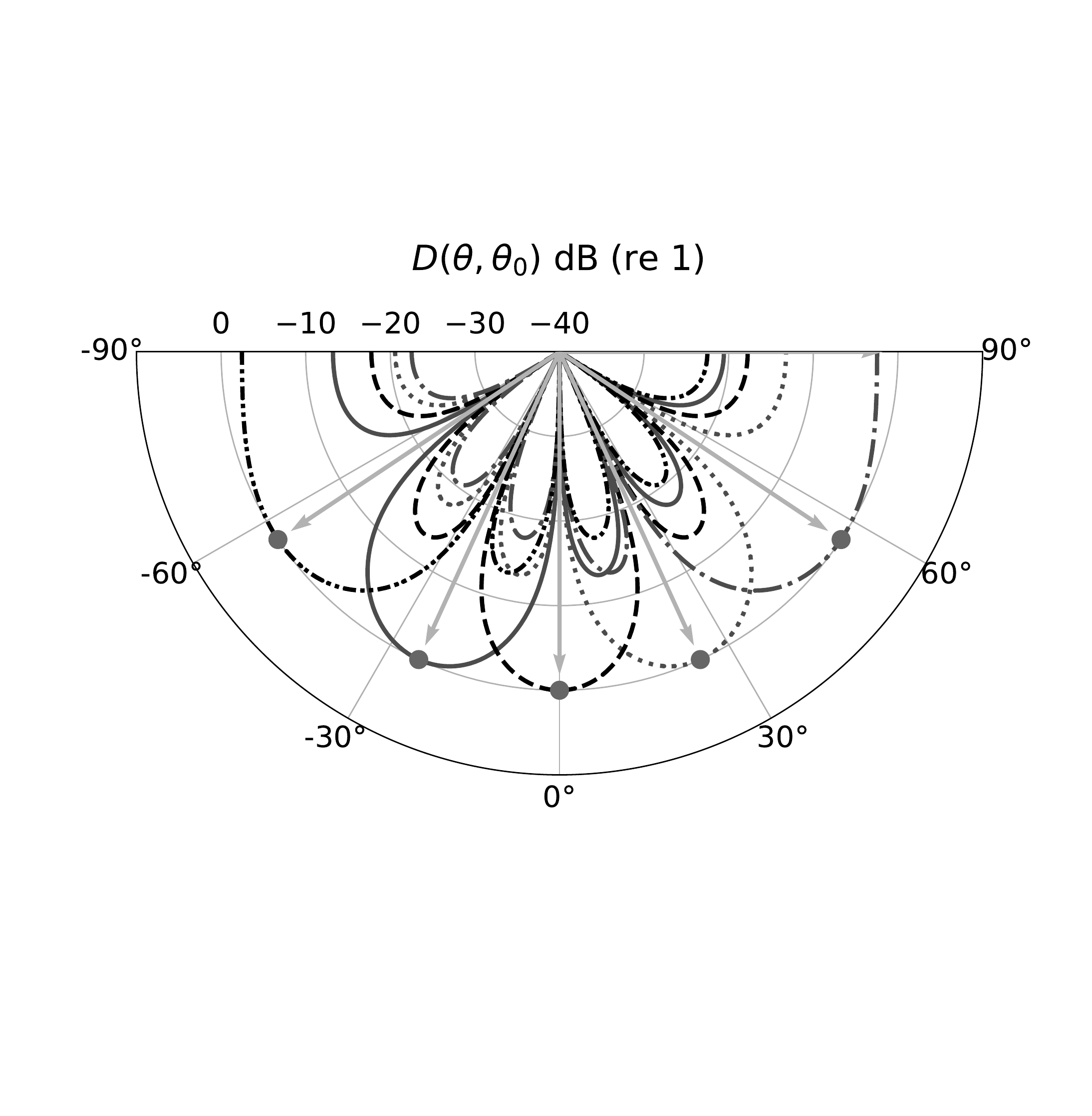}
		\caption{}
		\label{fig:ULA-M-Symmetric-3435Hz}
		\vspace{5mm}
	\end{subfigure}
	\begin{subfigure}[t]{0.49\linewidth}
		\centering
		\includegraphics[trim={11mm 60mm 11mm 45mm},clip,width=1\linewidth]{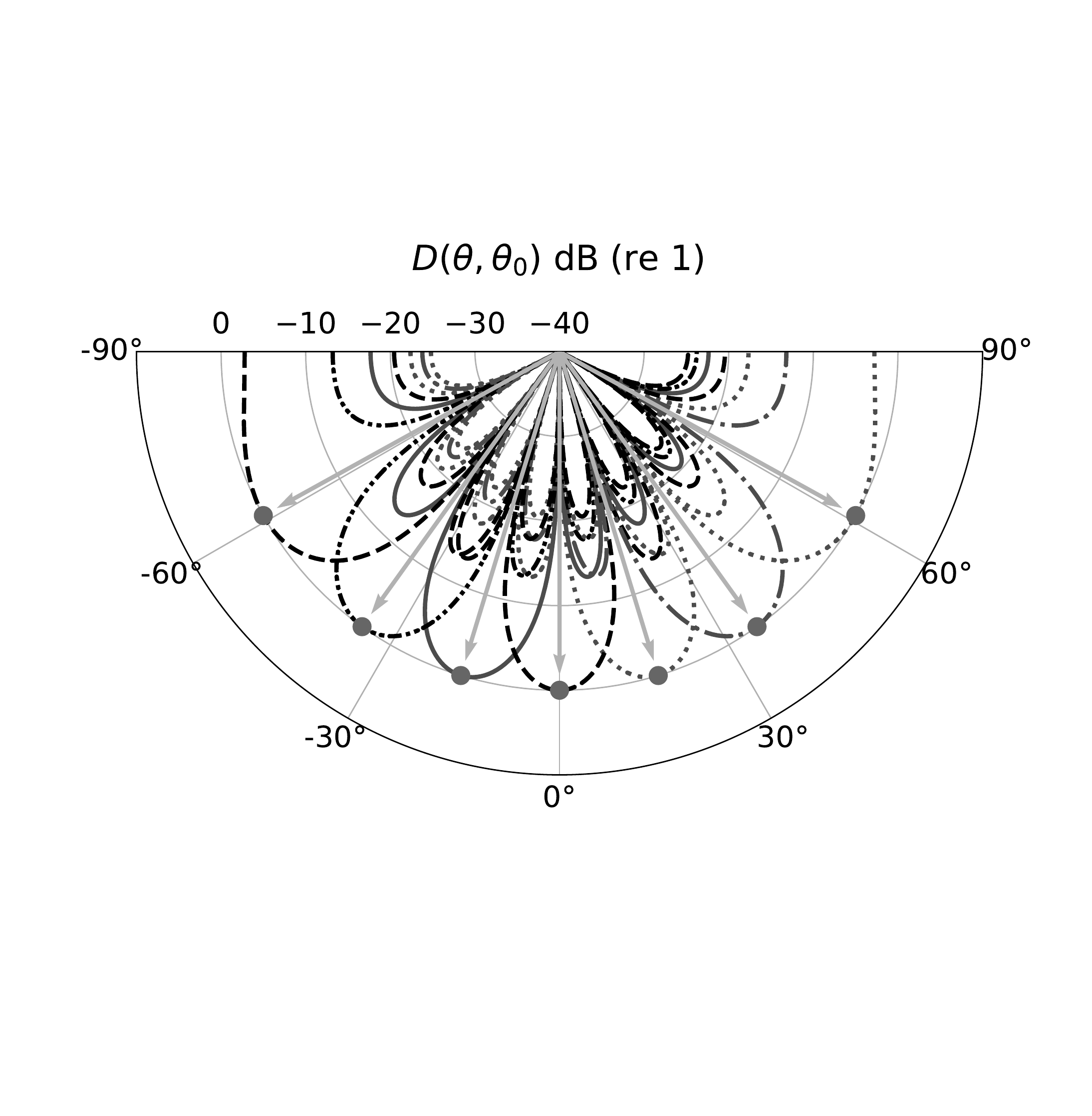}
		\caption{}
		\label{fig:ULA-M-Symmetric-4899Hz}
	\end{subfigure}
	\caption{Far-field ULA super ideal focusing when the control points (grey circles) are symmetric about zero degrees and frequency is increased: a) $M=3$ and $f\approx 1484$ Hz; b) $M=5$ and $f\approx 3435$ Hz; c) $M=7$ and $f\approx 4899$ Hz.}
	\label{fig:ULA-M-Symmetric}
\end{figure}

The analysis in \cref{sec:Symmetric 3 Points Centred About 0} is now extended to ideal focusing at $M$ symmetric points centred around $\theta = 0^\circ$. It is assumed that one control point is positioned at $\theta = 0^\circ$. This implies an odd number of control points. Since the control points greater or less than zero degrees differ only by a sign only half of them need be considered (the point at zero degrees is included as well). With this in mind, the following relationship ensures that ideal focusing can be achieved at the $M$ points:
\begin{equation}
\sin\theta_i = (i-1) \alpha,
\end{equation}
where $i = 1, \dots, (M+1)/2$, with the constraint
\begin{equation}
(i-1)\alpha\leq 1.
\end{equation}
Thus, the number of odd control points is limited by
\begin{equation}
\label{eq:J3-ula-ideal-focusing-control-point-inequality}
M\leq \frac{2L\Delta x}{\Lambda}+1,
\end{equation}
where $M\geq 3$. \cref{eq:J3-ula-ideal-focusing-control-point-inequality} states that, for a fixed array length, the number of allowable control points decreases as the wavelength increases. In fact, when the wavelength equals $L\Delta x$ the maximum number of controls points allowed is three. On the other hand, if $\Lambda\ll L\Delta x$ then more control points can be added. This result can be seen in \cref{fig:ULA-M-Symmetric} which shows examples of the changing directivity pattern as frequency increases for the same fixed length ULA used in \cref{fig:ula-3-point-ideal-focusing-asymmetric2} and \cref{fig:ula-3-point-ideal-focusing-symmetric}. It can be seen that the number of control points increases as frequency increases due to the fixed array length. When the frequency is lowest at approximately 1484 Hz, as shown in \cref{fig:ULA-M-Symmetric-1484Hz}, the number of symmetric control points allowed within the $180^\circ$ control point span is three. At 3435 Hz, as shown in \cref{fig:ULA-M-Symmetric-3435Hz}, the number of control points has increased to five. At 4899 Hz, as shown in \cref{fig:ULA-M-Symmetric-4899Hz}, the number of control points has increased to seven.

Finally, the exact relationship between the array length and the control point angles can be found from
\begin{equation}
L\Delta x = \frac{(i-1)\Lambda}{\sin\theta_i},
\end{equation}
where $0^\circ<\theta_i\leq 90^\circ$.
As seen in \cref{sec:Symmetric 3 Points Centred About 0}, the maximum control point angle is $90^\circ$ and occurs when $\Lambda_{low} =L\Delta x$.
\section{Conclusion}
\label{sec:J3-Conclusion}
An analysis of the linear discrete inverse problem in the context of sound-field reproduction has been given. The concept of focusing crosstalk was formalised and shown to be tied to the linear independence of the system. It was shown that the system ideally minimises the focusing crosstalk to zero at all points which leads to the ideal focusing state. Ideal focusing was shown, by way of Hadamard's inequality, to be the state wherein the determinant of the Gram matrix is maximised. The nature of the ideal focusing state was examined further with a modal analysis. The derived singular system rigorously linked the focusing behaviour to the plant matrix condition number and amplification factor of its pseudoinverse. The amplification factor of the plant matrix pseudoinverse is minimised in the ideal focusing state, however, the system is not necessarily optimally-conditioned. To achieve optimal conditioning it was shown that the focusing efficacy at all control points must be equal, which corresponds to the special case of equal eigenvalues while in the ideal focusing state. Thus, it was established that pairwise orthogonality between rows is a necessary but insufficient requirement for optimally-conditioned underdetermined systems, which can only occur when the length of the plant row vectors are equal.

The application of the analysis framework was demonstrated with two theoretical cases studies. In the first case study, the acoustic crosstalk cancellation system was examined and it was shown that the stability and behaviour of the pseudoinverse is dependent on focusing crosstalk at just one point in space. The two-channel CTC system was examined and the general conditions for ideal focusing were established. It was shown that to maintain ideal focusing as a function of frequency there must be a proportional change in the loudspeaker and control point geometry. The relationship between ideal focusing and the system geometry was clarified further by introducing a monopole source in free-field model assuming a shadowless head. It was shown that the path length differences are critical in determining ideal focusing, namely that the difference of path length differences must sum to odd multiples of half the acoustic wavelength. Subsequently, it was shown that OSD is a special case of super ideal focusing. The analysis framework was shown to generalise the findings of OSD and it was shown that there are other asymmetric two-channel geometries that can achieve super ideal focusing, i.e., geometries that allow for head rotation and asymmetric loudspeaker placement. Furthermore, the physical conditions for the symmetric UPDA to achieve super ideal focusing were given, under the same acoustics assumptions, thus providing an example of how super ideal focusing occurs in the multichannel setting.

The second case study examined the application of sound-field reproduction in the far-field using a ULA. The requirements for super ideal focusing at $M$  irregularly spaced directions in the far-field were given. As in the CTC case, the ULA must undergo a frequency-dependent change in geometry in order to maintain super ideal focusing. In turn, the length of the array also determines the number of control point directions in which super ideal focusing can occur. This case study extended the previous findings by Nelson and Kim on optimally-conditioned systems and shows that there exists more general physical requirements that need be satisfied in order to maintain super ideal focusing in this specific case.

While the results presented here have been given in the context of sound-field reproduction, the general results can be applied to other problems, especially those that share the same mathematical models of the wave physics. The hope is that this work will inspire future design considerations and interpretations in applications of the linear discrete inverse problem, especially in sound-field reproduction.
\section*{Acknowledgements}
This work was supported by the Engineering and Physical Sciences Research Council (EPSRC) through the University of Southampton's Doctoral Training Partnership grant 1786079.
\appendix
\section{Spatially Matched Filters at $M$ Points}
\label{sec:J3-Temporarily and Spatially Matched Filters} 
The concept of spatially matched filters originally established by Tanter et al.\ in \cite{Tanter2000} is extended to $M$ control points. Spatially matched filters at $M$ control points maximise
\begin{equation}
\label{eq:J3-M-point-spatially-matched-ratio}
J\left(\bv{H}_0(\omega)\right) = \frac{\norm{\plant(\omega)\bv{H}_0(\omega)}^2}{\norm{\bv{H}_0(\omega)}^2}\leq\norm{\plant(\omega)}^2,
\end{equation}
where $\bv{H}_0(\omega)$ are a chosen set of filters. Thus, spatially matched filters lead to an equality of the above inequality. Substituting in the filters given by \cref{eq:normalised-focusing-filters} into \cref{eq:J3-M-point-spatially-matched-ratio} yields 
\begin{equation}
\label{eq:J3-M-point-spatially-matched-inequality}
J\left(\filters_{\mathrm{I}}(\omega)\right) = \mathrm{min}\left(\lambda_m\right) \leq \mathrm{max}\left(\lambda_m\right).
\end{equation}
\cref{eq:J3-M-point-spatially-matched-inequality} can only be an equality if $\filters_0(\omega) = \filters_{opt}(\omega)$, i.e., all the eigenvalues are equal. The physical interpretation of this result is that the system is most efficient when the Gram is super ideal and given by \cref{eq:J3-unit-gram}. Thus, when super ideal focusing occurs the filters given by \cref{eq:normalised-focusing-filters} become \cref{eq:J3-super-ideal-focusing-pseudoinverse} and are spatially matched filters at $M$ points. This result means that optimally conditioned solutions, such as OSD, consist of spatially matched filters at multiple points.
\section{Derivation of UPDA Focusing Crosstalk}
\label{sec:Derivation of UPDA Focusing Crosstalk}
The following derives \cref{eq:UPDA-focusing-xt-simplified} from \cref{eq:UPDA-focusing-xt-sum}. \cref{eq:UPDA-focusing-xt-sum} is restated here as
\begin{equation}
\label{eq:J2-UPLD-cos-sum-appendix}
\focusxt \approx \frac{1}{R^2}\sum_{l=-\frac{L-1}{2}}^{\frac{L-1}{2}}\cos\left(kl\Delta\eta\right),
\end{equation}
where $\Delta\eta \defined 2\pld{max}/(L-1)$ and $\pld{max}\defined \mathrm{max}(\abs{\pld{i}}), ~i=1, \dots, L$. Let $x = k\Delta\eta$ and $A=(L-1)/2$. \cref{eq:J2-UPLD-cos-sum-appendix} can be written as a power series:
\begin{align}
\focusxt &\approx \frac{1}{R^2}\sum_{l=-A}^{A}\cos\left(lx\right),\\[5pt]
& = \frac{1}{R^2}\Re\left\lbrace\sum_{l=-A}^{A}\left(\euler^{-\Ij x}\right)^l\right\rbrace,\\[5pt]
%&= \frac{1}{R^2}\Re\left\lbrace\frac{\euler^{\Ij xA}-\euler^{-\Ij x\left(A+1\right)}}{1-\euler^{-\Ij x}}\right\rbrace\\[5pt]
&= \frac{1}{R^2}\Re\left\lbrace\frac{\euler^{\Ij x\left(A+\frac{1}{2}\right)}-\euler^{-\Ij x\left(A+\frac{1}{2}\right)}}{\euler^{\Ij \frac{x}{2}}-\euler^{-\Ij \frac{x}{2}}}\right\rbrace,\\[5pt]
&= \frac{\sin\left(x\left(A+\frac{1}{2}\right)\right)}{R^2\sin\left(\frac{x}{2}\right)},\\[5pt]
\label{eq:J2-UPDA-focusing-crosstalk-derivation-result}
&= \frac{\sin\left(\frac{Lk\pld{max}}{L-1}\right)}{R^2\sin\left(\frac{k\pld{max}}{L-1}\right)}.
\end{align}
\cref{eq:J2-UPDA-focusing-crosstalk-derivation-result} gives \cref{eq:UPDA-focusing-xt-simplified}.
\bibliographystyle{elsarticle-num} 
\bibliography{C:/Users/Eric/Dropbox/PhD/Bibtek_Library/Eric_Hamdan_Phd_Bibtek_Library}
\end{document}

%% file: MathCommands.tex
\newcommand{\plant}{\mathbf{G}}
\newcommand{\pressures}{\mathbf{p}}
\newcommand{\desired}{\mathbf{d}}
\newcommand{\solution}{\mathbf{q}}

\newcommand{\herm}{^\mathrm{H}}
\newcommand{\trans}{^\mathrm{T}}

\newcommand{\norm}[1]{||#1||_2}
\newcommand{\gM}[1]{\mathbf{g}_{#1}}

\newcommand{\ggh}{\mathbf{GG\herm}}

\newcommand{\filters}{\mathbf{H}}

\newcommand{\inCv}[1]{\in\mathbb{C}^{#1}}
\newcommand{\inCm}[2]{\in\mathbb{C}^{#1\times#2}}
\newcommand{\inRv}[1]{\in\mathbb{R}^{#1}}
\newcommand{\inRm}[2]{\in\mathbb{R}^{#1\times#2}}

\newcommand{\x}[1]{\mathbf{x}_{#1}}

\newcommand{\fronorm}[1]{||#1||_F}

\newcommand{\lsvectors}{\mathbf{U}}
\newcommand{\rsvectors}{\mathbf{V}}
\newcommand{\svalues}{\mathbf{\Sigma}}

\newcommand{\svd}{\lsvectors\svalues\rsvectors\herm}

\newcommand{\diag}[1]{\mathrm{diag}\left(#1\right)}
\newcommand{\rank}[1]{\mathrm{rank}\left(#1\right)}
\newcommand{\inv}[1]{(#1)^{-1}}

\newcommand{\abs}[1]{\left|#1\right|}

\newcommand{\bv}[1]{\mathbf{#1}}
\newcommand{\pinv}{\plant^{+}}
\newcommand{\adj}[1]{\mathrm{adj}\left(#1\right)}
\newcommand{\determ}[1]{\mathrm{det}\left(#1\right)}
\newcommand{\xM}[1]{\bv{x}_{#1}}
\newcommand{\sL}[1]{\bv{s}_{#1}}

\newcommand{\dggh}{\bv{\Lambda}}

\newcommand{\range}[1]{\mathcal{R}(#1)}
\newcommand{\arraygain}[2]{D(\bv{x}_{#1},\bv{x}_{#2})}

\newcommand{\cosTheta}[1]{\cos\Theta_{#1}}
\newcommand{\Rml}[2]{R_{#1#2}}
\newcommand{\uladirect}{D(\theta_i, \theta_j)}
\newcommand{\Ij}{\mathrm{j}}
\newcommand{\inC}[1]{\in\mathbb{C}^{#1}}

\newcommand{\J}{\mathrm{j}}
\newcommand{\delay}{\mathrm{e}^{-\J \omega\Delta t}}

\newcommand{\focusxt}{\xi_{21}}

\newcommand{\uplda}{\text{UPDA} }

\newcommand{\pld}[1]{\eta_{#1}}

\newcommand{\ieq}{\overset{!}{=}}
\newcommand{\defined}{\overset{\mathrm{def}}{=}}

\newcommand{\gramtwo}{\xi_{12}}
\newcommand{\gramthree}{\xi_{21}}

\newcommand{\Gram}{\bv{\Xi}}
\newcommand{\gramian}{\Gamma}
\newcommand{\gram}[2]{\xi_{#1 #2}}
\newcommand{\euler}{\mathrm{e}}

\newcommand{\plp}[1]{r_{#1}}

%% file: Hamdan_Fazi_Focusing_Phenomena_in_Linear_Inverse_Problems_in_Acoustics_v2.bbl
\begin{thebibliography}{10}
\expandafter\ifx\csname url\endcsname\relax
  \def\url#1{\texttt{#1}}\fi
\expandafter\ifx\csname urlprefix\endcsname\relax\def\urlprefix{URL }\fi
\expandafter\ifx\csname href\endcsname\relax
  \def\href#1#2{#2} \def\path#1{#1}\fi

\bibitem{Veronesi1989}
W.~A. Veronesi, J.~D. Maynard, Digital holographic reconstruction of sources
  with arbitrarily shaped surfaces, The Journal of the Acoustical Society of
  America 85~(2) (1989) 588--598.
\newblock \href {https://doi.org/10.1121/1.397583}
  {\path{doi:10.1121/1.397583}}.

\bibitem{Williams2001}
E.~G. Williams, \href{https://doi.org/10.1121/1.1404381}{Regularization methods
  for near-field acoustical holography}, The Journal of the Acoustical Society
  of America 110~(4) (2001) 1976--1988.
\newblock \href {http://arxiv.org/abs/https://doi.org/10.1121/1.1404381}
  {\path{arXiv:https://doi.org/10.1121/1.1404381}}, \href
  {https://doi.org/10.1121/1.1404381} {\path{doi:10.1121/1.1404381}}.
\newline\urlprefix\url{https://doi.org/10.1121/1.1404381}

\bibitem{Chardon2012}
G.~Chardon, L.~Daudet, A.~Peillot, F.~Ollivier, N.~Bertin, R.~Gribonval,
  \href{https://doi.org/10.1121/1.4740476}{Near-field acoustic holography using
  sparse regularization and compressive sampling principles}, The Journal of
  the Acoustical Society of America 132~(3) (2012) 1521--1534.
\newblock \href {http://arxiv.org/abs/https://doi.org/10.1121/1.4740476}
  {\path{arXiv:https://doi.org/10.1121/1.4740476}}, \href
  {https://doi.org/10.1121/1.4740476} {\path{doi:10.1121/1.4740476}}.
\newline\urlprefix\url{https://doi.org/10.1121/1.4740476}

\bibitem{Nelson2000}
P.~A. Nelson, S.~H. Yoon,
  \href{http://www.sciencedirect.com/science/article/pii/S0022460X99928377}{{Estimation
  of Acoustic Source Strength by Inverse Methods: Part I, Conditioning of the
  Inverse Problem}}, Journal of Sound and Vibration 233~(4) (2000) 639 -- 664.
\newblock \href {https://doi.org/https://doi.org/10.1006/jsvi.1999.2837}
  {\path{doi:https://doi.org/10.1006/jsvi.1999.2837}}.
\newline\urlprefix\url{http://www.sciencedirect.com/science/article/pii/S0022460X99928377}

\bibitem{Kim2004}
Y.~Kim, P.~A. Nelson,
  \href{http://www.sciencedirect.com/science/article/pii/S0022460X03010642}{Optimal
  regularisation for acoustic source reconstruction by inverse methods},
  Journal of Sound and Vibration 275~(3) (2004) 463 -- 487.
\newblock \href {https://doi.org/https://doi.org/10.1016/j.jsv.2003.06.031}
  {\path{doi:https://doi.org/10.1016/j.jsv.2003.06.031}}.
\newline\urlprefix\url{http://www.sciencedirect.com/science/article/pii/S0022460X03010642}

\bibitem{Holland2012}
K.~R. Holland, P.~A. Nelson,
  \href{http://www.sciencedirect.com/science/article/pii/S0022460X12003574}{An
  experimental comparison of the focused beamformer and the inverse method for
  the characterisation of acoustic sources in ideal and non-ideal acoustic
  environments}, Journal of Sound and Vibration 331~(20) (2012) 4425 -- 4437.
\newblock \href {https://doi.org/https://doi.org/10.1016/j.jsv.2012.05.005}
  {\path{doi:https://doi.org/10.1016/j.jsv.2012.05.005}}.
\newline\urlprefix\url{http://www.sciencedirect.com/science/article/pii/S0022460X12003574}

\bibitem{Holland2013}
K.~R. Holland, P.~A. Nelson,
  \href{http://www.sciencedirect.com/science/article/pii/S0022460X13005245}{The
  application of inverse methods to spatially-distributed acoustic sources},
  Journal of Sound and Vibration 332~(22) (2013) 5727 -- 5747.
\newblock \href {https://doi.org/https://doi.org/10.1016/j.jsv.2013.06.009}
  {\path{doi:https://doi.org/10.1016/j.jsv.2013.06.009}}.
\newline\urlprefix\url{http://www.sciencedirect.com/science/article/pii/S0022460X13005245}

\bibitem{Bauck1992}
J.~Bauck, D.~H. Cooper,
  \href{http://www.aes.org/e-lib/browse.cfm?elib=6733}{{Generalized Transaural
  Stereo}}, in: Audio Engineering Society Convention 93, San Francisco, CA,
  USA, 1992.
\newline\urlprefix\url{http://www.aes.org/e-lib/browse.cfm?elib=6733}

\bibitem{Kirkeby1993}
O.~Kirkeby, P.~A. Nelson, \href{https://doi.org/10.1121/1.407330}{Reproduction
  of plane wave sound fields}, The Journal of the Acoustical Society of America
  94~(5) (1993) 2992--3000.
\newblock \href {https://doi.org/10.1121/1.407330}
  {\path{doi:10.1121/1.407330}}.
\newline\urlprefix\url{https://doi.org/10.1121/1.407330}

\bibitem{Nelson1995}
P.~A. Nelson, F.~Orduna-Bustamante, H.~Hamada, Inverse filter design and
  equalization zones in multichannel sound reproduction, IEEE Transactions on
  Speech and Audio Processing 3~(3) (1995) 185--192.
\newblock \href {https://doi.org/10.1109/89.388144}
  {\path{doi:10.1109/89.388144}}.

\bibitem{Fazi2010}
F.~M. Fazi, \href{https://eprints.soton.ac.uk/158639/}{Sound field
  reproduction}, Ph.D. thesis, University of Southampton (February 2010).
\newline\urlprefix\url{https://eprints.soton.ac.uk/158639/}

\bibitem{Olivieri2016}
F.~Olivieri, F.~M. Fazi, P.~A. Nelson, M.~Shin, S.~Fontana, L.~Yue,
  \href{http://www.sciencedirect.com/science/article/pii/S0022460X16002340}{Theoretical
  and experimental comparative analysis of beamforming methods for loudspeaker
  arrays under given performance constraints}, Journal of Sound and Vibration
  373 (2016) 302 -- 324.
\newblock \href {https://doi.org/https://doi.org/10.1016/j.jsv.2016.03.005}
  {\path{doi:https://doi.org/10.1016/j.jsv.2016.03.005}}.
\newline\urlprefix\url{http://www.sciencedirect.com/science/article/pii/S0022460X16002340}

\bibitem{Galvez2019}
M.~F. Sim{\'o}n~G{\'a}lvez, D.~Menzies, F.~M. Fazi,
  \href{http://www.aes.org/e-lib/browse.cfm?elib=20451}{{Dynamic Audio
  Reproduction with Linear Loudspeaker Arrays}}, J. Audio Eng. Soc 67~(4)
  (2019) 190--200.
\newline\urlprefix\url{http://www.aes.org/e-lib/browse.cfm?elib=20451}

\bibitem{Hoffmann2019}
F.-M. Hoffmann, E.~G. Williams, F.~M. Fazi, S.~Fontana,
  \href{http://www.sciencedirect.com/science/article/pii/S0022460X18306369}{An
  analytical model for wedge-shaped acoustic arrays}, Journal of Sound and
  Vibration 439 (2019) 56 -- 76.
\newblock \href {https://doi.org/https://doi.org/10.1016/j.jsv.2018.09.041}
  {\path{doi:https://doi.org/10.1016/j.jsv.2018.09.041}}.
\newline\urlprefix\url{http://www.sciencedirect.com/science/article/pii/S0022460X18306369}

\bibitem{Hoffmann2015}
F.~{Hoffmann}, F.~M. {Fazi}, {Theoretical Study of Acoustic Circular Arrays
  With Tangential Pressure Gradient Sensors}, IEEE/ACM Transactions on Audio,
  Speech, and Language Processing 23~(11) (2015) 1762--1774.
\newblock \href {https://doi.org/10.1109/TASLP.2015.2449083}
  {\path{doi:10.1109/TASLP.2015.2449083}}.

\bibitem{Nelson1992a}
P.~A. Nelson, H.~Hamada, S.~J. Elliott, Adaptive inverse filters for
  stereophonic sound reproduction, IEEE Transactions on Signal Processing
  40~(7) (1992) 1621--1632.
\newblock \href {https://doi.org/10.1109/78.143434}
  {\path{doi:10.1109/78.143434}}.

\bibitem{Nelson1996multichannel}
P.~A. Nelson, F.~Orduna-Bustamante, H.~Hamada,
  \href{http://www.aes.org/e-lib/browse.cfm?elib=7877}{{Multichannel Signal
  Processing Techniques in the Reproduction of Sound}}, J. Audio Eng. Soc
  44~(11) (1996) 973--989.
\newline\urlprefix\url{http://www.aes.org/e-lib/browse.cfm?elib=7877}

\bibitem{House2020}
C.~House, J.~Cheer, S.~Daley,
  \href{http://www.sciencedirect.com/science/article/pii/S0003682X20305405}{An
  experimental investigation into active structural acoustic cloaking of a
  flexible cylinder}, Applied Acoustics 170 (2020) 107436.
\newblock \href
  {https://doi.org/https://doi.org/10.1016/j.apacoust.2020.107436}
  {\path{doi:https://doi.org/10.1016/j.apacoust.2020.107436}}.
\newline\urlprefix\url{http://www.sciencedirect.com/science/article/pii/S0003682X20305405}

\bibitem{Pignier2017}
N.~J. Pignier, C.~J. O'Reilly, S.~Boij,
  \href{http://www.sciencedirect.com/science/article/pii/S0022460X17300937}{Identifying
  equivalent sound sources from aeroacoustic simulations using a numerical
  phased array}, Journal of Sound and Vibration 394 (2017) 203 -- 219.
\newblock \href {https://doi.org/https://doi.org/10.1016/j.jsv.2017.01.051}
  {\path{doi:https://doi.org/10.1016/j.jsv.2017.01.051}}.
\newline\urlprefix\url{http://www.sciencedirect.com/science/article/pii/S0022460X17300937}

\bibitem{Leclere2017}
Q.~Leclère, A.~Pereira, C.~Bailly, J.~Antoni, C.~Picard,
  \href{https://doi.org/10.1177/1475472X17718883}{A unified formalism for
  acoustic imaging based on microphone array measurements}, International
  Journal of Aeroacoustics 16~(4-5) (2017) 431--456.
\newblock \href {http://arxiv.org/abs/https://doi.org/10.1177/1475472X17718883}
  {\path{arXiv:https://doi.org/10.1177/1475472X17718883}}, \href
  {https://doi.org/10.1177/1475472X17718883}
  {\path{doi:10.1177/1475472X17718883}}.
\newline\urlprefix\url{https://doi.org/10.1177/1475472X17718883}

\bibitem{Tanter2000}
M.~Tanter, J.~Thomas, M.~Fink, \href{https://doi.org/10.1121/1.429459}{Time
  reversal and the inverse filter}, The Journal of the Acoustical Society of
  America 108~(1) (2000) 223--234.
\newblock \href {http://arxiv.org/abs/https://doi.org/10.1121/1.429459}
  {\path{arXiv:https://doi.org/10.1121/1.429459}}, \href
  {https://doi.org/10.1121/1.429459} {\path{doi:10.1121/1.429459}}.
\newline\urlprefix\url{https://doi.org/10.1121/1.429459}

\bibitem{Tanter2001}
M.~Tanter, J.-F. Aubry, J.~Gerber, J.-L. Thomas, M.~Fink,
  \href{https://doi.org/10.1121/1.1377051}{{Optimal focusing by spatio-temporal
  inverse filter. I. Basic principles}}, The Journal of the Acoustical Society
  of America 110~(1) (2001) 37--47.
\newblock \href {http://arxiv.org/abs/https://doi.org/10.1121/1.1377051}
  {\path{arXiv:https://doi.org/10.1121/1.1377051}}, \href
  {https://doi.org/10.1121/1.1377051} {\path{doi:10.1121/1.1377051}}.
\newline\urlprefix\url{https://doi.org/10.1121/1.1377051}

\bibitem{Wu2008}
S.~F. Wu, \href{https://doi.org/10.1121/1.2977731}{Methods for reconstructing
  acoustic quantities based on acoustic pressure measurements}, The Journal of
  the Acoustical Society of America 124~(5) (2008) 2680--2697.
\newblock \href {http://arxiv.org/abs/https://doi.org/10.1121/1.2977731}
  {\path{arXiv:https://doi.org/10.1121/1.2977731}}, \href
  {https://doi.org/10.1121/1.2977731} {\path{doi:10.1121/1.2977731}}.
\newline\urlprefix\url{https://doi.org/10.1121/1.2977731}

\bibitem{Nelson2001a}
P.~A. Nelson, {A Review of Some Inverse Problems in Acoustics}, International
  Journal of Acoustics and Vibration 6~(3) (2001) 118--134.

\bibitem{Hansen1987}
P.~C. Hansen,
  \href{https://www.amazon.com/Rank-Deficient-Discrete-Ill-Posed-Problems-Mathematical/dp/0898714036?SubscriptionId=0JYN1NVW651KCA56C102&tag=techkie-20&linkCode=xm2&camp=2025&creative=165953&creativeASIN=0898714036}{{Rank-Deficient
  and Discrete Ill-Posed Problems: Numerical Aspects of Linear Inversion
  (Monographs on Mathematical Modeling and Computation}}, Society for
  Industrial and Applied Mathematics, 1987.
\newline\urlprefix\url{https://www.amazon.com/Rank-Deficient-Discrete-Ill-Posed-Problems-Mathematical/dp/0898714036?SubscriptionId=0JYN1NVW651KCA56C102&tag=techkie-20&linkCode=xm2&camp=2025&creative=165953&creativeASIN=0898714036}

\bibitem{Meyer2001}
C.~D. Meyer, Matrix analysis and applied linear algebra, SIAM: Society for
  Industrial and Applied Mathematics, 2001.

\bibitem{AdiBen-Israel2003}
T.~N. E.~G. Adi Ben-Israel,
  \href{https://www.ebook.de/de/product/5228785/adi_ben_israel_thomas_n_e_greville_generalized_inverses.html}{{Generalized
  Inverses}}, Springer New York, 2003.
\newline\urlprefix\url{https://www.ebook.de/de/product/5228785/adi_ben_israel_thomas_n_e_greville_generalized_inverses.html}

\bibitem{Fink1997}
M.~Fink, Time-reversed acoustics, Physics Today 20 (1997) 34.

\bibitem{Theodoridis2013}
S.~Theodoridis, R.~Chellappa, {Academic Press Library in Signal Processing,
  Volume 3: Array and Statistical Signal Processing}, 1st Edition, Academic
  Press, Inc., Orlando, FL, USA, 2013.

\bibitem{Yan2019}
S.~Yan,
  \href{https://www.ebook.de/de/product/35350473/shefeng_yan_broadband_array_processing.html}{{Broadband
  Array Processing}}, Springer-Verlag GmbH, 2019.
\newline\urlprefix\url{https://www.ebook.de/de/product/35350473/shefeng_yan_broadband_array_processing.html}

\bibitem{Turin1960}
G.~{Turin}, An introduction to matched filters, IRE Transactions on Information
  Theory 6~(3) (1960) 311--329.
\newblock \href {https://doi.org/10.1109/TIT.1960.1057571}
  {\path{doi:10.1109/TIT.1960.1057571}}.

\bibitem{Montaldo2004}
G.~Montaldo, M.~Tanter, M.~Fink, \href{https://doi.org/10.1121/1.1636462}{Real
  time inverse filter focusing through iterative time reversal}, The Journal of
  the Acoustical Society of America 115~(2) (2004) 768--775.
\newblock \href {http://arxiv.org/abs/https://doi.org/10.1121/1.1636462}
  {\path{arXiv:https://doi.org/10.1121/1.1636462}}, \href
  {https://doi.org/10.1121/1.1636462} {\path{doi:10.1121/1.1636462}}.
\newline\urlprefix\url{https://doi.org/10.1121/1.1636462}

\bibitem{Vignon2006}
F.~Vignon, J.-F. Aubry, A.~Saez, M.~Tanter, D.~Cassereau, G.~Montaldo, M.~Fink,
  \href{https://doi.org/10.1121/1.2161452}{{The Stokes relations linking time
  reversal and the inverse filter}}, The Journal of the Acoustical Society of
  America 119~(3) (2006) 1335--1346.
\newblock \href {http://arxiv.org/abs/https://doi.org/10.1121/1.2161452}
  {\path{arXiv:https://doi.org/10.1121/1.2161452}}, \href
  {https://doi.org/10.1121/1.2161452} {\path{doi:10.1121/1.2161452}}.
\newline\urlprefix\url{https://doi.org/10.1121/1.2161452}

\bibitem{Hamdan2021}
E.~C. Hamdan, F.~M. Fazi,
  \href{http://www.sciencedirect.com/science/article/pii/S0022460X20305733}{A
  modal analysis of multichannel crosstalk cancellation systems and their
  relationship to amplitude panning}, Journal of Sound and Vibration 490 (2021)
  115743.
\newblock \href {https://doi.org/https://doi.org/10.1016/j.jsv.2020.115743}
  {\path{doi:https://doi.org/10.1016/j.jsv.2020.115743}}.
\newline\urlprefix\url{http://www.sciencedirect.com/science/article/pii/S0022460X20305733}

\bibitem{Gauthier2011}
P.-A. Gauthier, C.~Camier, Y.~Pasco, A.~Berry, E.~Chambatte, R.~Lapointe, M.-A.
  Delalay,
  \href{http://www.sciencedirect.com/science/article/pii/S0022460X11006031}{Beamforming
  regularization matrix and inverse problems applied to sound field measurement
  and extrapolation using microphone array}, Journal of Sound and Vibration
  330~(24) (2011) 5852 -- 5877.
\newblock \href {https://doi.org/https://doi.org/10.1016/j.jsv.2011.07.022}
  {\path{doi:https://doi.org/10.1016/j.jsv.2011.07.022}}.
\newline\urlprefix\url{http://www.sciencedirect.com/science/article/pii/S0022460X11006031}

\bibitem{Fink1993}
M.~Fink,
  \href{https://doi.org/10.1088%2F0022-3727%2F26%2F9%2F001}{Time-reversal
  mirrors}, Journal of Physics D: Applied Physics 26~(9) (1993) 1333--1350.
\newblock \href {https://doi.org/10.1088/0022-3727/26/9/001}
  {\path{doi:10.1088/0022-3727/26/9/001}}.
\newline\urlprefix\url{https://doi.org/10.1088%2F0022-3727%2F26%2F9%2F001}

\bibitem{Prada1994}
{Prada, C. and Fink, M.}, {Eigenmodes of the time reversal operator: A solution
  to selective focusing in multiple target media}, {Wave Motion} 20 (1994) 151.

\bibitem{Prada1996}
C.~Prada, S.~Manneville, D.~Spoliansky, M.~Fink, {Decomposition of the time
  reversal operator: Detection and selective focusing on two scatterers}, {J.
  Acoust. Soc. Am.} 99 (1996) 2067.

\bibitem{Nelson1999}
P.~A. Nelson, in: Proceedings of 6th International Congress on Sound and
  Vibration, 1999, pp. 7--32.

\bibitem{Nelson2001b}
P.~A. Nelson, Y.~Kim, {Optimal conditioning of inverse problems in acoustic
  radiation}, in: Proceedings of 17th International Congress on Acoustics,
  2001, p.~94.

\bibitem{Kim2003}
Y.~Kim, P.~A. Nelson,
  \href{http://www.sciencedirect.com/science/article/pii/S0022460X02014529}{Spatial
  resolution limits for the reconstruction of acoustic source strength by
  inverse methods}, Journal of Sound and Vibration 265~(3) (2003) 583 -- 608.
\newblock \href {https://doi.org/https://doi.org/10.1016/S0022-460X(02)01452-9}
  {\path{doi:https://doi.org/10.1016/S0022-460X(02)01452-9}}.
\newline\urlprefix\url{http://www.sciencedirect.com/science/article/pii/S0022460X02014529}

\bibitem{Hirono2018}
F.~Hirono, {Far-Field Microphone Array Techniques for Acoustic Characterisation
  of Aerofoils}, Ph.D. thesis (10 2018).

\bibitem{Takeuchi2002}
T.~Takeuchi, P.~A. Nelson, Optimal source distribution for binaural synthesis
  over loudspeakers, The Journal of the Acoustical Society of America 112~(6)
  (2002) 2786--2797.
\newblock \href {https://doi.org/10.1121/1.1513363}
  {\path{doi:10.1121/1.1513363}}.

\bibitem{Takeuchi2007}
T.~Takeuchi, P.~A. Nelson,
  \href{http://www.aes.org/e-lib/browse.cfm?elib=14181}{{Subjective and
  Objective Evaluation of the Optimal Source Distribution for Virtual Acoustic
  Imaging}}, J. Audio Eng. Soc 55~(11) (2007) 981--997.
\newline\urlprefix\url{http://www.aes.org/e-lib/browse.cfm?elib=14181}

\bibitem{Takeuchi2008}
T.~Takeuchi, P.~A. Nelson, {Extension of the Optimal Source Distribution for
  Binaural Sound Reproduction}, Acta Acustica united with Acustica 94~(6)
  (2008) 981--987.
\newblock \href {https://doi.org/10.3813/aaa.918114}
  {\path{doi:10.3813/aaa.918114}}.

\bibitem{Asano1999}
F.~{Asano}, Y.~{Suzuki}, D.~C. {Swanson}, {Optimization of control source
  configuration in active control systems using Gram-Schmidt
  orthogonalization}, IEEE Transactions on Speech and Audio Processing 7~(2)
  (1999) 213--220.
\newblock \href {https://doi.org/10.1109/89.748126}
  {\path{doi:10.1109/89.748126}}.

\bibitem{Koyama2020}
S.~{Koyama}, G.~{Chardon}, L.~{Daudet}, {Optimizing Source and Sensor Placement
  for Sound Field Control: An Overview}, IEEE/ACM Transactions on Audio,
  Speech, and Language Processing 28 (2020) 696--714.
\newblock \href {https://doi.org/10.1109/TASLP.2020.2964958}
  {\path{doi:10.1109/TASLP.2020.2964958}}.

\bibitem{Shafarevich2012}
A.~O.~R. Igor R.~Shafarevich,
  \href{https://www.ebook.de/de/product/19950791/igor_r_shafarevich_alexey_o_remizov_linear_algebra_and_geometry.html}{{Linear
  Algebra and Geometry}}, Springer-Verlag GmbH, 2012.
\newline\urlprefix\url{https://www.ebook.de/de/product/19950791/igor_r_shafarevich_alexey_o_remizov_linear_algebra_and_geometry.html}

\bibitem{Golub2013}
G.~H. Golub,
  \href{https://www.ebook.de/de/product/20241149/gene_h_golub_matrix_computations.html}{{Matrix
  Computations}}, J. Hopkins Uni. Press, 2013.
\newline\urlprefix\url{https://www.ebook.de/de/product/20241149/gene_h_golub_matrix_computations.html}

\bibitem{Burrus2013}
C.~S. Burrus, {Basic Vector Space Methods in Signal and Systems Theory},
  \url{http://cnx.org/contents/037514b9-80e0-4f2a-8c4e-770950181202@5.2/} (10
  Jan. 2013).

\bibitem{Nelson1992}
P.~A. Nelson, F.~Orduna-Bustamante, H.~Hamada,
  \href{http://www.aes.org/e-lib/browse.cfm?elib=6165}{{Multi-Channel Signal
  Processing Techniques in the Reproduction of Sound}}, in: Audio Engineering
  Society Conference: UK 7th Conference: Digital Signal Processing (DSP),
  London, UK, 1992.
\newline\urlprefix\url{http://www.aes.org/e-lib/browse.cfm?elib=6165}

\bibitem{Rozanski2017}
M.~Różański, R.~Wituła, E.~Hetmaniok,
  \href{http://www.sciencedirect.com/science/article/pii/S0024379517304135}{{More
  subtle versions of the Hadamard inequality}}, Linear Algebra and its
  Applications 532 (2017) 500 -- 511.
\newblock \href {https://doi.org/https://doi.org/10.1016/j.laa.2017.07.003}
  {\path{doi:https://doi.org/10.1016/j.laa.2017.07.003}}.
\newline\urlprefix\url{http://www.sciencedirect.com/science/article/pii/S0024379517304135}

\bibitem{Scharnhorst2001}
K.~Scharnhorst, Acta Applicandae Mathematicae 69~(1) (2001) 95–103.
\newblock \href {https://doi.org/10.1023/a:1012692601098}
  {\path{doi:10.1023/a:1012692601098}},
  \href{http://dx.doi.org/10.1023/A:1012692601098}{[link]}.
\newline\urlprefix\url{http://dx.doi.org/10.1023/A:1012692601098}

\bibitem{Futura1971}
T.~Futura, {A elementary proof of Hadamard's Theorem}, Mat. Vesnik 8 (23)
  (1971) 267--269.

\bibitem{Fazi2007}
F.~M. Fazi, P.~A. Nelson,
  \href{http://www.aes.org/e-lib/browse.cfm?elib=14302}{{The Ill-Conditioning
  Problem in Sound Field Reconstruction}}, in: Audio Engineering Society
  Convention 123, New York, NY, USA, 2007.
\newline\urlprefix\url{http://www.aes.org/e-lib/browse.cfm?elib=14302}

\bibitem{Gover2010}
E.~Gover, N.~Krikorian,
  \href{http://www.sciencedirect.com/science/article/pii/S0024379510000443}{Determinants
  and the volumes of parallelotopes and zonotopes}, Linear Algebra and its
  Applications 433~(1) (2010) 28 -- 40.
\newblock \href {https://doi.org/https://doi.org/10.1016/j.laa.2010.01.031}
  {\path{doi:https://doi.org/10.1016/j.laa.2010.01.031}}.
\newline\urlprefix\url{http://www.sciencedirect.com/science/article/pii/S0024379510000443}

\bibitem{Williams1999}
E.~G. Williams, {Fourier Acoustics: Sound Radiation and Nearfield Acoustical
  Holography}, ACADEMIC PR INC, 1999.

\bibitem{Takeuchi2001}
T.~Takeuchi, M.~Teschl, P.~Nelson,
  \href{http://www.aes.org/e-lib/browse.cfm?elib=10065}{{Subjective evaluation
  of the Optimal Source Distribution system for virtual acoustic imaging}}, in:
  Audio Engineering Society Conference: 19th International Conference: Surround
  Sound - Techniques, Technology, and Perception, 2001.
\newline\urlprefix\url{http://www.aes.org/e-lib/browse.cfm?elib=10065}

\bibitem{Takeuchi2001a}
T.~Takeuchi, P.~Nelson,
  \href{http://www.aes.org/e-lib/browse.cfm?elib=9954}{Optimal source
  distribution system for virtual acoustic imaging.}, in: Audio Engineering
  Society Convention 110, 2001.
\newline\urlprefix\url{http://www.aes.org/e-lib/browse.cfm?elib=9954}

\end{thebibliography}
